\pdfoutput=1
\documentclass[prb,twocolumn,amsmath,amssymb,superscriptaddress]{revtex4}

\usepackage{graphicx}
\usepackage{dcolumn}
\usepackage{bm}

\begin{document}

\title{Analytic solutions to the central spin problem for NV centres in diamond}

\author{L. T. Hall}
 \email{lthall@physics.unimelb.edu.au}
\affiliation{Centre for Quantum Computation and Communication Technology, School of Physics, University of Melbourne, Victoria 3010, Australia}%
\author{J. H. Cole}
\affiliation{Chemical and Quantum Physics, School of Applied Sciences, RMIT University, Melbourne, 3001, Australia}
\author{L. C. L. Hollenberg}
\affiliation{Centre for Quantum Computation and Communication Technology, School of Physics, University of Melbourne, Victoria 3010, Australia}%

\begin{abstract}
Due to interest in both solid state based quantum computing architectures and the application of quantum mechanical systems to nanomagnetometry, there has been considerable recent attention focused on understanding the microscopic dynamics of solid state spin baths and their effects on the coherence of a controllable, coupled central electronic spin. Many analytic approaches are based on simplified phenomenological models in which it is difficult to capture much of the complex physics associated with this system. Conversely, numerical approaches reproduce this complex behaviour, but are limited in the amount of theoretical insight they can provide. Using a systematic approach based on the spatial statistics of the spin bath constituents, we develop a purely analytic theory for the NV central spin decoherence problem that reproduces the experimental and numerical results found in the literature, whilst correcting a number of limitations and inaccuracies associated with existing analytical approaches.
\end{abstract}


\maketitle




\section{Introduction}\label{SBAIntro}
\begin{figure}
\centering
  \includegraphics[width=\columnwidth]{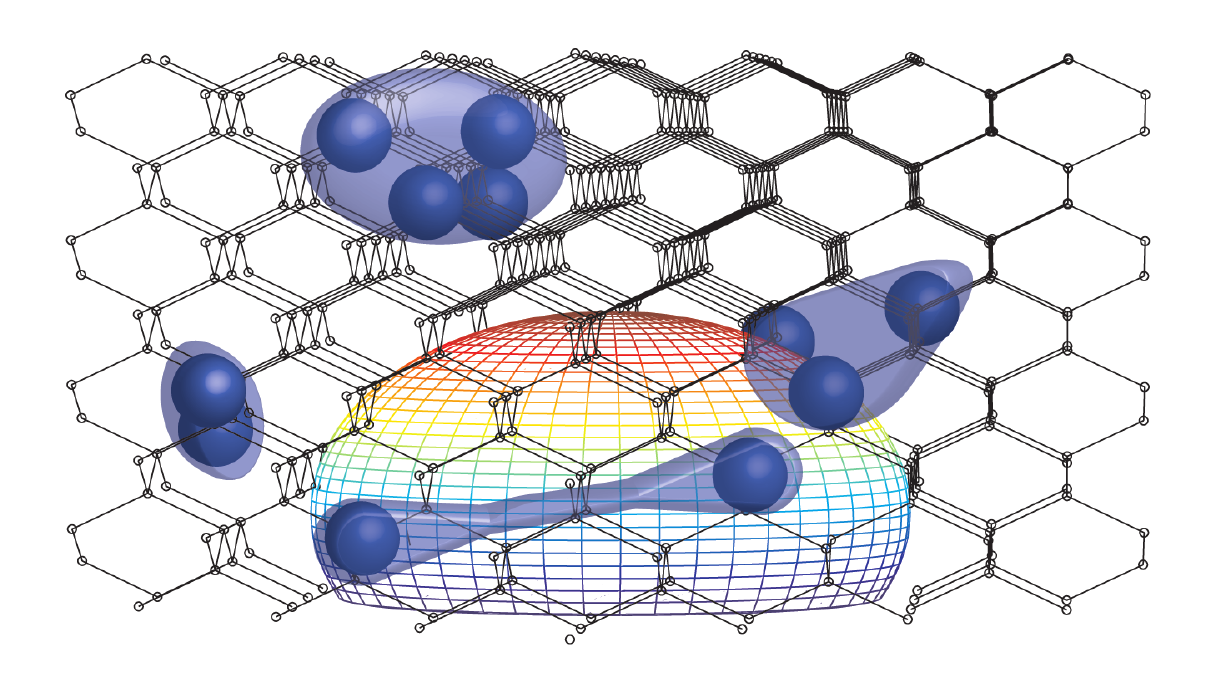} \\
   \caption[Schematic of the central spin problem]{Schematic of the central spin problem showing a central spin coupled to a number of adjacent environmental spins.
 }\label{SBAFig1}
\end{figure}

The central spin problem refers to a special class of open quantum systems, in which a central spin (S) interacts with a large number of strongly coupled spins in the environment (E), as depicted in Fig.\,\ref{SBAFig1}. This results in an irreversible loss of quantum information from the central system, as quantified by the decay of phase coherence between its corresponding basis states, such that certain elements of the system are no longer able to interfere with each other. In the case of pure dephasing processes, for which there is no energy transfer between the system and the environment, this results in a damping of the off-diagonal terms of the associated reduced density matrix of S,
\begin{eqnarray}
  \mathrm{Tr}_\mathrm{E}\left\{ \rho(t) \right\}_{ij}  &=& L_{ij}\mathrm{Tr}_\mathrm{E}\left\{ \rho(0) \right\}_{ij},
\end{eqnarray}where $\mathrm{Tr}_\mathrm{E}\left\{\ldots \right\}$ denotes a trace over the environmental degrees of freedom of the density matrix $\rho$, $L_{ij} = \exp\left(-\Lambda_{ij}(t)\right)$ is often referred to as the decoherence envelope, and $\Lambda_{ij}$ as the decoherence function. It is the derivation of these quantities in the context of the central spin problem with which we concern ourselves in this work.

This problem has received renewed attention over the last decade, due in no
small part to localised electrons in solids being promising candidates for
qubits in quantum computation, metrology and communication systems; a result
of their long coherence times, ease of quantum control and already well
established fabrication techniques. In the context of quantum information processing,
the utility of these systems hinges upon the requirement for
spin coherence times to be sufficiently long to ensure that the necessary
number of quantum operations \cite{Pre98} can be performed within the associated
coherence time. Examples of such systems that have been
suggested as building blocks for quantum computer architectures include
spins qubits in quantum dots\cite{Los98,Ba010}, donor impurities\cite{Kan98,Sou04,Hol06,Mor11}, and Nitrogen-Vacancy (NV)
centers in diamond\cite{Wra06}.
In the context of metrology, with particular regard to parameter estimation, the NV centre has emerged as a unique physical platform for nanoscale magnetometry\,\cite{Deg08,Tay08,Bal08,Maz08,Hal10a}, nano-NMR\cite{Sta13,Mam13,Per13}, bio-imaging\,\cite{Hal10b,McG11,Hal12}, electrometry\cite{Dol11}, thermometry\,\cite{Neu13}, and decoherence imaging\,\cite{Col09,Hal09,Ste12,Kau13,McG13}. In each case, the associated sensitivity is ultimately limited by the coherence properties of the NV spin that arise from the strong coupling to the surrounding bath of electron and/or nuclear spins. In all of these applications and platforms, a comprehensive understanding of the central
spin problem is therefore necessary to make accurate predictions of the
quantum properties and behaviour of the central spin arising from the
material properties of the surrounding environment.

The first modern approach to this problem, in the context of phosphorus donors in silicon, involved treating the combined effect of the spin bath environment on the central spin as a semi-classical magnetic field whose dynamic properties were intended to mimic the magnetic dipole flip-flop processes taking place amongst the environmental spins\cite{Sou03,Sou03b}. Despite not accounting for the effect of the central spin on the surrounding environment, this approach still finds considerable application today, particularly in the NV community\cite{Tay08,Han08,Maz08b,Dob09,Lan10,Lan11,Wan12}. In order to account for the full interaction between the central spin and its environment, quantum cluster expansion\cite{Wit05,Wit06,Sai07}, nuclear pair-wise\cite{Yao06,Yao07,Liu07}, correlated cluster expansion\cite{Yan08,Yan09}, and disjoint cluster expansion\cite{Maz08b} models have been developed, in which the environment is systematically clustered into groups of strongly interacting spins, with each order of the cluster hierarchy corresponding to successively weaker, and hence less important, interactions. In addition, master equation approaches in which all hyperfine coupling constants are assumed to be identical have been developed\,\cite{Bar11}, with subsequent developments accounting for non-uniform couplings\,\cite{Bar12}.

\begin{figure}
\centering
  \includegraphics[width=\columnwidth]{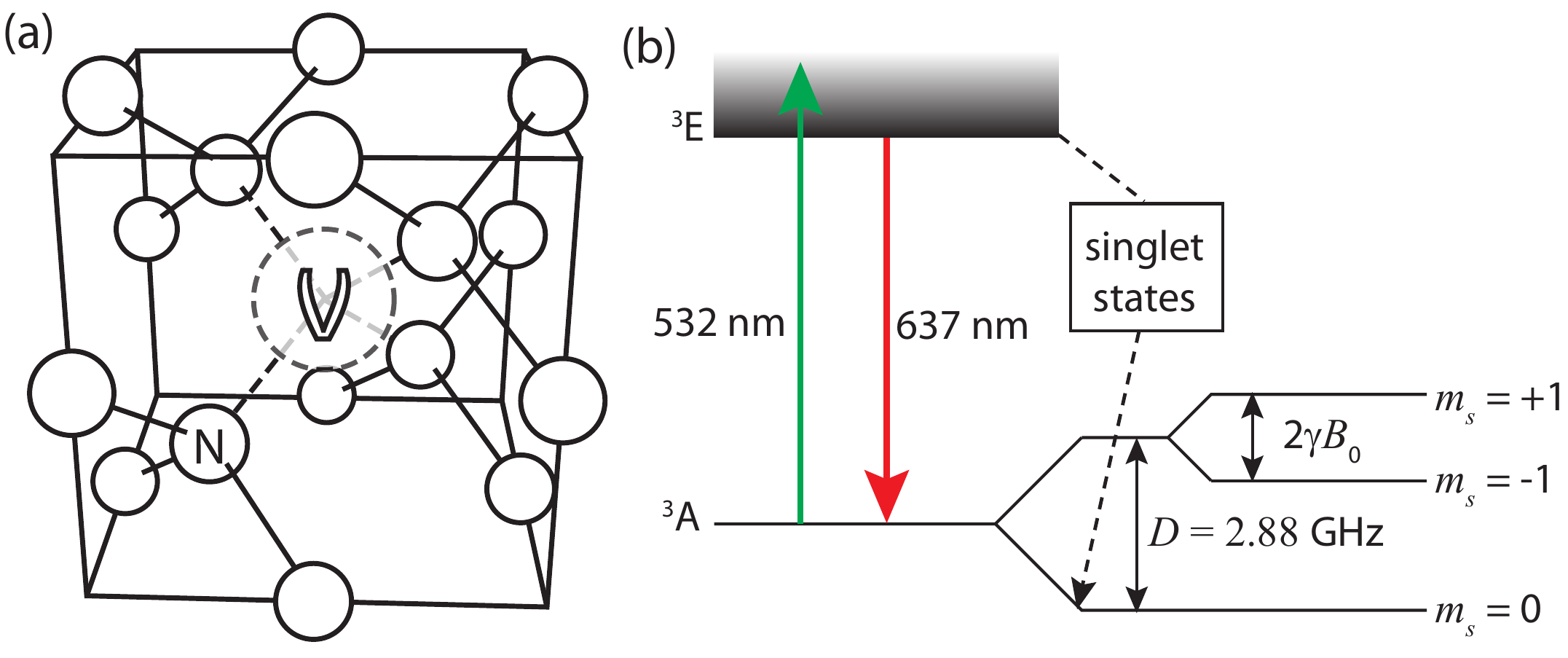} \\
   \caption[Overview of the Nitrogen Vacancy (NV) centre in the context of the central spin problem.]
   {(a) The Nitrogen Vacancy (NV) centre point defect in a diamond lattice,
comprised of a substitutional atomic nitrogen impurity (N) and an adjacent
crystallographic vacancy (V). (b) The NV ground state spin sublevels are separated by $D=2.87$\, GHz. Upon optical excitation at 532 nm, the
population of the $\bigl|0\bigr\rangle$ state may be readout by monitoring the intensity of the emitted red light.
 }\label{NVFig1}
\end{figure}

The Nitrogen Vacancy (NV) centre (see ref.\,\onlinecite{Doh13} for an extensive review) is a point defect in a diamond lattice
comprised of a substitutional atomic nitrogen impurity and an adjacent
crystallographic vacancy (Fig.\,\ref{NVFig1}\,(a)). The energy level scheme of the C$_{3v}$-symmetric NV system
(Fig.\,\ref{NVFig1}\,(b)) consists of ground ($^3$A), excited ($^3$E) and
meta-stable ($^1$A) electronic states. The ground state spin-1 manifold has 3
spin sub-levels $\left(\bigl|\,0\bigr\rangle,\bigl|\pm1\bigr\rangle\right)$,
which in zero field are split by 2.88\,GHz. An important property of the NV
system is that under optical excitation the spin levels are distinguishable
by a difference in fluorescence, hence spin-state readout is achieved by
purely optical means\cite{Jel02,Jel06}. The degeneracy
between the $\bigl|\pm1\bigr\rangle$ states may be lifted with the
application of a background field, with a corresponding separation of
17.6\,MHz\,G$^{-1}$ permitting all three states to be accessible via
microwave control, however the $\bigl|\pm1\bigr\rangle$ states are not
directly distinguishable from one another via optical means. By isolating
either the $\bigl|\,0\bigr\rangle\leftrightarrow\bigl|+1\bigr\rangle$ or
$\bigl|\,0\bigr\rangle\leftrightarrow\bigl|-1\bigr\rangle$ transitions, the NV spin system constitutes a controllable, addressable spin qubit.

The large zero field splitting is fortuitous in the present context, in that it renders typical NV-environment spin-spin couplings (of roughly MHz) unable to cause transitions in the ground-state spin triplet manifold. Whilst the weak spin-orbit coupling to crystal phonons (and low phonon density, owing to the large Debye temperature of diamond) leads to longitudinal relaxation of the spin state on timescales of roughly $T_1\sim10$\,ms at room temperature, the lateral relaxation ($T_2$) of the NV spin is determined by the dipole-dipole coupling to other spin impurities in the diamond crystal and can occur on timescales of 0.1-1\,$\mu$s in naturally occurring type-1b diamond. Such coherence times may be greatly extended with the use of higher grade diamond crystal, or with the application of spin-echo\cite{Jel02a} or higher order dynamic decoupling sequences to suppress the effect of the surrounding spin bath on the NV spin. For example, the use of `ultra-pure' diamond crystal with parts-per-billion (ppb) concentrations of nitrogen electron donor impurities (as compared with naturally occurring concentrations of parts-per-million or more) leads to $^{13}$C nuclear spin limited spin-echo coherence times of $T_2\sim 100\,\mu\mathrm{s}-1$\,ms. Such results are by no means fundamental however, and may be further improved with the use of isotopically pure diamond crystal with reduced $^{13}$C concentrations. As such, $T_2$ may, at least in principle, be as long as the longitudinal relaxation time, $T_1$.
These relatively narrow
spectral properties of the NV$^{-}$ ground state, together with its room temperature operation and optical readout, make it an ideal candidate
for both single spin-based magnetometry and electrometry. As such, the NV spin coupled to a surrounding bath of $^{13}$C nuclear spins of natural abundance (1.1\%) is the primary physical system with which we concern ourselves for this study.

 Traditionally, the theoretical analysis of the central spin problem has been based on phenomenological assumptions regarding the self-interaction dynamics of the surrounding spin bath, namely by replacing the bath with a classical Ornstein-Uhlenbeck noise source\,\cite{And53,Kla62}. This noise gives rise to fluctuations in the Larmor frequency of the central spin and leads to an eventual dephasing between its initially coherent basis states, a
process referred to as spectral diffusion. Such approaches involve making an ad-hoc assumption of a decaying exponential form
for the autocorrelation function of the effective magnetic dipole field, with the decay rate being deduced from the effective environmental flip-flop rates\,\cite{Sou03,Sou03b}. Despite the extensive application this theory has found throughout the
literature\,\cite{Tay08,Maz08b,Dob09,Lan10,Lan11,Wan12}, three major problems
persist: 1.\,\,There is no theroetical/experimental/numerical reason to suggest
      the phenomenological assumption of an exponential form for the
      autocorrelation function should be made; 2.\,\,Upon assuming a particular form for the autocorrelation function,
      the task of determining the correlation time (assuming the decay can, in fact, be
      described by a single time constant) still remains; and, 3.\,\,Inherent in this is
the assumption that the central spin has no influence on the evolution of the
environment constituents.

With regard to the first problem, the assumption of an exponential correlation function leads to a cubic exponential decay in the coherence of the central spin under a spin-echo pulse sequence for times shorter than the autocorrelation time (see ref.\,\onlinecite{Sou09} for a review). However, there have been a significant number of
theoretical results in the literature suggesting that this decay may in fact show a quartic exponential
dependence\cite{Wit05,Yao06,Wit06,Sou09,Ren12}. Such behaviour can only arise from a Gaussian
shaped autocorrelation function (or at least quadratic on relevant timescales), casting some doubt on the assumed
exponential form. In addition, the numerical work in ref.\,\onlinecite{Wit05} shows that numerical computation of the combined effect of many randomly distributed clusters leads to an approximately Gaussian decay of coherence, the likes of which have been observed experimentally\cite{Bal09}. These problems are addressed and clarified in this work, as our results show that the spatial distribution of spins around the central spin has a significant effect on the analytic form of the autocorrelation function.
This is critical to the development of spin-based quantum technologies, as there have been many quantitative predictions of better performance with the use of pulse-based microwave control schemes\,\cite{Tay08,Lee08,Hal10a}, and the exact analytic form of the spectral cutoff was shown to directly
affect the performance of such schemes\,\cite{Cyw08,Uhr08}.

Also outstanding is the determination of the environmental autocorrelation time. One of the earliest modern attempts at deriving this timescale from
microscopic physical processes was given in ref.\,\onlinecite{Sou03b}, in which each
magnetic dipole coupled nuclear spin pair (consisting of spins $m$ and $n$) was treated as a bistable fluctuator, where the
number of transitions between states in a given time interval $t$ is treated
as a Poissonian variable with parameter $t/T_{mn}$. The effective flip-flop
rate of the pair, $1/T_{mn}$ could then be calculated using their mutual
dipolar coupling strength via perturbation theory, resulting in a linear exponential decay of the autocorrelation function. However, this still
requires certain phenomenological assumptions to be made about the associated
density of states, and does not address the microscopic reasons behind how
the $T_{mn}$ quantities are distributed. Adopting this approach in the context of quantum sensing applications would mean that, for a given $T_2$ measurement, one would be lead to infer that the associated correlation time of the environment is three orders of magnitude longer than its true value. As an example, using the treatment of the nuclear spin bath in ref.\,\onlinecite{Tay08}, typical coherence times of $T_2^* = 1\,\mu$s and $T_2 = 300\,\mu$s would imply a correlation time of $T_c = {T_2^3}/\left[{6\left(T_2^*\right)^2}\right] = 4.5\,$s. This is in stark contrast to what would be expected from an examination of the average nuclear-nuclear coupling strength of $1/nb\sim 40$\,ms, based on an average impurity density $n \approx 2\,\mathrm{nm}^{-3}$, and indeed the correlation times of $T_c\approx10\,$ms calculated in this work. In fact, as we will show here, a
magnetisation conserving two spin flip-flop model must have an
autocorrelation function with zero derivative at $t=0$, and hence cannot
produce the linear behaviour exhibited by a pure exponential decay.

In contrast to the ad-hoc fitting of data to phenomenological models which do not account for the influence of the central spin on its surrounding environment, fully quantum mechanical approaches to the problem have been developed over
the last 5 years using cluster expansion\,\cite{Wit05,Wit06,Sai07} and correlated cluster expansion\,\cite{Yan08,Yan09} methods. Here the randomly
distributed spins are aggregated into small, strongly
interacting groups, with the latter showing better convergence in cases where the decoherence time of the central spin is comparable to, or longer than, the autocorrelation time of the environment, as is the case with a bath of electron spins. In the opposite regime, as would the case for an electron spin coupled to a nuclear spin bath, these two approaches agree, and to lowest non-trivial
order they are in accordance with earlier nuclear pair-correlation approaches\,\cite{Yao06,Yao07,Liu07}. In this limit, at least for short times, all of these approaches are shown to be consistent with a
quartic-exponential decay, the likes of which may also be deduced using a generalised semi-classical argument\,\cite{Hal09}.

In cases of relatively strong hyperfine interactions resulting from low magnetic field regimes, direct dipole-dipole couplings are either treated as a perturbation\,\cite{Cyw09b,Sai07}, or ignored completely\,\cite{Cyw09,Cyw10,Bar11,Bar12}. Here, the dominant interaction between environmental constituents is due to hyperfine-mediated flip-flops, resulting from environmental spins becoming increasingly coupled to the lateral components of the central spin as the magnetic field strength decreases. Such effects are negligible in the case of NV centres in diamond, owing to its 2.88\,GHz zero-field splitting. This approach is well suited to cases where the central spin always has a non-zero projection along its quantisation axis ($m_s = \pm\frac{1}{2},\,\pm\frac{3}{2},\ldots$), as is the case of the spin-$\frac{1}{2}$ Si:P and Ga:As systems, thus causing many of the bath spins to be off-resonance and hence unable to flip with each other, but is not valid in integer spin systems where the $m_s = 0$ state is appreciably populated, as is predominantly the case with the NV centre. Under these conditions, environmental spins are free to evolve exclusively under their mutual couplings, and information encoded onto them by the central is free to propagate throughout the environment. As such, these theories do not account for the irreversible leakage of quantum information from the central
spin to distant environmental components. This approach is a reasonable approximation for times much shorter than the autocorrelation time of the environment, but not on timescales over which environmental interactions are appreciable. In the effective Hamiltonian models above\,\cite{Cyw09b,Cyw10} or master equation approaches\,\cite{Bar11} all hyperfine coupling constants are assumed to be identical, meaning the effects of the hyperfine distribution on the decoherence behaviour have not been addressed. Non-uniform hyperfine couplings were treated in ref.\,\onlinecite{Bar12}, however the assumption of non-interacting nuclei renders this approach unsuitable for NV centres.

 Finally, we remark that any short time expansion is only valid for times shorter than the reciprocal of the strongest dipolar coupling frequencies in the system, and of particular concern is that any
two spins can be found arbitrarily close together (or effectively so on the
length scales of the system), making an expansion in low orders of
these couplings diverge. In this limit, it is the dipolar interaction between environmental spins that sets their quantisation axis, not the Zeeman interaction, invalidating the assumption of each cluster's magnetisation being conserved with respect to the global $z$ axis. This is another instance where consideration of the spatial distribution of spin impurities becomes important, and despite being able to describe the decoherence in the compact forms given by the works described above, no
discussion has been made regarding the statistical distributions of the spin-spin coupling strengths. Instead, one is forced to resort to Monte-Carlo based numerics at this point.
The possible outcomes for various
realisations of spatial distributions of spin impurities for the case of an
NV centre coupled to a nuclear spin bath have been numerically investigated in refs.\,\onlinecite{Maz08b,Ren12,Wit12}, and that for electron donors and quantum dots in silicon in ref.\,\onlinecite{Wit12}. An extensive numerical study of the magnetic dependence of the coherence time of an NV centre on the strength of the applied background magnetic field was conducted in ref.\,\onlinecite{Ren12}, taking both realistic hyperfine distributions and environmental spin-spin interactions into account. A fully analytic, quantum mechanical description of the effects of the entire range of magnetic field strengths on a central spin coupled to a completely randomly distributed spin bath is presented in this work. In what follows, we focus primarily on the case of an NV centre interacting with its native 1.1\% $^{13}$C nuclear spin bath.

\section{Theoretical background}\label{SecTheorBG}
  The Hamiltonian describing this system is given by
  \begin{eqnarray}
 \mathcal{H} = \mathcal{H}_\mathrm{S} + \mathcal{H}_\mathrm{SE} + \mathcal{H}_\mathrm{EZ} + \mathcal{H}_\mathrm{EE}\label{HamilFull},
  \end{eqnarray}
where $\mathcal{H}_\mathrm{S}$ is the self-Hamiltonian of the central
electron spin, which may include the coupling of the NV spin to its proximate nitrogen nuclear spin, as well as zero field and Zeeman splittings. The hyperfine interaction between the central spin (S) and the
environment (E) is described by $\mathcal{H}_\mathrm{SE}$, which in the present
context is a point-dipole interaction, but may also include Fermi-contact
interactions in other systems. This is described by
\begin{eqnarray}
  \mathcal{H}_\mathrm{SE} &=& \sum_i\frac{a}{R_i^3}\left[ \vec{\mathcal{S}}\cdot\vec{\mathcal{E}}_i-3\frac{ \left(\vec{\mathcal{S}}\cdot\mathbf{R}_i\right)\left(\mathbf{R}_i\cdot\vec{\mathcal{E}}_i\right)}{R_i^2}\right]\label{HyperfineCoupling} ,
\end{eqnarray}
where $\vec{\mathcal{S}}$ and $\vec{\mathcal{E}}_i$ are the spin-vector operators for the NV spin and the
  $i^\mathrm{th}$ environmental spin, $\mathbf{R}_i$ is their mutual separation, and $a = \frac{\mu_0}{4\pi\hbar}\mu_\mathrm{S}\mu_\mathrm{E}$. The magnetic moments of the NV and environmental spins are denoted $\mu_\mathrm{S}$ and $\mu_\mathrm{E}$ respectively. The large zero-field splitting is some three orders of magnitude greater than any other coupling in this system, allowing us to ignore any coupling to the lateral components $\left(\mathcal{S}_x\,\,\mathrm{and}\,\,\mathcal{S}_y\right)$ of the NV spin.

  The Zeeman (Z) interaction of the environmental spins is described by $\mathcal{H}_{\mathrm{EZ}} = \sum_i\vec{\mathcal{E}}_i\cdot\vec{\omega}_i$, where $\vec{\omega}_i = \gamma_\mathrm{E}\mathbf{B}_0$ describes the Zeeman field felt by spin $i$, having gyromagnetic ratio $\gamma_\mathrm{E} = \mu_\mathrm{E}/\hbar$, due to a background field $\mathbf{B}_0$.

  The nuclear spin-spin interactions (E) are described by
 \begin{eqnarray}
  \mathcal{H}_\mathrm{EE} &=& \sum_{j<i}\frac{b}{r_{ij}^3}\left[ \vec{\mathcal{E}}_i\cdot\vec{\mathcal{E}}_j-3\frac{ \left(\vec{\mathcal{E}}_i\cdot\mathbf{r}_{ij}\right)\left(\mathbf{r}_{ij}\cdot\vec{\mathcal{E}}_j\right)}{r_{ij}^2}\right] ,
\end{eqnarray}
where $\mathbf{r}_{ij}$ is the mutual separation of spins $i$ and $j$, and $b = \frac{\mu_0}{4\pi\hbar}\mu^2_\mathrm{E}$.

In the case of large Zeeman couplings, some transitions between environmental spin states due to the hyperfine and dipolar interactions will be disallowed due to energy conservation, ensuring that the total axial magnetisation of the spins involved in the interaction is conserved. However, at low fields, the energy cost for these transitions may be easily paid for by these interactions, meaning that axial magnetisation need not be conserved. In what follows, we refer to (non)axial magnetisation conserving transitions as being `(non)secular'.

\begin{figure}
  \includegraphics[width=8cm]{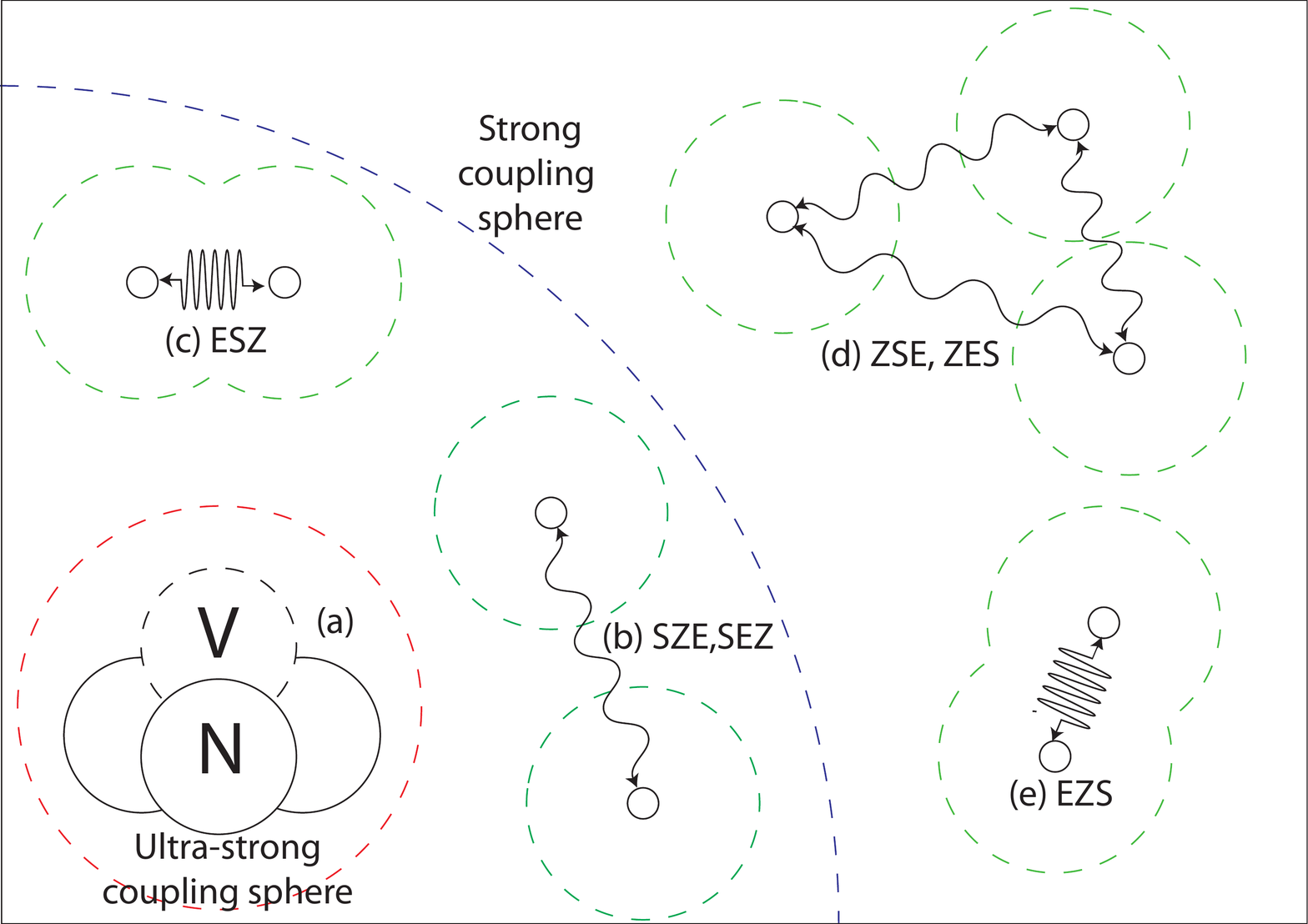}\\
  \caption[Schematic showing the parameter regimes relevant to the central spin problem.]
  {Schematic showing the parameter regimes relevant to the central spin problem.
  (a) Ultra strong coupling region, in which the interaction between the NV centre and an adjacent
  spin is stronger than its 2.88\,GHz zero field splitting. This region is not considered in this
  work. (b) and (c) Strong coupling region, in which the coupling of the spins to the NV centre is
  stronger than their coupling to a background field. In (b), the spins are weekly coupled to each
  other and is representative of two possible regimes: SEZ and SZE. In (c), the spins are strongly
  coupled to each other and thus represent the ESZ regime. (d) and (e) Weak coupling region, in which
  the coupling of the spins to a background field is greater than their coupling to the NV. In (d),
  the spins are weakly coupled to each other and hence represent the ZSE and ZES regimes. In (e),
  the spins are strongly coupled to each other and represent the EZS regime. }\label{SBARegions}
\end{figure}

\begin{figure}
  \includegraphics[width=8cm]{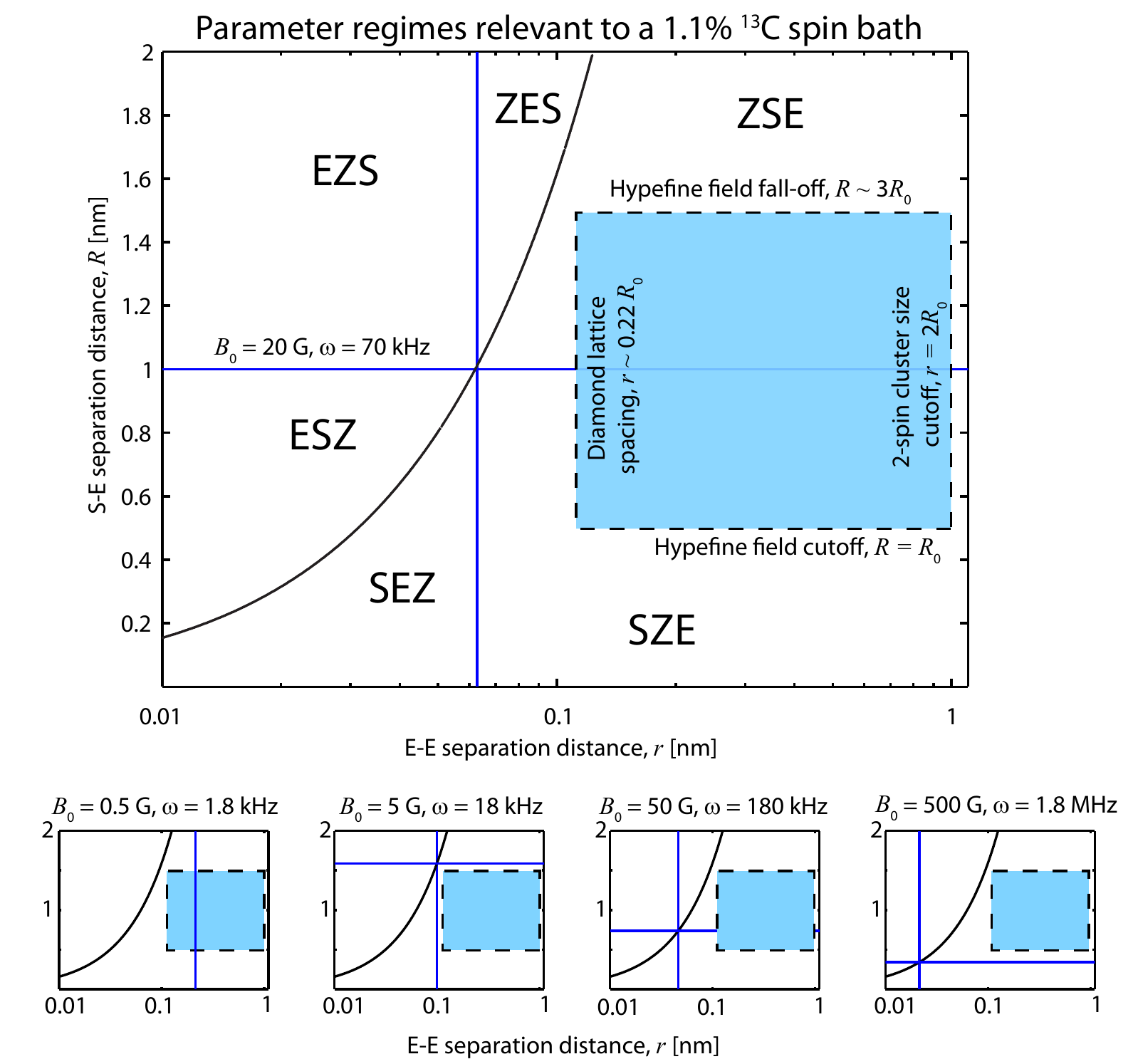}\\
  \caption{Plot showing the locations of the 6 parameter regimes in $R-r$ space for a naturally occurring
1.1\% $^{13}$C nuclear spin bath. The location of the intersection point (blue cross-hairs) changes along
the S=E coupling line for different magnetic field strengths (the main plot depicts the case of 20\,G, and the sequence below
depicts the 0.5\,G,\,\,5\,G,\,\,50\,G,\,\,and\,\,500\,G cases). For this spin bath, we see that only the regimes where $S\gg E$ (ie, SEZ,
  SZE and ZSE) are relevant. }\label{SBANuclearRegionsP}
\end{figure}

\begin{figure*}
  \includegraphics[width=12cm]{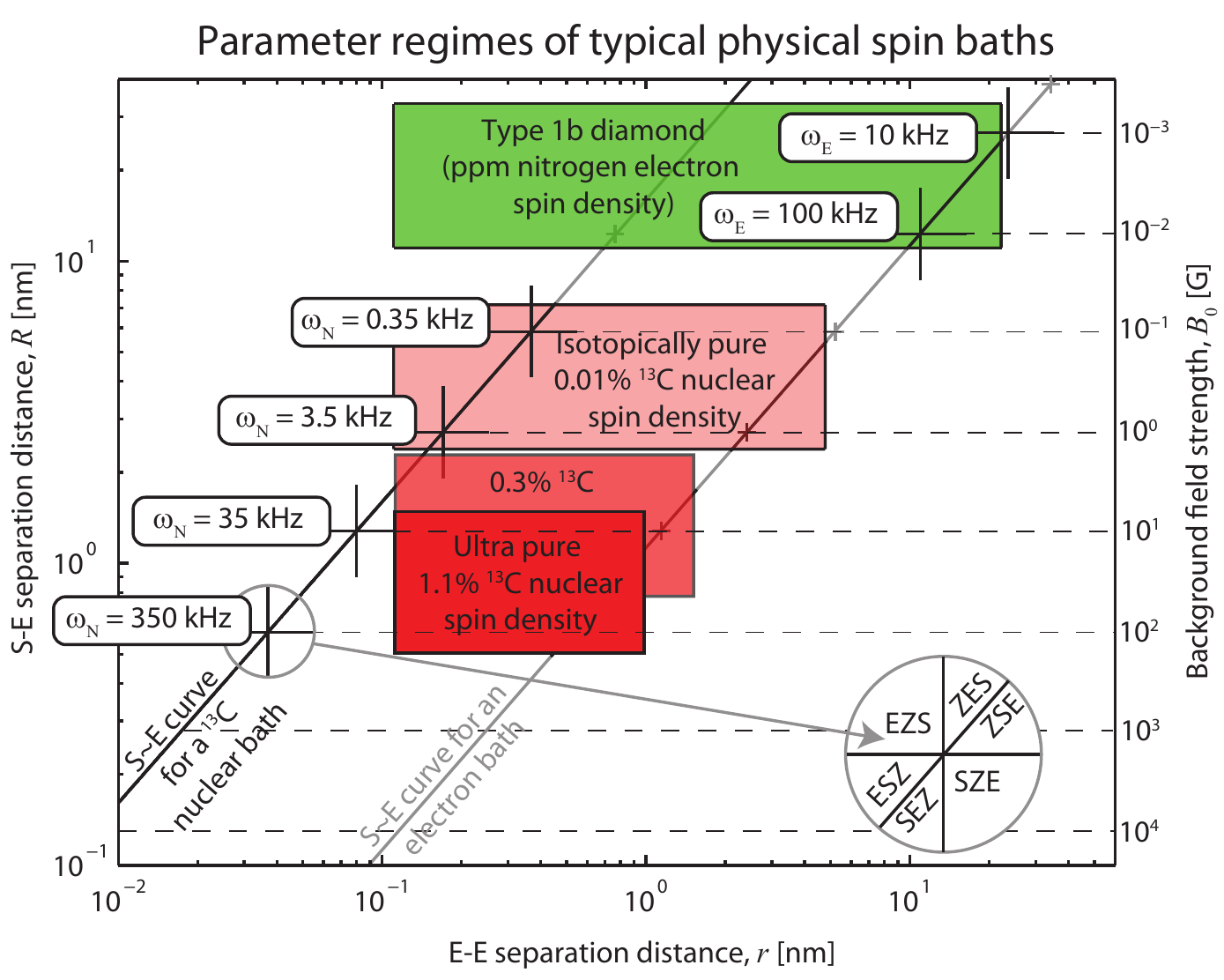}\\
  \caption{Plot showing the locations of the 6 parameter regimes in $R-r$ space for
1.1\%, 0.3\% and 0.01\% $^{13}$C nuclear spin baths and a naturally occurring electron spin bath arising from
 nitrogen donor spins at ppm concentrations. For the latter example, we see that only the regimes where $E\gg S$ (ie, ESZ,
  EZS and ZES) are relevant.}\label{SBANuclearRegionsManyBathsP}
\end{figure*}

For a given spin,
$\mathcal{E}_i$, we may classify its parameter regime in
terms of the relative strengths of the energy scales considered above:
spin-environment coupling (S), environment self coupling (E), and Zeeman splitting (Z),
as determined by the Hamiltonian components, $\mathcal{H}_\mathrm{SE}$, $\mathcal{H}_\mathrm{ZE}$,
and $\mathcal{H}_\mathrm{EE}$ respectively. This gives rise to six distinct parameter regimes, as summarised below, and depicted schematically in
Fig.\,\ref{SBARegions}, and parametrically in
Figs.\,\ref{SBANuclearRegionsP} and \ref{SBANuclearRegionsManyBathsP} for
various examples of physical systems.

\begin{figure}
\centering
  \includegraphics[width=6cm]{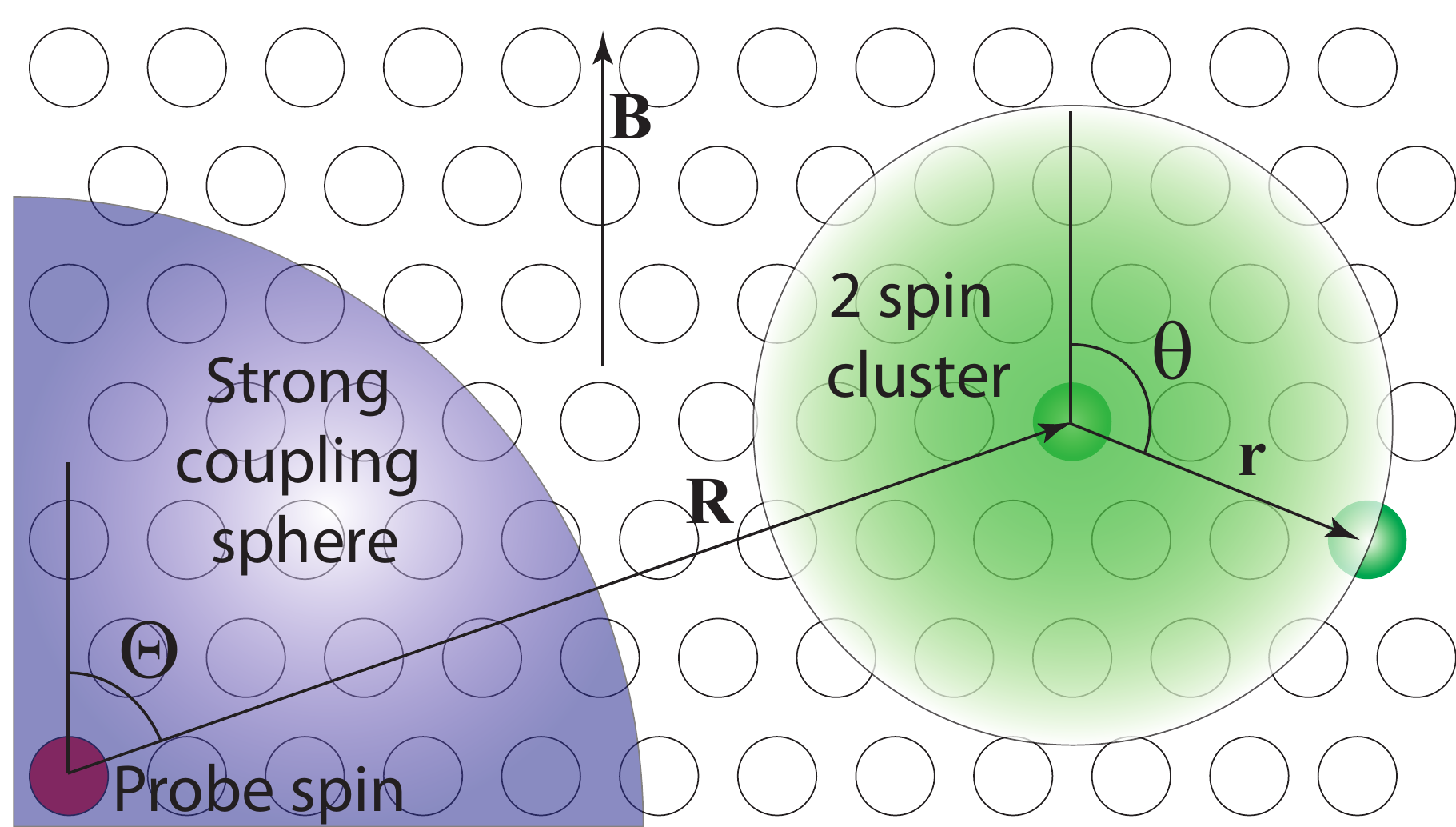}\\
  \caption[Schematic representation of a two spin cluster coupled to a central
  spin.]{Schematic representation of a two spin cluster coupled to a central
  spin. The separation vector between the cluster and the central spin is
  defined by the location of the closest spin, $\mathbf{R}$. The structure of
  the cluster is defined by the separation vector(s) of the cluster
  constituents, $\mathbf{r}$. Evaluation of the ensemble averaged quantities
  requires integration over $\mathbf{R}$, and averaging over $\mathbf{r}$.}\label{ClusterSchem}
\end{figure}

For the sake of brevity, we label these six regions according to
the relative strengths of the environmental couplings. For example, a label
of ZSE (read Z$>$S$>$E) would imply that both S-E and E-E
couplings are secular (a consequence of their quantisation axis being set by
the Zeeman field), and that the spins couple more strongly to the NV than
to each other. Conversely, a label of ESZ would imply that both S-E
and E-E couplings are non secular, and that the spins couple more
strongly to each other than to the NV (see fig.\,\ref{ClusterSchem}). The
geometric boundaries on these regimes are summarised below.

In the \textbf{ZSE} regime, the nuclei are sufficiently far from both the NV and each other that the Zeeman interaction dominates over both the hyperfine and dipolar interactions, ensuring that both classes of interactions must conserve axial magnetisation.
   The $\langle \mathcal{H}_\mathrm{ZE}^2\rangle\gg \langle \mathcal{H}_\mathrm{SE}^2\rangle\gg \langle \mathcal{H}_\mathrm{EE}^2\rangle$ condition yields the following constraints on the geometry of the cluster:
      \begin{eqnarray}
          \left(\frac{a}{\omega}\right)^{1/3}\leq &R&  \leq r \left(\frac{a}{b}\right)^{1/3},\nonumber\\
          R\left(\frac{b}{a}\right)^{1/3}\leq &r& <\infty. \nonumber
      \end{eqnarray}
Clusters in the \textbf{SZE} regime are sufficiently close to the NV to ensure that the hyperfine coupling dominates over the Zeeman interaction, however, the associated nuclei are still far enough apart to ensure that the Zeeman interaction is larger than their mutual dipolar coupling. The $\langle \mathcal{H}_\mathrm{SE}^2\rangle\gg \langle \mathcal{H}_\mathrm{ZE}^2\rangle\gg \langle \mathcal{H}_\mathrm{EE}^2\rangle$ condition ensures that
      \begin{eqnarray}
          0\leq &R&  \leq\left(\frac{a}{\omega}\right)^{1/3},\nonumber\\
          \left(\frac{b}{\omega}\right)^{1/3}\leq &r& <\infty. \nonumber
      \end{eqnarray}
Clusters in the \textbf{SEZ} regime are both sufficiently tightly bound and close to the NV to ensure that both hyperfine and dipolar couplings dominate over the Zeeman interaction, however, the associated nuclei are still far enough apart to ensure that the hyperfine interaction is larger than their mutual dipolar coupling. The $\langle \mathcal{H}_\mathrm{SE}^2\rangle\gg \langle \mathcal{H}_\mathrm{EE}^2\rangle\gg \langle \mathcal{H}_\mathrm{ZE}^2\rangle$ condition ensures that
      \begin{eqnarray}
          0 \leq &R&  \leq r \left(\frac{a}{b}\right)^{1/3},\nonumber\\
         R\left(\frac{b}{a}\right)^{1/3}  \leq &r& \leq \left(\frac{b}{\omega}\right)^{1/3}. \nonumber
      \end{eqnarray}

The remaining three regimes, ZES, EZS and ESZ, may be quantified in an equivalent manner,
however the physical
constraints placed on $R$ and $r$ due to the diamond lattice render
these regimes impossible for a naturally occurring 1.1\% $^{13}$C
nuclear spin bath. This is illustrated in
Fig.\,\ref{SBANuclearRegionsP}, where the possible physical locations an environmental spin
may occupy are shown in the shaded region. The constraints on $r$ arise from
the fact that no two spins may be within a distance of less than one lattice
site from each other; whereas having a large separation
means that there is little chance of the two spins in question being part of the
same cluster (this will be quantified in section \ref{SpatStats}). Similarly, the
constraints on $R$ arise from the lattice spacing,
and the fact that the hyperfine
field vanishes at large $R$.
In particular, we see that, whilst changing the background field
strength changes the relative number of spins in the ZSE, SZE and SEZ
regimes, a 1.1\% $^{13}$C nuclear spin bath will never occupy any regime for
which E$\gg$S. That is, in solving for this particular physical system, we need not
consider any of the ZES, EZS or ESZ regimes.

This will not be true for all spin baths however, as shown by Fig.\,\ref{SBANuclearRegionsManyBathsP}, in which 1.1\%, 0.3\% and 0.01\%,$^{13}$C
nuclear spin baths are considered, together with naturally occurring type-1b
diamond containing an electron spin bath due to nitrogen donor impurities at
parts-per-million (ppm) concentrations. The latter example presents a stark
contrast to the 1.1\%,$^{13}$C case, as the only appreciable regimes that
need be considered here are ZES, EZS or ESZ, a consequence of the
comparatively strong electron-electron coupling of the environmental spins, however electron spin baths are not the focus of this work.

In the present context, we define the
decoherence of the NV as the loss of coherence between the $\bigl|0\bigr\rangle$ and $\bigl|+1\bigr\rangle$ states of
the NV spin. This corresponds to the decay of the off diagonal terms in the
corresponding density matrix, and may be computed directly from the lateral
(in the $x-y$ plane) projection of the NV magnetisation vector,
  $S = \bigl\langle \mathcal{S}_x + i\mathcal{S}_y \bigr\rangle$.
As the decoherence generally leads to a decay of this signal, we
define a `decoherence function', $\Lambda(t)$, such that we may write
  $L = e^{-\Lambda(t)}$,
and we refer to the time taken to reach $\Lambda(t)=1$ as the
`coherence time.' Our task is then to determine the functional form of
$\Lambda$ in response to the separate parameter regimes discussed above.

In order to determine the full decoherence behaviour due to all spins in the environment, we may break up the environment into separate clusters consisting of strongly
interacting spins and ignore the comparatively weak interactions between
adjacent clusters. By virtue of the large zero-field splitting, and the maximum hyperfine coupling for an adjacent $^{13}C$ being of order 40\,MHz, the NV spin exists in a `pure dephasing' regime in which only the relative phases of the spin states change and the respective populations do not. This will not be true for electron spin baths in the high density limit, however such systems are beyond the scope of this work. We may then write the Hamiltonian as $  \mathcal{H} =
\sum_k\mathcal{H}_k,$ where $\mathcal{H}_k$ acts only on the $k^\mathrm{th}$
cluster.
Since all of the $\mathcal{H}_k$s commute, the time evolution
operator may be factorised as $ \mathcal{U}(t) = \prod_k\mathcal{U}_k(t)$. This implies that the full decoherence function is then simply a sum over all geometric configurations and locations of the environmental clusters. We note that this result will break down near the anti-crossing of the NV spin states at roughly 1024\,Gauss, as the nuclear spins will be able to exchange energy with the NV spin. However, as the linewidth of the spin bath is of order kHz, this effect corresponds to a very narrow magnetic field interval of roughly 0.1\,Gauss and is therefore ignored.

As we will show, incorporation of higher-order clustering has no effect on the leading order behaviour of the decoherence function, and is thus not important for the decoherence behaviour in the presence of low-order pulse sequences such as FID and spin-echo. In the following, we examine the decoherence functions associated with individual clusters of environmental spins, and then move on to discuss the statistics associated with how the environmental spins are distributed spatially. These distributions will then be used to compute the full decoherence behaviour due to all clusters in the environment.

\section{Single cluster dynamics and decoherence}\label{SingleClusterQuantities}
As mentioned above, in analysing the decoherence due to the native 1.1\% $^{13}$C spin bath, we need only consider the three `strong coupling' regimes (ZSE, SZE and SEZ), in which the interaction between the central spin and the environment,
$\mathcal{H}_\mathrm{SE}$, dominates over the spin-spin interactions within the environment, $\mathcal{H}_\mathrm{EE}$.
It is important to note that, due to the secular approximation imposed on the NV centre, this
dominant S-E interaction is only apparent when the NV spin is in the
$\bigl|+1\bigr\rangle$ state. In this case the large hyperfine interaction results in the nuclei having a large mismatch in their respective transition frequencies, meaning their comparatively weak mutual dipole interaction will be unable to cause a mutual flip-flop (this effect will we discussed in detail in section\,\ref{SingleClusterCorrFns}). This is somewhat advantageous, as the exponentiation of the full Hamiltonian,
inclusive of all $\mathcal{H}_\mathrm{SE}$, $\mathcal{H}_\mathrm{ZE}$ and $\mathcal{H}_\mathrm{EE}$ terms, is not analytically possible in
general. On the other hand, when the NV spin is in the $\bigl|0\bigr\rangle$ state, there will be no hyperfine coupling, and the environmental
evolution will be self governed. This means that the environmental spins are free to
evolve unperturbed according to $\mathcal{H}_{\mathrm{ZE}}$ and
$\mathcal{H}_{\mathrm{EE}}$.

Since the evolution is so heavily dependent on the NV spin state, we can
project this Hamiltonian along both basis states.
Thus,
\begin{eqnarray}
\mathcal{H}
  &\equiv&  \bigl|1\bigr\rangle \bigl\langle 1 \bigr|\mathcal{H}_1 + \bigl|0\bigr\rangle\bigl\langle 0 \bigr|\mathcal{H}_0.
\end{eqnarray}
Because no hyperfine coupling exists when the NV is in the
$\bigl|0\bigr\rangle$ state, projection onto the distinct NV states allows us
to distinguish between the Hamiltonians associated with the
$\bigl|0\bigr\rangle$ and $\bigl|+1\bigr\rangle$ states, namely
$\mathcal{H}_{0}$ and $\mathcal{H}_{1}$ respectively.

An experiment in which the central spin is left to evolve under the action of
the environment alone is referred to as a Free-Induction-Decay (FID), and the
majority of the associated dephasing may be attributed to inhomogeneous
broadening from quasi static components of the spin bath. In the case of an
NV centre in either an electron or a nuclear spin bath, this broadening is
typically of the order of a few MHz, equating to an effective magnetic field
of a few $\mu$T. As such, coherence times are typically of the order of
$T_2^*\sim1\,-1\,0\,\mu$s, depending on the sample at hand. For such an
experiment, the time evolution operator is given by,
\begin{eqnarray}
  \mathcal{U}_\mathrm{fid}(t)
    &=& \bigl|1\bigr\rangle \bigl\langle 1 \bigr|\otimes\exp\left(-i\mathcal{H}_1 t\right) + \bigl|0\bigr\rangle \bigl\langle 0 \bigr|\otimes\exp\left(-i\mathcal{H}_0 t\right)\nonumber\\
     &\equiv& \bigl|1\bigr\rangle \bigl\langle 1 \bigr|\otimes\mathcal{U}_1(t) + \bigl|0\bigr\rangle \bigl\langle 0 \bigr|\otimes\mathcal{U}_0(t),
\end{eqnarray}where $\mathcal{U}_1(t)$ and $\mathcal{U}_0(t)$ are the
projections of the time-evolution operator onto the $\bigl|+1\bigr\rangle$
 and $\bigl|0\bigr\rangle$ states of the NV spin respectively.

In general, we wish to consider the effect of different pulse
sequences, which involve periods of free evolution followed by applied pulses
at particular times. A general time evolution operator will contain exponents
of the above Hamiltonians, however these exponents will appear as different
components of the 2$\times$2 matrix describing the central spin, depending on the
pulse sequence considered. To keep things general, we write
\begin{eqnarray}
  \mathcal{U}(t) &=& \left(
                       \begin{array}{cc}
                         \mathcal{K}_{11}(t) & \mathcal{K}_{10}(t) \\
                         \mathcal{K}_{01}(t) & \mathcal{K}_{00}(t) \\
                       \end{array}
                     \right),
\end{eqnarray}however, just what the $\mathcal{K}_{mn}(t)$ are will depend on
the pulse sequence employed. For the FID case just mentioned, we just simply
have $\mathcal{K}_{11}(t) = \mathcal{U}_{1}(t)$, $\mathcal{K}_{00}(t) =
\mathcal{U}_{0}(t)$, $\mathcal{K}_{10}(t) = \mathcal{K}_{01}(t) = 0$.

The relatively short coherence times of a FID experiment may be extended by
2-4 orders of magnitude by applying an appropriate sequence of $\pi$ pulses
(or `bit-flips', denoted $\mathcal{F}$), under which the quantum amplitudes
of the $|1\rangle$ and $|0\rangle$ states are swapped. In the simplest
instance, we consider a Hahn-echo, or spin-echo pulse sequence, involving a
single $\pi$ pulse applied at time $t/2$. The effect of this sequence is to
refocus any static components of the bath, thereby extending coherence times
by roughly 2 orders of magnitude, with typical times of 400\,$\mu$s-1\,ms. The
time evolution operator for a spin echo experiment is
\begin{eqnarray}
  \mathcal{U}_\mathrm{se}(t)      &=& \mathcal{U}_\mathrm{fid}(t/2)\,\mathcal{F}\,\mathcal{U}_\mathrm{fid}(t/2),
\end{eqnarray}hence we make the identification
\begin{eqnarray}
  \mathcal{K}_{10}(t) &=& \mathcal{U}_{1}(t/2)\mathcal{U}_{0}(t/2)\nonumber\\
  \mathcal{K}_{01}(t) &=& \mathcal{U}_{0}(t/2)\mathcal{U}_{1}(t/2)\nonumber\\
  \mathcal{K}_{11}(t) &=& \mathcal{K}_{00}(t) = 0.
\end{eqnarray}
The density matrix, $\rho(t)$, at $t=0$ is given by
\begin{eqnarray}
  \rho(0) &=& \biggl[\bigl|1\bigr\rangle \bigl\langle 1 \bigr|
  +\bigl|1\bigr\rangle \bigl\langle 0 \bigr| + \bigl|0\bigr\rangle \bigl\langle 1 \bigr|
  +\bigl|0\bigr\rangle \bigl\langle 0 \bigr|\biggr]\otimes\mathcal{M}_\mathrm{E},\nonumber
\end{eqnarray}where $\mathcal{M}_\mathrm{E}$ denotes a purely mixed environmental state. The in-plane magnetisation at time $t$ is found from
\begin{eqnarray}
  L &=& \mathrm{Tr}\bigl\{\left(\mathcal{S}_x + i\mathcal{S}_y\right)\rho(t)\bigr\} .
\end{eqnarray}From this, we see that the FID and spin echo signals are given by
\begin{eqnarray}
  L_\mathrm{fid}       &=& \frac{1}{2^k}\mathrm{Tr}_\mathrm{E}\biggl\{\mathcal{U}_{0}(t)\mathcal{U}_{1}^\dag(t) \biggr\}\nonumber\\
    L_\mathrm{se}       &=& \frac{1}{2^k}\mathrm{Tr}_\mathrm{E}\biggl\{\mathcal{U}_{0}(t/2)\mathcal{U}_{1}(t/2)\mathcal{U}^\dag_{0}(t/2)\mathcal{U}^\dag_{1}(t/2)\,\biggr\}\label{TraceDecoherenceEnvelope}
\end{eqnarray}respectively, where $k$ is the number of spins in the cluster. The exact forms of the propagators will be determined by the regime in question,
allowing us to make asymptotic expansions in terms of the relative coupling scales, such as
$an/\omega$, $bn/\omega$ and $a/b$, where $n$ is the density of the spins in the bath.

\subsection{Environmental autocorrelation functions and frequency spectra}\label{SingleClusterCorrFns}
Before deriving the relevant decoherence functions, we take a brief detour to
examine the dynamic behaviour of the nuclear spin bath environment as described by the effective
semiclassical magnetic field felt at an arbitrary point in the lattice due to
the interacting environmental spins. Whilst the existence of such a field is not sufficient to describe
the induced decoherence behaviour of the central spin, due to the omission
of the hyperfine couplings, it does give us an insight into the natural dynamic behaviour of the spin bath, and how it changes with the background magnetic field strength. In this section, we derive the autocorrelation functions of the
effective magnetic field due to 2 and 3 spin clusters in the environment, for
both secular (high-field) and non-secular (low-field) flip-flop regimes. We conclude this discussion of autocorrelation functions with an analysis of the effect of the hyperfine coupling on the nuclear dynamics. This analysis justifies why we may ignore the dipole-dipole coupling between nuclei when the NV spin is in either of the $\bigl|\pm1\bigr\rangle$ states.

\subsubsection{Secular nuclear dynamics}
When a background field of sufficient strength to set the quantisation axis
of the spins in the cluster is applied, some of the terms in the Hamiltonian
describe spin transitions that are no longer energy conserving and are hence
disallowed. In this case, we make the secular approximation in which all non-magnetisation conserving
transitions are ignored, giving the following secular Hamiltonian,
\begin{eqnarray}
  \mathcal{H}_\mathrm{sec} = B_{12}\vec{\mathcal{E}}_1\cdot\vec{\mathcal{E}}_2+ \omega_1\mathcal{E}_{1,z} + \omega_2\mathcal{E}_{2,z}
\end{eqnarray}
where $B_{12} = b/r_{12}^3[1 - 3\cos(\theta)]$.
The effective magnetic field operator as felt by the central
spin is due to the axial components of the hyperfine interaction,
\begin{eqnarray}
  \mathcal{B}_{2} &=& \sum_{j=1}^{N_k}\left(A_{zx}^{(j)}\mathcal{E}_x^{(j)}+A_{zy}^{(j)}\mathcal{E}_y^{(j)}+A_{zz}^{(j)}\mathcal{E}_z^{(j)}\right),
\end{eqnarray}
where $N_k$ is the number of spins in the $k^\mathrm{th}$ cluster. For
$n_k=2$, this leads to an autocorrelation function of
\begin{widetext}
\begin{eqnarray}
\bigl\langle \mathcal{B}_{2}(t)\mathcal{B}_{2}(0) \bigr\rangle_\mathrm{S}
&=&   A_{z,1}^2 + A_{z,2}^2+\left(A_{x,1}^2+A_{x,2}^2+A_{y,1}^2+A_{y,2}^2\right)\cos (t \omega )-\left[\Delta_z ^2+\left(\Delta _x^2+\Delta _y^2\right)\cos (t \omega )\right] \sin ^2\left(\frac{B_{12} t}{2}\right)\nonumber\\
\label{SecAutoCorrFn}
\end{eqnarray}
\end{widetext}
where $\Delta_{x,y,z} \equiv \left|A_{x,y,z,1} - A_{x,y,z,2}\right|$. Since we are only concerned with couplings to the axial ($z$) component of the NV spin, we have adopted the short hand notation of $A_{zx}^{(j)} \equiv A_{x,j}$, $A_{zy}^{(j)} \equiv A_{y,j}$ and $A_{zz}^{(j)} \equiv A_{z,j}$.

 Eq.\,\ref{SecAutoCorrFn} shows that there is always a static component of the secular autocorrelation function present regardless of the geometric arrangement of the spins in the cluster. 
The total axial magnetisation for a given cluster is constant, and hence the
NV only sees a fluctuating field if the two spins have different hyperfine
coupling strengths ($A_{z,1}$ \& $A_{z,2}$). The larger this difference, the
greater the strength of the effective fluctuating field, however the axial
flipping rate, $B_{12}$ decreases with their spatial separation. If the spins
are sufficiently close together, such that their energy scale is dictated by
their mutual interaction, non-energy conserving transitions become
permissable, and the secular condition is violated. This case is dealt with below in section\,\ref{SingleClusterCorrFnsNonSec}.

These methods may be extended to obtain corrections for three spin
interactions and higher. However, despite being interested in the short-time
and relatively weak coupling to the next-nearest-neighbour, we cannot use
perturbation theory, as the couplings strengths still become infinite as the
next-nearest-neighbour separation goes to zero. A short-time expansion of
Eq.\,\ref{SecAutoCorrFn} would diverge as $r$ approaches 0, hence to use
perturbation theory at a given order for all possible geometric
configurations (particularly when $B_{12}^2\gg A_{z1}^2+A_{z1}^2$, which
defines the high frequency, and hence short-time behaviour of the dynamics),
we require the leading order of the relevant probability density function to
be at least $\mathcal{O}\left\{r^4\right\}$. As we will see in section\,\ref{SpatStats}, this
corresponds to the third nearest neighbour and above. Hence, perturbation
theory cannot be applied until cluster sizes of four or greater are
considered.

In analysing the dynamics of a 3 spin cluster, we initially assume that a
strongly coupled pair exists, and introduce a third impurity whose coupling
to the initial two is comparatively weak. We assume the two couplings
involving the third spin are of similar order and make small perturbations
about this condition. This is justified by the rapid fall-off of the
dipole-dipole coupling, which ensures that any large deviation from this
condition will yield a 2 spin cluster and an effectively separate, uncoupled
spin. From this, we find the autocorrelation function of a single 3 spin
cluster to be
\begin{widetext}
\begin{eqnarray}
  \bigl\langle \mathcal{B}_3(t)\mathcal{B}_3(0) \bigr\rangle_\mathrm{S} &=&
     \bigl\langle \mathcal{B}_2(t)\mathcal{B}_2(0) \bigr\rangle_z + A_{\text{z3}}^2-\frac{4}{9}\left[ \Delta _{13} \sin ^2\left(\frac{3 B_{13} t}{4}\right)+\Delta
   _{23} \sin ^2\left(\frac{3 B_{23} t}{4}\right)\right] + \mathrm{Larmor\,\,terms}.
\end{eqnarray}
\end{widetext}
This result exhibits almost identical properties to the 2 spin cluster case, with a persistent static component, and fluctuating components whose amplitudes are again proportional to the respective hyperfine coupling differences.

\subsubsection{Non-secular nuclear dynamics}\label{SingleClusterCorrFnsNonSec}
In the opposite limit, where the quantisation axis of the nuclear spins is set by their mutual coupling, we cannot ignore the non-magnetisation conserving terms in the dipole tensor describing their interaction. The Hamiltonian describing the dipolar coupling between two spins when all
possible terms are included is given by
\begin{eqnarray}
  \mathcal{H}_{\mathrm{N}} = \frac{b}{r^3}\left[\vec{\mathcal{E}}_1\cdot\vec{\mathcal{E}}_2-\frac{3}{r^2}\left(\mathbf{r}\cdot\vec{\mathcal{E}}_1\right)\left(\mathbf{r}\cdot\vec{\mathcal{E}}_2\right)\right],
\end{eqnarray}
which yields the following non-secular autocorrelation function of the axial
magnetic field,
\begin{widetext}
\begin{eqnarray}
 \bigl\langle \mathcal{B}_{2}(t)\mathcal{B}_{2}(0)\bigr\rangle_\mathrm{N} &=&\left(A_{{z1}}^2+A_{{z2}}^2\right) \left[1-\frac{4}{3} \sin ^2\left(\frac{3 B_{12} t}{4}\right)\right]- \frac{2}{3}\Delta _z^2\left[ \sin ^2\left(\frac{B_{12}
   t}{4}\right)+ \frac{1}{2}\sin ^2\left(\tfrac{B_{12} t}{2}\right)- \sin ^2\left(\frac{3 B_{12} t}{4}\right)\right]. \label{NonSecACF}
\end{eqnarray}
\end{widetext}
Note that, where the secular autocorrelation function only had
fluctuating components proportional to differences in prob-spin couple
strengths $\left(\Delta_z\right)$ within a given cluster, the non-secular
function also contains terms that are present regardless of the geometric
arrangement of the cluster constituents. This is a consequence of the fact
that, for a non-secular cluster, the background field does not set the
quantisation axis of the spins, hence the magnetisation component along the
background field direction is not constant. As we will see in section\,\ref{EnsAutoCorrFns}, when the contributions to the full autocorrelation functions are summed over all clusters in the environment, we see very large differences between the dynamic behaviour of spins in secular and non-secular flip-flop regimes.

\begin{figure*}
  \includegraphics[width=\textwidth]{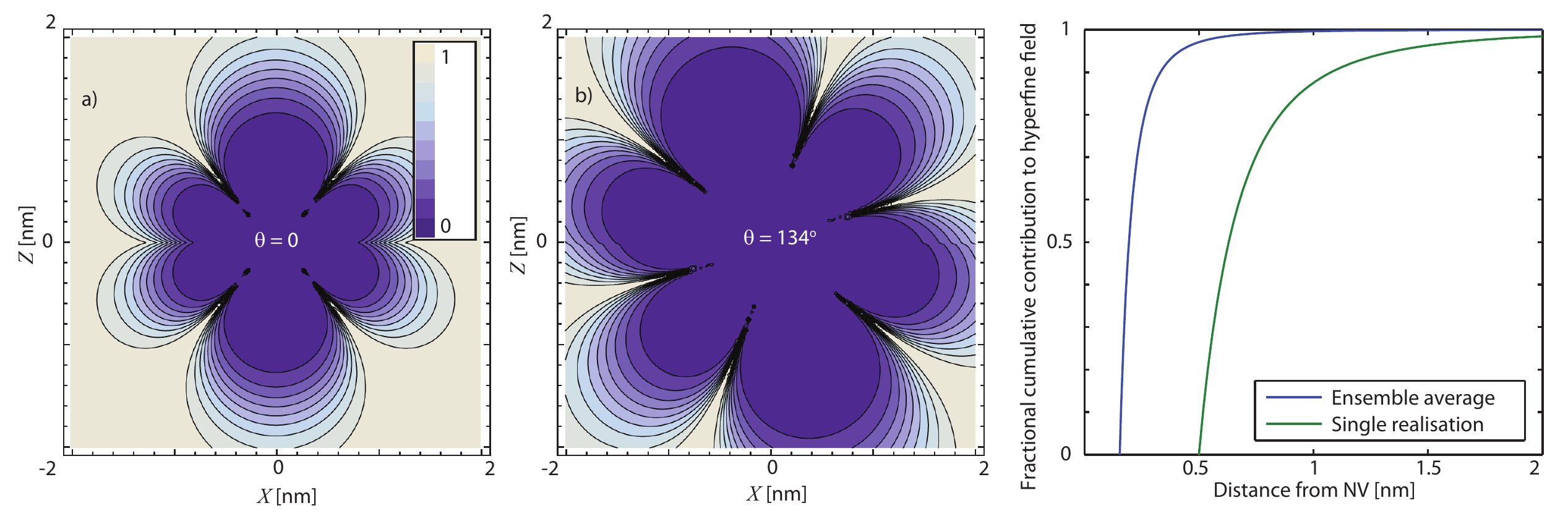}\\
  \caption{Plot showing the suppression of dipole mediated nuclear spin dynamics due to the hyperfine field of the NV centre. Values of the suppression constant, $K$ (see eq.\,\ref{SuppConstant}) are plotted in a) and b) for cases where the two nuclei occupy adjacent lattice sites, with relative orientations of $\theta=0$ and $\theta = 134^o$ respectively. The fractional cumulative contribution of all nuclei within a distance $R$ to the total hyperfine field is shown in c) for both the ensemble case (blue) and that of a typical realisation of the environmental spin distribution (green). This plot show that effectively all nuclei making an appreciable contribution to the hyperfine field also reside in the suppression region. }\label{SuppConstPlot}
\end{figure*}

\subsubsection{Suppression of nuclear dynamics due to hyperfine fields}
In cases where a strong magnetic field gradient exists, two nuclear spins will posses a mutual detuning between their respective Zeeman energies given by $\delta_z = \left|\omega_1-\omega_2\right|$. Solving for the autocorrelation function in this case, we find
\begin{widetext}
\begin{eqnarray}
\bigl\langle \mathcal{B}_{2}(t)\mathcal{B}_{2}(0) \bigr\rangle_\mathrm{S}
&=&  A_{z,1}^2 + A_{z,2}^2-\Delta_z ^2\frac{B_{12}^2}{B_{12}^2+\delta_z^2}\sin^2\left(\frac{t}{2} \sqrt{B_{12}^2+\delta_z^2}\right) + \mathrm{Larmor\,\,terms}\label{BBtCorrFnWithDetuning},\end{eqnarray}
\end{widetext}showing a modulation in the fluctuation amplitude by a factor of ${B_{12}^2}/\left({B_{12}^2+\delta_z^2}\right)$, which becomes significantly damped as the magnitude of the detuning approaches that of the mutual dipolar coupling strength. We would not expect such a situation to arise as the result of inhomogeneities in an applied background as the associated detunings are simply not large enough over the distance of a few angstroms, which would require a magnetic field gradient of $\sim (b/l^3)/(l\gamma_{\mathrm{E}})\approx1\mathrm{\,mT\,nm^{-1}}$. Where significant detunings can arise however, are as the result of the hyperfine field generated by the central NV spin. For nuclear spins up to a few nanometres from the NV centre (which are responsible for the decoherence of the NV spin, as shown in Figs.\,\ref{SBANuclearRegionsP}\,\,\,and\,\,\,\ref{SBANuclearRegionsManyBathsP}), the difference in hyperfine couplings between two adjacent lattice sites is much greater than the associated dipolar coupling between them, leading to a complete suppression of the nuclear spin dynamics. We make this statement more precise as follows.

When the detuning between the Zeeman energies of two coupled nuclear spins is the result of the NV hyperfine field, we have that $\delta^2_z = \Delta_z^2 = \left(A_{z,1}-A_{z,2}\right)^2$. Eq.\,\ref{BBtCorrFnWithDetuning} shows this leads to a suppression of the associated fluctuation amplitude by a factor of
\begin{eqnarray}
  K &=& \frac{B^2}{B^2+\Delta_z^2}.\label{SuppConstant}
\end{eqnarray}Consider the cluster arrangement depicted in Fig.\,\ref{ClusterSchem}, where the separation vectors between the NV and two coupled nuclear spins is
\begin{eqnarray}
  \mathbf{R_1} &=& R\biggl(\cos\left(\Phi\right)\sin\left(\Theta\right),\sin\left(\Phi\right)\sin\left(\Theta\right),\cos\left(\Theta\right)\biggr)\nonumber\\
  \mathbf{R_2} &=& \mathbf{R_1} + \mathbf{r},
\end{eqnarray}where $\mathbf{r}$ is the separation vector between the two spins, as given by
\begin{eqnarray}
  \mathbf{r} &=& r\biggl(\cos\left(\phi\right)\sin\left(\theta\right),\sin\left(\phi\right)\sin\left(\theta\right),\cos\left(\theta\right)\biggr).
\end{eqnarray}As the largest coupling strength comes from nuclei that occupy adjacent lattice sites, we take $r = l$, where $l=1.54$\,{\AA} is the lattice constant for diamond. From Eq.\,\ref{HyperfineCoupling}, the axial hyperfine coupling strengths are given by
\begin{eqnarray}
  A_{z,i} &=& \frac{a}{R_i^3}\left[1-3\cos^2\left(\Theta_i\right)\right],
\end{eqnarray}and the nuclear dipolar coupling strength is
\begin{eqnarray}
  B &=& \frac{b}{l^3}\left[1-3\cos^2\left(\theta\right)\right].
\end{eqnarray}Using these quantities, we plot the magnitude of the suppression constant, $K$ (Eq.\,\ref{SuppConstant}), in Fig.\,\ref{SuppConstPlot}. These results depict the worse case scenario (where the dipolar coupling is maximal and the hyperfine detuning is minimal) for the two possible cluster orientations of $\theta = 0$ (Fig.\,\ref{SuppConstPlot} a)) and $\theta = 134^o$ (Fig.\,\ref{SuppConstPlot} a)), and show that the nuclear dynamics are still strongly suppressed for NV-nuclear separations greater than 1\,nm, and as great as 2\,nm in the $\theta = 134^o$ case.

Naturally, as the NV-nuclear separation distance increases, both the hyperfine field and the corresponding hyperfine field detuning between adjacent lattice sites will decrease. For large enough NV-$^{13}$C separations, the dipolar coupling will eventually dominate over the hyperfine detuning, however the reduced hyperfine coupling implies that spins in these regions will necessarily be too weakly coupled to the NV to have any effect on its evolution. To make this statement precise, consider the fractional contribution of the hyperfine field from a lower cutoff, $R_{0}$, to an arbitrary radial distance $R$ as given by
\begin{eqnarray}
    \int_{R_{0}}^R  n A_z^2 \,\mathrm{d}^3\mathbf{R}\biggl/ \int_{R_{0}}^\infty n A_z^2 \,\mathrm{d}^3\mathbf{R}
   &=& 1-\frac{R_{0}^3}{R^3}\label{Rcutoff},
\end{eqnarray}where $n$ is the average density of $^{13}$C spin in the lattice. The choice of $R_{0}$ will depend on the diamond sample at hand. In an ensemble average over many environmental distributions, all lattice sites will be equally populated, meaning that we must choose $R_0 = l$ as our lower cutoff. On the other hand, in a single realisation of the environmental distribution, we would not expect to find a nuclear spin within a distance of $R_0 = (3/4\pi n)^{1/3} = 5.0\,$nm, which we take as our lower cutoff. We plot Eq.\,\ref{Rcutoff} for these two cases in Fig.\,\ref{SuppConstPlot} c), showing that there is effectively no contribution from spins residing more than a nanometre from the NV centre.
It is for this reason that nuclear-nuclear dipolar couplings may be ignored for cases where the NV spin state is in either of its $\bigl|\pm1\bigr\rangle$ basis states. Furthermore, as the NV spin must be in either of these states to feel the effect of the dipole field, this shows that a semi-classical mean-field approach cannot reproduce the decoherence behaviour of an NV centre coupled to a nuclear spin bath. This will be explored further in section\,\ref{NoClassical}.

\subsection{Single spin clusters and free-induction decay}\label{SingleSpinDecoherence}
Having discussed the environmental dynamics of the nuclear spin flip-flops as unperturbed by the presence of the central spin, we now discuss the exclusive hyperfine dynamics of environmental spins coupled to the NV without considering their mutual dipolar couplings.
Again, this is not sufficient to explain the full decoherence behaviour under spin-echo and higher order pulse sequences, however it does gives us an insight into how the hyperfine dynamics transition from non-secular to secular behaviour with an increasing magnetic field strength. Furthermore, given that FID timescales are of the order of a few $\mu$s and thus too fast to see the effects of dipolar couplings between environmental spins, non interacting spins are sufficient to explain all FID effects.

Single spin clusters, by definition, do not include any interaction with
adjacent spins. As we will show, such a simplified arrangement is not
sufficient to describe any true decoherence in this system, however, it does
serve as a useful exercise in demonstrating how some of the limiting
parameter regimes emerge. The hyperfine and Zeeman coupling components of the Hamiltonian as projected onto the
$\bigl|0\bigr\rangle$ and $\bigl|+ 1\bigr\rangle$ states of the NV spin
are given by
\begin{eqnarray}
  \mathcal{H}_1 &=& A_x\mathcal{E}_x+A_y\mathcal{E}_y+\left(A_z+\omega\right)\mathcal{E}_z\nonumber\\
  \mathcal{H}_0 &=& \omega\mathcal{E}_z,
  \end{eqnarray}from which we determine the FID and spin-echo envelopes using Eq.\,\ref{TraceDecoherenceEnvelope},
\begin{eqnarray}
  L_\mathrm{fid} &=&\cos \left(\frac{t \lambda }{2}\right) \cos \left(\frac{t \omega }{2}\right)
  + \frac{ \Omega  }{\lambda }\sin \left(\frac{t \lambda }{2}\right) \sin \left(\frac{t \omega }{2}\right)\nonumber\\
  L_\mathrm{se} &=&1-2\frac{ A_x^2+A_y^2 }{\lambda ^2}\sin ^2\left(\frac{t \lambda }{2}\right) \sin ^2\left(\frac{t \omega }{2}\right),
\end{eqnarray} where $\Omega = A_z+\omega$ and
$\lambda = \sqrt{A_x^2+A_y^2 + \Omega^2}$. In this section, we will
examine the behaviour of these expressions in cases of high and low magnetic
fields, however one can immediately see that there is no spin-echo
decoherence at both $\omega\rightarrow0$ and $\omega\rightarrow\infty$
limits. This is in direct contrast with experimental observations, where the
decoherence rate is maximal at zero field, and decreases to a final, constant
value at sufficiently high magnetic fields. This implies that we must
introduce more complex spin-spin interactions to be able to explain this
discrepancy. Higher order clusters are considered in the following sections,
hence in this section we focus solely on FID behavior.

Expanding the above result for $\omega\gg A_z$, we find the contribution to
the FID from a single spin to be
\begin{widetext}
\begin{eqnarray}
 L^{(1)}_\mathrm{fid}\biggl|_{\omega\gg A}  &\sim& \cos \left(\frac{A_z t }{2}\right)-\frac{A_x^2+A_y^2}{2 \omega ^2} \sin \left(\frac{ \omega t }{2}\right) \sin \left(\frac{1}{2}  \left(A_z+\omega \right)t\right),\label{FIDSingleClusterLHigh}
\end{eqnarray}and in the low field limit ($\omega\ll A_z$) we find
\begin{eqnarray}
  L_\mathrm{fid}\biggl|_{\omega\ll A} &\sim& \cos \left(\frac{A t}{2}\right) \cos \left(\frac{\omega t  }{2}\right)
  +\left[\frac{A_z }{A}+\frac{\omega   \left(A_x^2+A_y^2\right) }{A^3} -\frac{3 \omega ^2 \left(A_x^2+A_y^2\right) A_z }{2 A^5}\right]\sin \left(\frac{A t}{2}\right) \sin \left(\frac{\omega t    }{2}\right),\label{FIDSingleClusterLLow}
\end{eqnarray}
  where $A = \sqrt{A_x^2+A_y^2 + A_z^2}$.
\end{widetext}

There are a number of points worthy of discussion here, particularly with the regard to the effect of the magnetic field strength the effective hyperfine coupling strength. In the infinite magnetic field limit this coupling is completely determined by the axial hyperfine component, $A_z$, alone. This is because the Zeeman coupling is responsible for setting the quantisation axis of the nuclei, hence the NV spin is unable to drive transitions in the nuclear spins. On the other hand, in the zero field case, it is the hyperfine coupling thats sets the quantisation axis of the nuclei, meaning that their magnetisation need not be conserved with respect to the background magnetic field. This leads to a greater effective hyperfine coupling, owing to the inclusion of $A_x$ and $A_y$ terms. This is an important effect that carries over into the analysis of higher order pulse sequences, as it distinguishes the ZSE regime from the SZE and SEZ regimes.

\begin{figure}
  \includegraphics[width=\columnwidth]{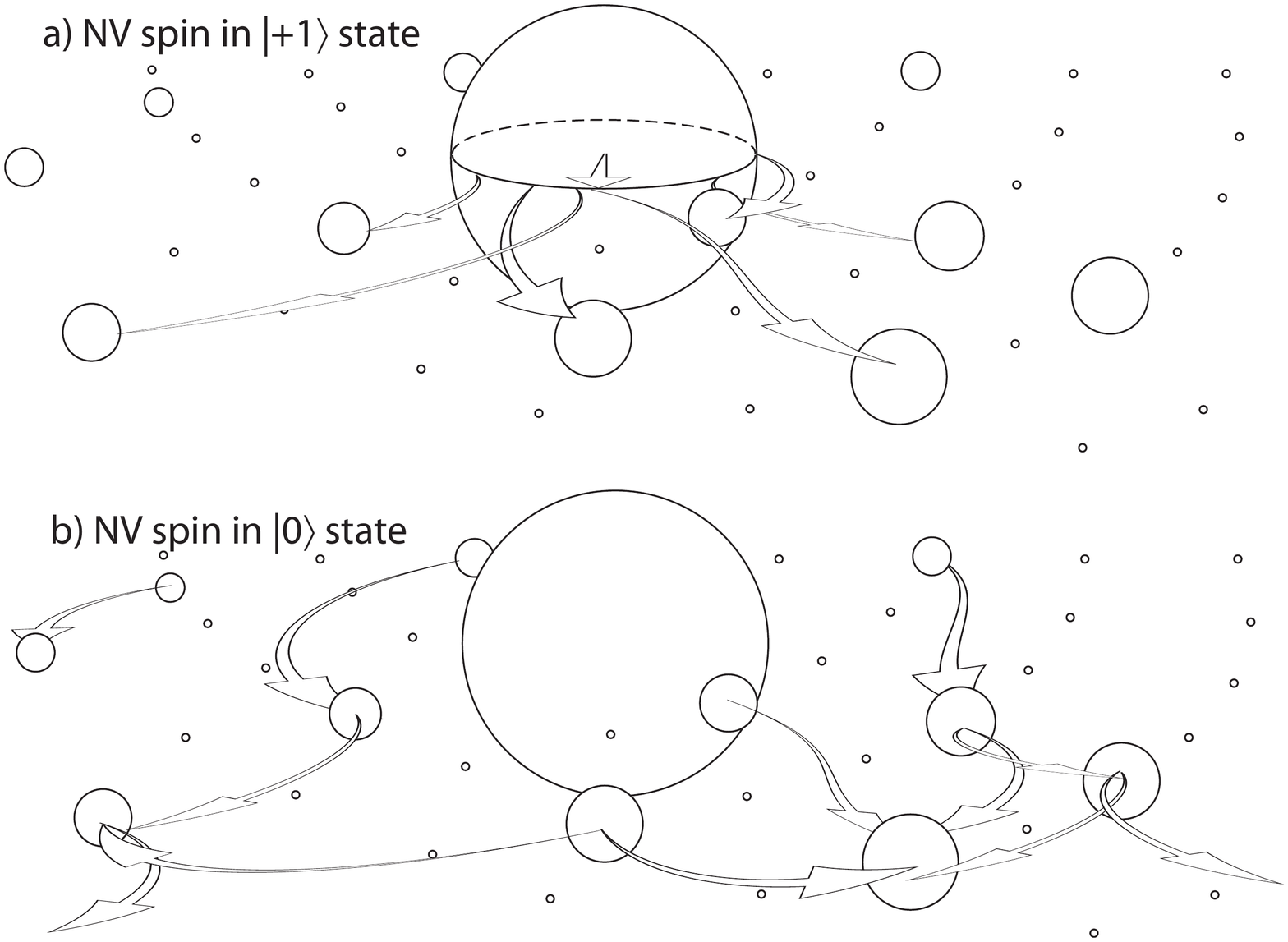}\\
  \caption{Schematic depicting the two step process of NV spin decoherence. In the first step (a), the NV spin is in its $\bigl|+1\bigr\rangle$ state, and quantum information regarding the NV spin state is imparted onto the environmental nuclear spins via the hyperfine interaction. This process is effectively reversible, as the nuclei cannot interact due to the strong hyperfine field of the NV. When the NV spin is flipped into its $\bigl|0\bigr\rangle$ state (b), the hyperfine coupling is turned off and this information is free to propagate throughout the lattice via the nuclear dipole-dipole interaction, rendering its loss irreversible.}\label{SBABathPropHyperfine}
\end{figure}

\subsection{Two-spin clusters and spin-echo decay}
In the previous sections, we discussed how treating the dipolar-dipole coupled nuclear spin bath as a fluctuating magnetic field does not explain the decoherence of the NV spin, as the NV can only sense the effect of the nuclei if its hyperfine field is simultaneously suppressing their activity. On the other hand, treatment of the hyperfine interaction exclusively, to the exclusion of the nuclear dipolar interaction, only shows periodic entanglement between the NV spin and the nuclei, with no permanent decay of NV spin coherence on long timescales. These results imply that the NV spin coherence is essentially two-part process (see Fig.\,\ref{SBABathPropHyperfine}), in which quantum information of the NV spin is first imparted to the independent nuclei via the hyperfine interaction when the NV spin is in the $\bigl|+1\bigr\rangle$ state. This information may then be propagated throughout the crystal via the nuclear dipole-dipole interaction whilst the NV spin is in the $\bigl|0\bigr\rangle$ state. As such, we must incorporate both interactions in order to be able to analyse the true decoherence behaviour of the NV spin.

  We begin by discussing how the full Hamiltonian (Eq.\,\ref{HamilFull}) may be
  simplified according to the six parameter regimes in question to solve for
  the corresponding evolution.

\subsubsection{Strong, secular hyperfine coupling, secular dipole-dipole coupling
(ZSE)}\label{ZSE} As discussed earlier, when the NV spin is in the $\bigl|+1\bigr\rangle$ state,
the difference in the hyperfine couplings will yield sufficient detuning
to suppress any dipolar flip-flops, hence we may ignore the dipolar term in
the projection of the Hamiltonian on the $\bigl|+1\bigr\rangle$ spin state.
Furthermore, the fact that the Zeeman terms are much greater than the dipolar
terms allows us to ignore spin-spin interactions that do not conserve total
magnetisation with respect to the background field, $\vec{\omega}$, and make
the secular approximation for the dipole-dipole coupling. Thus the relevant Hamiltonians for the ZPS regime
are given by
  \begin{eqnarray}
         \mathcal{H}_1 &=& \sum_{k=1}^2\bigl[A_{x,k}\mathcal{E}_{x,k} + A_{y,k}\mathcal{E}_{y,k} + \left(A_{z,k}+\omega\right)\mathcal{E}_{z,k}\bigr]\nonumber\\
    \mathcal{H}_0 &=& B\vec{\mathcal{E}}_1\cdot \vec{\mathcal{E}}_2 + \omega\left({\mathcal{E}}_{z1}+{\mathcal{E}}_{z2}\right).
  \end{eqnarray}

Using Eq.\,\ref{TraceDecoherenceEnvelope}, and expanding to second order for small
$A_{x,y,z}/\omega$ (the full expression is given in Eq.\,\ref{APPESEEMLF}), we obtain the contribution to the spin echo decoherence of the central spin due to a 2 spin cluster,
\begin{widetext}
\begin{eqnarray}
  L_{\mathrm{ZSE}} &=& 1-\sin ^2\left(\frac{B t}{4}\right) \sin ^2\left(\frac{\Delta_zt}{4}   \right)-\frac{ \Delta _x^2+\Delta _y^2 }{\omega ^2}\sin ^2\left(\frac{B t}{4}\right)\sin ^2\left[  \left(A_{z,1}+\omega \right)\frac{t}{4}\right]\nonumber\\
    &&-4\frac{ A_{x,1}^2+A_{y,1}^2 }{\omega ^2}\sin ^2\left[  \left(A_{z,1}+\omega \right)\frac{t}{4}\right] \sin ^2\left(\frac{t \omega }{4}\right).\label{SEZPS}
\end{eqnarray}
\end{widetext}
We note that only the terms containing the dipole-dipole coupling, $B$, represent any actual decoherence, with the
presence of a finite magnetic field increasing the effect by a factor of $1 +
\frac{\Delta_x^2 + \Delta_y^2}{4\omega^2}$. The final term corresponds to the lateral dynamics (precession) of the nuclei, and hence does not
contribute any decoherence, for reasons analogous to those discussed in
section \ref{SingleSpinDecoherence}, however it does detail the emergence of
the decay/revival behaviour seen in spin echo experiments on electron spins
coupled to nuclear spin baths. Specifically, we see that the amplitude of the
revivals increase with decreasing magnetic field, as do their width.

Despite not contributing any true decoherence, the decays and revivals at the
Larmor frequency are susceptible to inhomogeneous broadening from the axial
couplings to all other spins in the bath, leading to an additional dephasing
component in the evolution of the central spin. Such a distinction is important,
as it explains the major difference between
numerically calculated and experimentally observed behaviour of this system. This effect will be considered in detail later in section\,\ref{ESEEMEns}.
Further corrections to the Larmor broadening due to larger cluster sizes may be calculated iteratively
by employing the spectral distribution 
when performing the ensemble average,
however these corrections will lead to terms with a dependence on $t$ beyond that of leading order and are thus not important.

\subsubsection{Strong, non-secular hyperfine coupling, secular dipole-dipole coupling (SZE)}
As with the ZSE regime, the $\bigl|+1\bigr\rangle$ state of the NV spin
yields sufficient detuning to suppress any dipolar flip-flops, hence we may
ignore the dipolar term in the projection of the Hamiltonian on the $\bigl|+1\bigr\rangle$ spin
state. We are working in a regime where the Zeeman terms are still much
greater than the dipolar terms, allowing us to ignore spin-spin interactions
that do not conserve total magnetisation with respect to the background
field. Thus the Hamiltonian, and hence the decoherence function, for the SZE regime are identical to that for the
ZSE regime, however we instead expand Eq.\,\ref{APPESEEMLF} for small $\omega/A^{1,2}_{x,y,z}$,
giving
\begin{widetext}
\begin{eqnarray}
  L_{\mathrm{SZE}} &=& 1-\sin ^2\left(\frac{B t}{4}\right) \sin ^2\left(\frac{\Delta }{4}\,t \right)
  -4\frac{
   A_{x,1}^2+A_{y,1}^2}{A_1^2}\left[1
   -\frac{2 \omega  A_{z,1}}{A_1^2}\right]\sin ^2\left(\frac{t \lambda _1}{4}\right) \sin ^2\left(\frac{t \omega }{4}\right)\nonumber\\
   &&
  +\frac{4  \left(A_{x,1}^2+A_{y,1}^2\right){}^2}{A_1^4}\left[1-\frac{4 \omega  A_{z,1}  }{A_1^2}\right]\sin ^4\left(\frac{t \lambda _1}{4}\right) \sin
   ^4\left(\frac{t \omega }{4}\right)
    \label{SEPZS},
\end{eqnarray}where $\Delta \equiv \left|A_1-A_2\right|$.
\end{widetext}
We note here that this expression is very similar to that of the ZSE regime,
however the effective hyperfine coupling strength has increased from $\Delta_z$ to
$\Delta$. This is a consequence of the quantisation axis of the spins being
set by their hyperfine coupling rather than their Zeeman coupling.

\subsubsection{Strong, non-secular hyperfine coupling, non-secular dipole-dipole coupling (SEZ)}
In this regime, we still have that the hyperfine couplings dominate when the
NV is in the $\bigl|+1\bigr\rangle$ state. When the NV is in the
$\bigl|0\bigr\rangle$, the dipolar couplings between the environmental spins
will dictate the evolution, as with the ZSE and SZE regimes, however in this
regime, the dipolar couplings dominate over the Zeeman terms. This means that
the quantisation axis of the spins are set by their mutual interaction, and
the cluster is thus not required to conserve magnetisation with respect to
the background field. Including all possible dipole interaction terms, we
have
  \begin{eqnarray}
         \mathcal{H}_1 &=& \sum_{k=1}^2\bigl[A_{x,k}\mathcal{S}_{x,k} + A_{y,k}\mathcal{S}_{y,k} + \left(A_{z,k}+\omega\right)\mathcal{S}_{z,k}\bigr]\nonumber\\
    \mathcal{H}_0 &=& B\left[ \vec{\mathcal{S}}_1\cdot\vec{\mathcal{S}}_2
  -3\left(\mathbf{n}\cdot\vec{\mathcal{S}}_1\right)\left(\mathbf{n}\cdot\vec{\mathcal{S}}_2\right) \right],
  \end{eqnarray}where $\mathbf{n}$ is the unit vector separating spins 1 and 2.
  The full spin echo envelope for the SEZ regime is too large to reproduce here, however we may simplify things immensely by averaging over the angular components of the cluster geometry ($\theta$, $\phi$), giving
\begin{eqnarray}
  L_{\mathrm{SEZ}} 
   &=&1- \frac{8}{15} \sin ^2\left(\frac{3 B t}{4}\right)\left[\sin ^2\left(\frac{A t}{2} \right)+\sin ^2\left(\frac{A t}{4} \right)\right] \label{SEPSZ}\nonumber\\
\end{eqnarray}where $A = \sqrt{A_{x}^{2} + A_{y}^{2} + A_{z}^{2}}$. Notice that
the hyperfine coupling now emerges as $A$ instead of $\Delta$, which is a
consequence of the magnetisation no longer being conserved with respect to
the background field. This results in a significantly larger fluctuation
amplitude, as $\left\langle A^2 \right\rangle = \left(\frac{4\pi n
a}{3}\right)^2$, whereas $\left\langle \Delta^2 \right\rangle = \left(2 n
a\right)^2$. The separation of hyperfine and dipolar processes also means that we need not distinguish between $A_1$ and $A_2$, as their relative locations are no longer important as far as the hyperfine component of the evolution is concerned. As the contributions of each spin will be summed over in an equivalent manner, we simply put $A_1 = A_2 = A$. This is in contrast to the ZSE and SZE cases, where the hyperfine couplings manifest as $\Delta_z$ and $\Delta$ respectively, as the treatment of spin 2 will depend on the location of spin 1.

In the following section we discuss the statistics associated with the random distribution of spin impurities in a spin bath environment. These statistics will be used to determine the combined effect on the coherence of the central spin from all clusters in the bath.

\section{Spatial statistics of randomly distributed impurities}\label{SpatStats}
In this section, we derive the probability density functions associated with
the distance between the nearest-neighbour (NN), next-nearest-neighbour
(NNN), and so forth, impurities in the environment. These distributions will
be used to determine the collective dynamic behaviour of the
environment and allow us to compare the contributions from the different
orders of clustering. We firstly consider the case of a continuum
distribution, in which spin impurities may adopt any position in the lattice
according to their spatial density. We then consider the specific case of NV
centres in diamond, in which carbon atoms are arranged in a tetrahedral
diamond lattice.

For a given lattice site density (or carbon atom number density) of $n_c$, the volume $V$ concentric on any one environmental spin impurity contains $N \approx n_cV
-1$ sites that may be
occupied by a second impurity. The probability of finding $X$ spins within $V$
is then a binomial distribution with $N$ independent trials, with
each site having a probability $\chi = 0.011$ of being occupied by a nucleus of non-zero spin,
\begin{eqnarray}
  \mathrm{P}(X|N,\chi) 
  &\approx& \frac{(V/V_0)!}{X!(V/V_0-X)!}\chi^X(1-\chi)^{V/V_0-X},\,\,\,\,\,\,
\end{eqnarray}
which, in the limit of low spin concentrations, $\chi\ll1$, approaches a
Poisson distribution,
\begin{eqnarray}
\mathrm{P}(X|V,\chi)  \approx\frac{1}{X!}\left(\zeta
r^3\right)^X\exp\left(-\zeta r^3\right),\label{PoissClust}
\end{eqnarray}
where $\zeta
\equiv \frac{4\pi \chi}{3V_0}$, implying an average spin impurity density of $n = \chi/V_0$. The probability that a sphere concentric on a
given environment spin contains at least one other spin is given by
which, by definition, is also the cumulative probability function. As such,
the probability of encountering a spin  \emph{at} $r$ (ie \emph{on} the shell
of $V$) is given by
\begin{eqnarray}
  \mathrm{P}(r) = \frac{d}{dr}  \mathrm{P}(X>0,r) = 4\pi n r^2\exp\left(-\frac{4 \pi nr^3}{3}\right).\label{SBANKdist}
\end{eqnarray}In other words, $\mathrm{P}(r)$ is the probability density function for the distance between two nearest neighbour spins.

This analysis may be extended to compute the probability distribution of the
distance to the $k^\mathrm{th}$ nearest neighbour.
Consider the region bounded by concentric spheres of radii $r_1$ and $r_0$,
the volume of which is $\frac{4}{3}\pi\left(r_1^3-r_0^3\right)$. As above,
the probability that at least one impurity exists in this region is
$1-\exp\left[-\zeta\left(r_1^3-r_0^3\right)\right]$,
which has the corresponding probability density function,
\begin{eqnarray}
  \mathrm{P}(r_1) &=&  3\zeta r_1^2\exp\left[-\zeta\left(r_1^3-r_0^3\right)\right].\nonumber
\end{eqnarray}Similarly, the probability density function for the distance to the $k^\mathrm{th}$ impurity is
\begin{eqnarray}
  \mathrm{P}(r_k) &=&  3\zeta r_k^2\exp\left[-\zeta\left(r_k^3-r_{k-1}^3\right)\right].\nonumber
\end{eqnarray}Taking $r_0=0$, the joint probability density function is
\begin{eqnarray}
  \mathrm{P}(r_1,\ldots,r_k) = \prod_{j=1}^kp_r\left(r_j\right)= (3\zeta)^kr^2_1\ldots r_k^2 \exp\left[-\zeta r_k^3\right]\nonumber.
\end{eqnarray}To obtain the distribution for each $r_j$, we successively integrate
over all $r_1,\ldots,r_{j-1}$, $r_{j+1},\ldots,r_k$ from 0 to $r_{j+1}$.

\begin{figure}
\centering
\includegraphics[width=8.5cm]{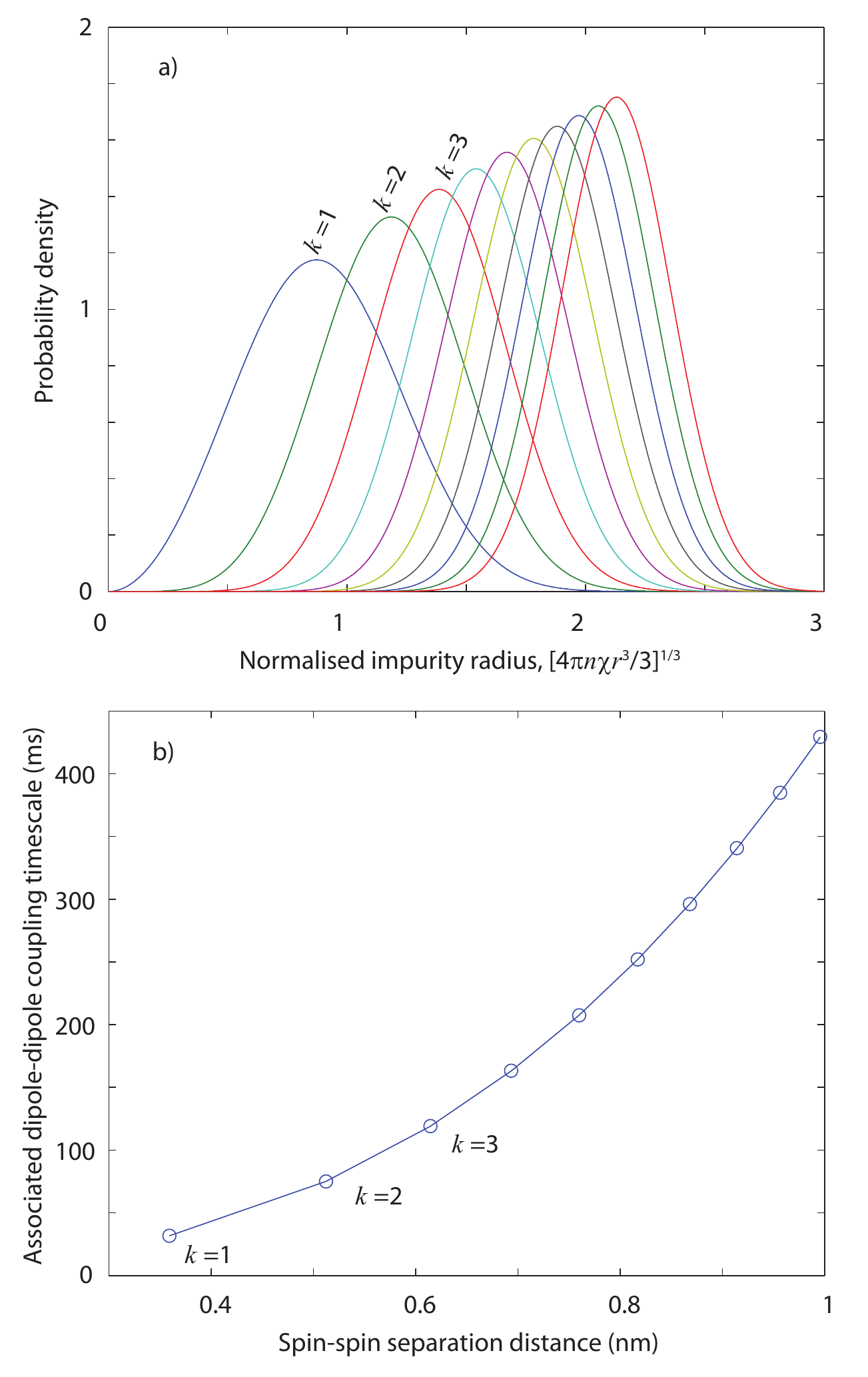}
  \caption{Probability distributions (a), and corresponding timescales and mean distance (b), associated with the first 10 nearest neighbours.
  }\label{SBAClusterStats}
\end{figure}

Thus, given the location of some environmental spin, the probability of finding its $k^\mathrm{th}$ nearest neighbour at a distance of $r_k$ is given by
\begin{eqnarray}
  \mathrm{P}_k(r_k) 
  &=& \frac{4\pi nr_k^2}{(k-1)!}\left(\frac{4\pi n r_k^3}{3}\right)^{k-1}\,\exp\left[-\frac{4\pi n r_k^3}{3} \right].
  \end{eqnarray}  Computing the first and second moments of this distribution, we find
\begin{eqnarray}
  \left\langle r_k\right\rangle &=&  \left(\frac{4\pi n}{3}\right)^{-\frac{1}{3}}\frac{\Gamma\left(k+\frac{1}{3}\right)}{(k-1)!},\\
\mathrm{and\,\,\,}  \left\langle r^2_k\right\rangle  &=& \left(\frac{4\pi n}{3}\right)^{-\frac{2}{3}}\frac{\Gamma\left(k+\frac{2}{3}\right)}{(k-1)!}.
\end{eqnarray} A plot of the mean distance to the first 10 nearest neighbours, $\left\langle r_k\right\rangle$ for $k = 1,\ldots,10$, is shown in Fig.\,\ref{SBAClusterStats}(b).
This quantity gives us an indication of how large the considered region may
be (and hence the timescale) before NNN interactions become important. As we
can see, for the case of an NV centre coupled to a $^{13}$C nuclear spin
bath, where $T_2<1\,$ms, we need only consider 2-spin interactions.

In the above analysis, we have assumed that a given impurity may adopt any
position within the environment, with the only constraint being the overall
average density with which the impurities are distributed. However, as our
primary focus is on the NV centre in diamond, this is not strictly correct,
as impurities may only occupy the atomic positions of a diamond cubic crystal
structure.

 Let $N(r)$ be the number of discreet lattice sites enclosed within a sphere
 of radius $r$, concentric on some impurity, and let $\nu_n \equiv\nu(r_n)$
 denote the number of discreet lattice sites at radius $r_n$.
 Again invoking a binomial distribution, the probability of encountering the nearest neighbour impurity 1 spin at radius $r_n$, is the joint probability that 1 or more impurities reside at $r_n$ and that there are no others within a sphere of this radius,
 \begin{eqnarray}
     \mathrm{P}_n &=& \left[1-(1-\chi)^{\nu_n}\right](1-\chi)^{N_{n-1}}.
 \end{eqnarray}

 The position vectors associated with the lattice sites in a cubic diamond unit cell of sidelength 4 are
$\bigl\{\mathbf{u}_k\bigr\} =
\bigl\{(0,0,0),(0,2,2)_\circlearrowleft,(3,3,3),(3,1,1)_\circlearrowleft\bigr\}$, where $\circlearrowleft$ denotes a cyclic permutation of vectorial components.
 If we let $(l,m,n)\,\in\mathbb{N}^3$ index each individual cell, then the cartesian coordinates of a given site are $\mathbf{U}_k = 4 (l,m,n) + \mathbf{u}_k.$
From this
we find that the squared distance to the $n^\mathrm{th}$ neighbour is $4n$ if
$n$ is even, and $4n-1$ if $n$ is odd.
Both $r_n$ and $\nu_n$ are given in
table\,\ref{SBATableOfDiscreteProperties}. Note that values of $r_n^2$ have been normalised, however the distance between adjacent lattice sites is given by $l=1.54\,${\AA}.
\begin{table}
  \centering
    \begin{tabular}{r|cccccccccccccccccccccccccccccccccc}
    neighbour&& 1 && 2  &&  3 && 4  && 5  && 6  &&\ldots&& odd $n$  && even $n$\\
        \hline\\
    $(\times l^2/3)\,\,\,r_n^2$    && 3 && 8  && 11 && 16 && 19 && 24 &&\ldots&&    4n-1  && 4n\\
    $\nu_n$ && 4 && 12 && 12 && 6  && 12 && 24 &&\ldots&&\\
  \end{tabular}
  \caption[Table of normalised squared-distances between crystal lattice sites, $r_n^2$ and the associated number of sites at that distance, $\nu_n$.]{Table of normalised squared-distances between crystal lattice sites, and the associated number of sites at that distance.}\label{SBATableOfDiscreteProperties}
\end{table}

This derivation of the discreet probability distribution allows us to
determine the extent to which the continuum approximation is valid when
computing ensemble averages of the various quantities that follow.
We now use these spatial distributions to analyse the behavior of a central NV spin as coupled to a 1.1\% $^{13}$C nuclear spin bath in ultra pure, single crystal diamond.

\section{Environmental autocorrelation functions and frequency spectra}\label{EnsAutoCorrFns}
In this section, we employ both the single cluster autocorrelation functions derived in section\,\ref{SingleClusterCorrFns} corresponding to secular (Eq.\,\ref{SecAutoCorrFn}) and non-secular (Eq.\,\ref{NonSecACF}) evolution of an individual cluster, together with the spatial statistics developed in the previous section, to determine the respective autocorrelation functions due to the sum of all clusters in the environment.

\begin{figure*}
  \centering
  \includegraphics[width=18cm]{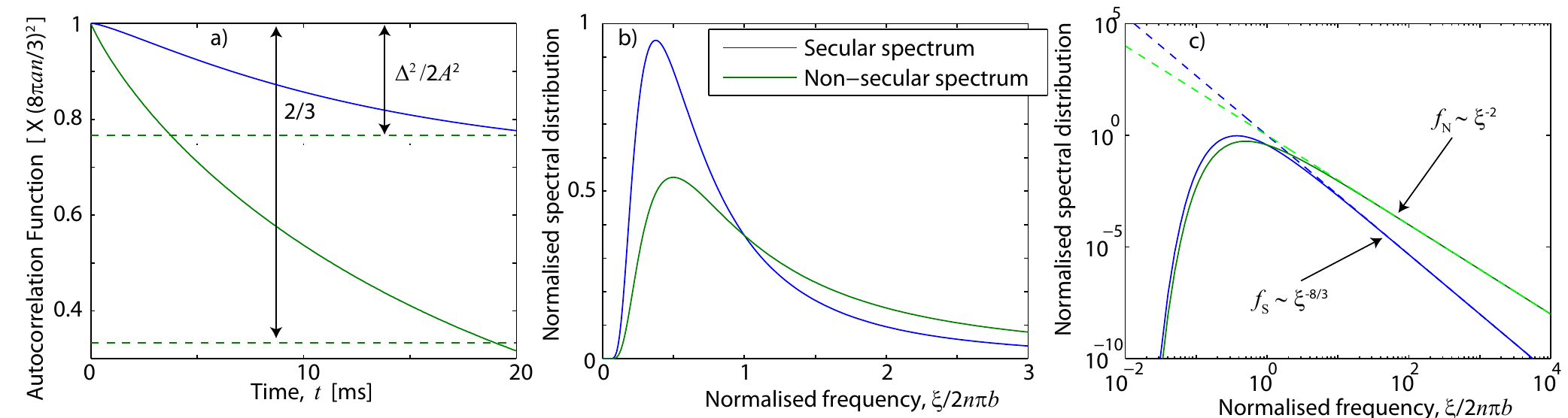}\\
  \caption{Normalised secular (blue) and non-secular (green) autocorrelation functions. (b) Spectral distributions associated with the secular (blue) and
  non-secular (green) spin-spin dynamics. (c) As in (b), but plotted on a log-log scale
  to show the high-frequency scaling of the spectra.}\label{AutoCorrFn}
\end{figure*}

         A comparison of the autocorrelation functions associated with the secular and non-secular regimes is plotted in Fig.\,\ref{AutoCorrFn}(a), using
   $ \left\langle A_z^2\right\rangle = \frac{4}{5}\left(\frac{4\pi}{3} an\right)^2$
   and $\left\langle\Delta_z^2\right\rangle \approx \frac{4}{5}(2an)^2$, from which we see that not only is the magnitude of the decay much greater in the non-secular case, but the non-secular decay is purely linear at $t=0$, indicating a self-similar, Markovian regime at all timescales. On the other hand, the secular case has zero-derivative at $t=0$, which is a consequence of the axial magnetisation of the cluster being conserved due to the dominant Zeeman coupling of the cluster constituents.

To obtain the
correct short time scaling of the secular function, we note that it is only the
small clusters ($r\ll R_0$) that contribute to short-time dynamics of the
system. The constituents of larger clusters communicate on much longer
timescales and hence manifest as an effectively DC signal. Another way to think of this is to view the ensemble averages taken over the spatial distributions (eq.\,\ref{SBANKdist}) as a
Fourier transform, with the conjugate (frequency) variable given by $\xi
\equiv 3br^{-3}/2$. The short time behaviour ($t\ll1/bn$) of the autocorrelation
function therefore corresponds to the high frequency behaviour ($\xi\gg bn$) of the spectral distribution. This is discussed further below. To
obtain the short time behaviour, we expand $\Delta_{x,y,z}^2$ about $r=0$, where, to
lowest order, where $\Delta_{x,y,z}^2\sim \mathcal{O}\left(r^2/R^8\right)$ (See Appendix\,\ref{AppMandN} for a complete description). Substituting this result, we find the collective autocorrelation function for the secular environment to be
\begin{eqnarray}
  \bigl\langle \mathcal{B}_{2}(t)\mathcal{B}_{2}(0) \bigr\rangle_\mathrm{S} &=& \frac{2}{5}\left(\frac{4}{3}\pi  a n\right)^2\biggl[  4+6 \cos ( \omega t )\nonumber\\
  &&-    \frac{1}{3}(7\cos ( \omega t )+4)M(t)\biggr]\label{SecAcn},
\end{eqnarray}where $M(t)$ is related to the secular magnetisation, as detailed in Appendix\,\ref{AppMandN}. This gives an autocorrelation time of
\begin{eqnarray}
  T_\mathrm{S} &=&  \frac{9}{4\pi ^2 b n  } \,\,\approx\,\, 9.6\,\mathrm{ms}.
\end{eqnarray}
 To leading order in $t$ we have
\begin{eqnarray}
  M(t) &\sim& \frac{4\pi \sqrt[3]{6} }{\Gamma \left(\frac{8}{3}\right)} (\pi b n t)^{5/3}-\frac{8 \pi  }{\sqrt{3} \Gamma \left(\frac{4}{3}\right)}\left(\pi b n t\right)^2. \end{eqnarray}

   On the other hand, the collective autocorrelation function for the non-secular environment may be computed exactly,
   \begin{eqnarray}
     \bigl\langle \mathcal{B}_{2}(t)\mathcal{B}_{2}(0) \bigr\rangle_\mathrm{N} &=&\frac{64}{9} \pi ^2 a^2 n^2\left[1-N(t)\right]\label{NonSecAcn}
   \end{eqnarray}where $N(t)$ is related to the non-secular magnetisation, as detailed in Appendix\,\ref{AppMandN}, giving the same autocorrelation time of
   \begin{eqnarray}
   T_\mathrm{N} &=& \frac{9}{4\pi ^2 b n }  \,\,\approx\,\, 9.6\,\mathrm{ms}.
   \end{eqnarray}
   To leading order, this results in a linear decay, given by
   \begin{eqnarray}
     N(t) &\sim& \frac{4}{9} \pi ^2 b n t.
   \end{eqnarray}

Whilst these regimes show an almost identical correlation time, the non-secular regime shows a much greater fluctuation magnitude (see Fig.\,\ref{AutoCorrFn}). This is a consequence of the fact that axial magnetisation must be conserved for a cluster in a secular regime, meaning that the central spin can only sense an effective field fluctuation if there is a difference in hyperfine couplings between the spins in that cluster. On the other hand, for the non-secular case, it is the cluster geometry that sets their quantisation axis, meaning that transitions that do not conserve axial magnetisation are now allowed.

It is important to note that whilst these results hold on timescales of order $T_\mathrm{S,N}$, they are not strictly correct for
timescales associated with cluster sizes smaller than the diamond lattice spacing, ie $T_\mathrm{min} \sim l^3/b \approx 300\,\mu$\,s. The $M(t)\sim t^{5/3}$ scaling at ultra-short timescales
is the result of the $\mathrm{p}(r)\sim r^2$ scaling of the probability density function associated with the distance between
neighbouring spins, which breaks down on length scales of $r\sim l$. By expanding Eqs.\,\ref{SecAutoCorrFn} and \ref{NonSecACF} on timescales of order $T_\mathrm{min}$, it is trivial to show
the initial quadratic scaling of both secular and non-secular autocorrelation functions.

Having obtained the autocorrelation functions of the effective magnetic
field, we can compute their Fourier transforms to give their corresponding
spectral distributions. We do this by noticing that the role of
 the conjugate frequency variable is played by $\xi\equiv B = 3br^{-3}/2$.
By transforming variables from $r$ to $\xi$, we identify the secular and
non-secular spectral distributions to be
\begin{eqnarray}
  f_\mathrm{S}(\xi) &=&  K_\mathrm{S}\left(\frac{3b}{2\xi}\right)^{2/3}\frac{b}{2\xi^2}\exp\left(-\frac{2n\pi  b}{\xi}\right),\nonumber\\
  f_\mathrm{N}(\xi) &=&  K_\mathrm{N}\frac{b}{2\xi^2}\exp\left(-\frac{2n\pi   b}{\xi}\right),
\end{eqnarray}respectively, where $K_\mathrm{S}$ and $K_\mathrm{N}$ are normalisation constants. The corresponding normalised spectra are plotted in
Figs.\,\ref{AutoCorrFn}\,(b)\,\&\,(c). The lack of any significant
spectral component near $\xi=0$ is symptomatic of the cutoff imposed by the
statistics associated with the size distribution of 2-spin clusters. That is,
since the exponential size cutoff associated with 2-spin clusters prohibits
arbitrarily large cluster sizes, there is no corresponding low frequency
region of the spectral density. Recall from the spatial statistics associated
with higher order cluster sizes (Eq.\,\ref{SBANKdist}), that each
successive $k^\mathrm{th}$ neighbour introduces an associated probability
distribution whose leading order behaviour scales as $r^{3k-1}/(k-1)!$. This,
in turn, contributes an additional factor of $1/\xi$ to the spectral
distribution for each successive order of clustering, with the modal
frequencies occurring at
\begin{eqnarray}
  \overline{{\xi}^{(k)}_\mathrm{S}} &=& \frac{2\pi n b}{k+\frac{5}{3}}\nonumber\\
   \overline{ {\xi}^{(k)}_\mathrm{N}} &=& \frac{2\pi n b}{k+1},
\end{eqnarray} for the secular and non-secular cases respectively. Incorporation of
successively higher orders of clustering will resolve the true low frequency
behaviour of the spectral distribution.

\section{Decoherence in ultra-pure single crystal diamond}
Having discussed the dynamic properties of the unperturbed spin bath environment, we move on to discussing the effect such an environment has on the coherence properties of a central NV spin. This analysis is performed by integrating the single cluster decoherence functions (see section\,\ref{SingleClusterQuantities}) over the $\mathbf{r}-\mathbf{R}$ domains as defined by the background field for the case of the naturally occurring 1.1\% $^{13}$C nuclear spin bath in ultra-pure single single crystal diamond. We initially discuss the free-induction behaviour of the NV spin in response to the influence of the surrounding spin bath for the limiting cases of high and low magnetic fields. We also mention the differences that arise between the FID behaviour of an ensemble of NV centres and that of a typical realisation of the surrounding environment.

We then move on to the analysis of the NV spin coherence in the presence of a spin-echo pulse sequence, noting
with reference to Figs.\,\ref{SBANuclearRegionsP}\,and\,\,\ref{SBANuclearRegionsManyBathsP} that the only regimes necessary for
consideration are, in order of decreasing magnetic field, ZSE, SZE and SEZ. We initially consider parameter regimes in which the decoherence is explained
exclusively by each of the respective decoherence functions, and then consider
the full dependence of the decoherence on the magnetic field strength. As with the FID case, we discuss the differences between the spin-echo decoherence of an NV ensemble and that of a single NV centre.

To determine the ensemble averaged decoherence behaviour, recall from that
the full spin echo envelope is given by the product of all envelopes due to
all clusters as weighted by the relevant spatial distributions,
\begin{eqnarray}
  L &=& \prod_iL_i,
\end{eqnarray}and taking the natural logarithm of both sides gives
\begin{eqnarray}
  \Lambda &=& -\sum_i\ln\left(L_i\right)\nonumber\\
  &\mapsto&  -\left\langle\ln\left(L_i\right)\right\rangle,
\end{eqnarray}where the final line above denotes the ensemble average taken
over all possible geometric cluster configurations. To
compute these averages, we employ a formal expansion for the natural
logarithm given by
\begin{eqnarray}
  \ln(1-x) &=& -\sum_{k=1}^\infty\frac{x^k}{k}\label{LogApprox}
\end{eqnarray}which holds for $-1\leq x<1$. This condition is automatically
satisfied, since $-1\leq L_\mathrm{ZS},\,L_\mathrm{SZ}\leq1$ and $0\leq L_\mathrm{ZSE},\,L_\mathrm{SZE},\,L_\mathrm{SEZ}\leq1$. For example, in the ZSE case, we have
\begin{eqnarray}
  \Lambda_\mathrm{ZSE} &=& \sum_{k=1}^\infty\frac{1}{k}\left[\sin ^2\left(\frac{B t}{4}\right)
   \sin ^2\left(\frac{\Delta_z}{4} t\right)\right]^k.
\end{eqnarray}
However, we are only interested in the leading order
behaviour of the ensemble averaged decoherence function, to which all terms
for $k\geq2$ do not contribute.

\subsection{Free-Induction Decay (FID)}\label{SectionCollFID}
In this section, we consider the FID behaviour of the NV spin due to the combined effect of all spin clusters in the environment. We firstly discuss the limiting regimes of both high and low magnetic fields as compared with the FID rate, and then move on to consider the full magnetic field dependence. We note that the timescales of the dipolar coupling in the environment are extremely slow (recall $T_\mathrm{S}=T_\mathrm{N}\sim10\,$ms) compared to the $T_2^* = 1-10\,\mu$s FID times discussed here. This allows us to ignore the dipolar evolution, meaning that there are only two regimes important to the study of FID behavior, depending on the relative strengths of their Zeeman coupling (Z) to the background field, and their hyperfine coupling to the NV spin (S). Consequently, quantities derived in a regime where the Zeeman coupling dominates are labeled `ZS' and similarly, quantities derived in a regime where the hyperfine coupling dominates are labeled `SZ'. Whilst this is a somewhat simplistic situation compared with the six possible regimes discussed in section\,\ref{SecTheorBG}, these considerations detail the transition from secular to non-secular hyperfine couplings with decreasing magnetic field, leading to faster decoherence, and are thus an important precursor to the spin-echo behaviour to be discussed in section \ref{SpinEchoDecoherence}.
\begin{figure}
  \includegraphics[width=\columnwidth]{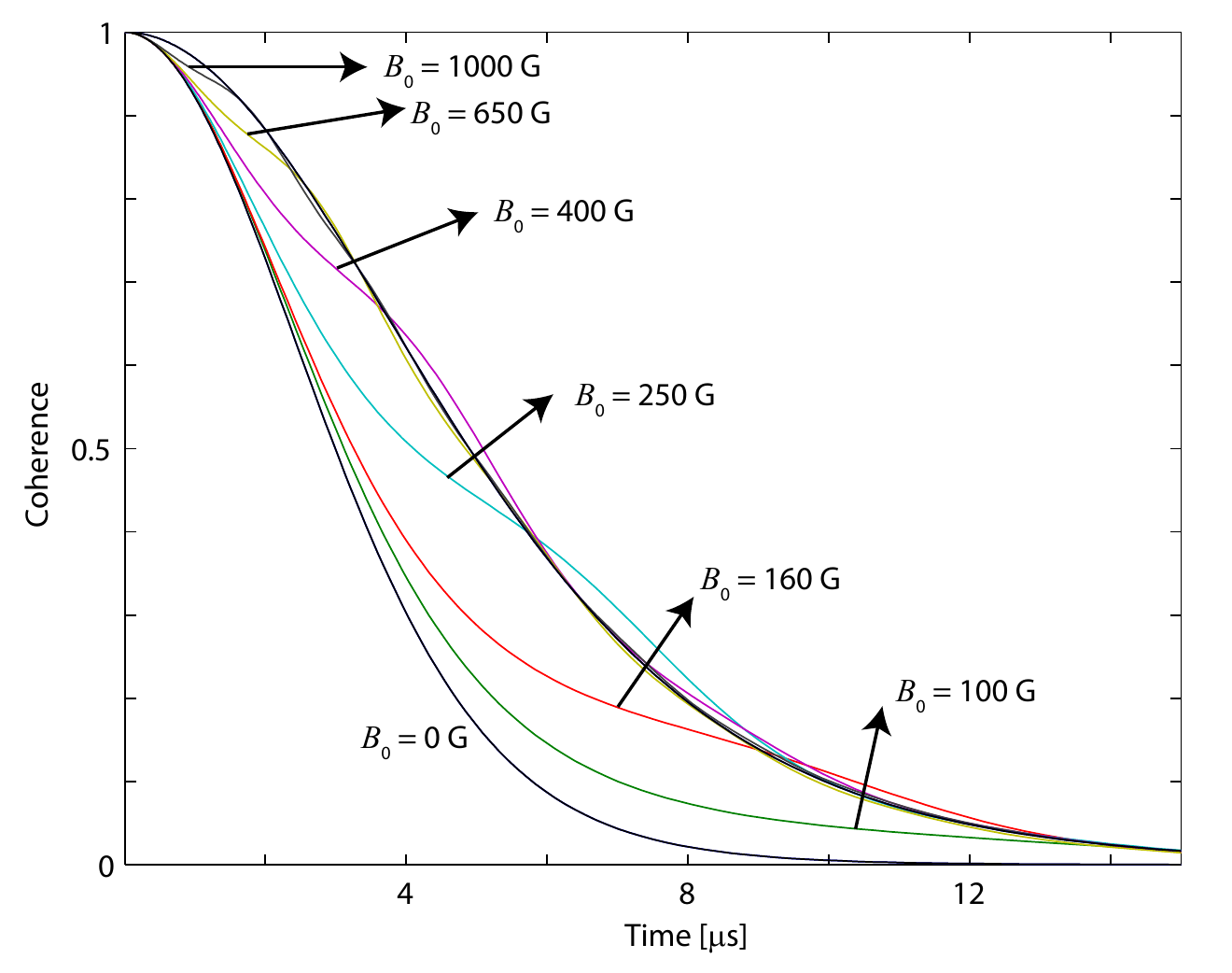}\\
  \caption{Plot showing the variation of the FID envelope with the strength of the background magnetic field. The transition from a SZ to a ZS regime occurs in the range of $100\,\mathrm{G}<B_0<1000\,\mathrm{G}$ at times associated with the Larmor period, $T_\omega\sim2\pi/\gamma_\mathrm{E}B_0$.}\label{SBAFIDEnvelopes}
\end{figure}

Fig.\,\ref{SBAFIDEnvelopes} shows the variation of the FID envelope with the strength of the axial background magnetic field, $B_0$, determined numerically, using a typical realisation for the spin bath distribution. From this, we see a monotonic increase of the FID time, $T_2^*$, with increasing $B_0$, which results in the transition of the effective hyperfine coupling strength from $A$ to $A_z$, as detailed in section\,\ref{SingleSpinDecoherence}. For magnetic fields of $100\,\mathrm{G}<B_0<1000\,\mathrm{G}$, we observe what is essentially a hybrid regime, in which the decoherence envelope looks like that of the pure ZS and SZ regimes for times above and below the Larmor period, $T_\omega\sim2\pi/\gamma_\mathrm{E}B_0$ respectively. This can be understood by recalling that spins in the SZ regime are those closest to the NV centre ($R\lesssim a/\omega$), and are thus responsible for the short time evolution of the NV spin. This contribution saturates beyond the Larmor period however, from which point onward, where the remaining time evolution is governed by the weaker coupling to spins in the ZS regime.

Using Eq.\,\ref{LogApprox}, the leading order behaviour of the FID
decoherence functions in the ZS and SZ regimes (Eqs.\,\ref{FIDSingleClusterLHigh} and \ref{FIDSingleClusterLLow}) is given by
\begin{eqnarray}
    \left\langle\Lambda_\mathrm{ZS}\right\rangle &\sim&  \left\langle 2 \sin ^2\left(\frac{ A_zt}{4}\right)\right\rangle\nonumber\\
      \left\langle \Lambda_\mathrm{SZ}\right\rangle &\sim&\left\langle 2 \sin ^2\left(\frac{A
   t}{4}\right)\right\rangle,\label{FIDDecoherenceFunctions}
\end{eqnarray}respectively.
\begin{figure*}
  \includegraphics[width=\textwidth]{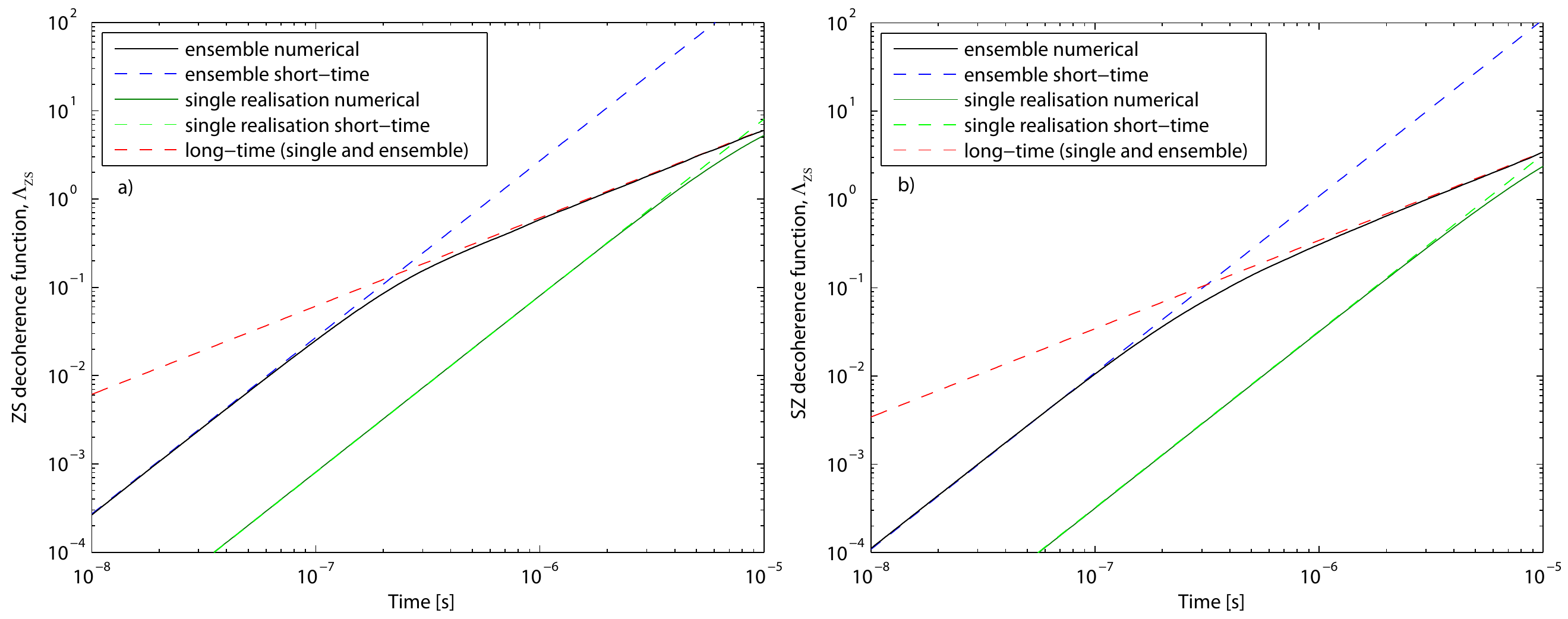}\\
  \caption{(a) Plot showing the behaviour of the FID decoherence function in the high field (ZS) regime, $\left\langle\Lambda_\mathrm{ZS}\right\rangle$. The solid black curve shows the numerical calculation over $10^6$ of the nuclear spin distribution, and the blue and red dashed curves depict the short/quadratic and long/linear analytic limits (see main text). The solid green curve shows the behaviour of a single realisation of the nuclear spin distribution, with the green dashed line showing the quadratic short time behaviour, that persists for times beyond $T_2^*$. The long-time limit of the single realisation behaviour is identical to that of the ensemble case. (b) As in (a), but for the low field (SZ) regime.}\label{SBAFIDDecoherenceFunction}
\end{figure*}
\subsubsection{FID at high magnetic fields (ZS)}
For magnetic fields greater than $\sim1000$\,G, every environmental spin will be in a regime where the Zeeman coupling is greater than the hyperfine coupling to the NV spin (SZ).
To visualise the FID behaviour in this regime, we firstly employ a numerical cluster expansion method (to zeroth order, since nuclear-nuclear interactions are not important), from which an ensemble average is performed over some $10^6$ realisation of the environmental spin distribution. The resulting ensemble averaged decoherence function, $\left\langle\Lambda_\mathrm{ZS}\right\rangle$ is plotted in Fig.\,\ref{SBAFIDDecoherenceFunction}(a). From this, we can see a linear scaling of the decoherence function for times approaching the FID time, $T_2^*$ (where $\left\langle\Lambda_\mathrm{ZS}\right\rangle\sim1$) and beyond, however a quadratic scaling is shown for times much shorter than this. Furthermore, the quadratic scaling is shown to persist for much longer in the case of individual realisations than for the ensemble averaged function. The analytic origins of these features are discussed in what follows.

To obtain the ensemble-averaged behaviour we must integrate over all possible outcomes of the environmental impurity distribution, which means that all lattice sites will be populated with equal likelihood. The long time behaviour arises from the low-frequency ($\sim1/T_2^*$) contributions to $\left\langle\Lambda_\mathrm{ZS}\right\rangle$, corresponding to spins more than a few lattice sites (roughly a nanometre) away from the NV, where the distribution effectively constitutes a continuum. The short-time behaviour, however, arises from spins that occupy the lattice sites surrounding the NV, where the bond-length of the diamond lattice, $l$, is important, ie on timescales of $T_\mathrm{d} = l^3/a\approx50$\,ns. These features may be reproduced by integrating $\Lambda_\mathrm{ZS}$ over $R$ from the diamond bond length, $l$, to $\infty$, and over $\Theta$ from 0 to $\pi$ (see Appendix\,\ref{AppFIDZS} for details). In the long time limit, where $t\gg t_d$, we find
\begin{eqnarray}
   \left\langle\Lambda_\mathrm{ZS}\right\rangle\biggl|_{t\sim T_2^*} &=& \frac{4 \pi ^2 }{9 \sqrt{3}}\, a n t,\label{FIDZSLong}
\end{eqnarray}showing a linear exponential free induction decay,
\begin{eqnarray}
   \left\langle L_\mathrm{ZS}\right\rangle &=& \exp\left(-\frac{t}{T_2^*}\right),
\end{eqnarray}where the free induction decay time is $T_2^* = \frac{9 \sqrt{3}}{4 \pi ^2 a n}  = 2.92\,\mu$s.
For times shorter than $T_\mathrm{d}$, we find a quadratic scaling in the decoherence function, given by
\begin{eqnarray}
   \left\langle\Lambda_\mathrm{ZS}\right\rangle\biggl|_{0<t\ll T_2^*} &\sim& \frac{2 \pi  a^2 n t^2}{15 l^3}\nonumber\\
   &=& \left(\frac{t}{960\,ns}\right)^2.
\end{eqnarray}Both the long and short-time analytic scalings are plotted together with the numerical results in Fig.,\,\ref{SBAFIDDecoherenceFunction}(a), showing excellent agreement.

Whilst this analysis accurately reproduces the experimental results for FID experiments conducted on NV ensembles (see, for example, ref.\,\onlinecite{Dob08}), experiments conducted on single NV centres exhibit a Gaussian shaped decay that typically persists as long as $T_2^*$. To reproduce this behaviour, we instead integrate $\left\langle\Lambda_\mathrm{ZS}\right\rangle$ from $R_0$ to $\infty$, as we would expect to find less than one impurity within a radius of $R_0$ from the NV centre. Following the same steps as in the ensemble case above, we find an initial quadratic scaling of
\begin{eqnarray}
  \left\langle\Lambda_\mathrm{ZS}\right\rangle\biggl|_{t\sim T_2^*}^\mathrm{(single)}  &=& \frac{8}{45} \pi ^2 a^2 n^2 t^2\nonumber\\
   &=& \left(\frac{t}{5.58\,\mu\mathrm{s}}\right)^2 ,
\end{eqnarray}followed by the same linear scaling as detailed in Eq.\,\ref{FIDZSLong}. The crossover point of these two regimes occurs at $t={5}/\left[{2 \sqrt{3} a n}\right]\approx11\,\mu$s, which is well past the point at which decoherence has occurred, showing that the FID behaviour of a single NV centre spin is dominated by a Gaussian decay.

\subsubsection{FID at low magnetic fields (SZ)}
Following on from the high-field limit of the previous section, we now move on to discussing the FID behaviour in the low field limit, the numerical result for which is shown in Fig.\,\ref{SBAFIDDecoherenceFunction}(b). This analysis is performed in an identical manner, save for the replacement of $A_z\mapsto A$, as dictated by Eq.\,\ref{FIDDecoherenceFunctions}. This leads to a slight increase in the FID rate but qualitatively identical behaviour as the ZS regime. For the long-time limit, we obtain
\begin{eqnarray}
   \left\langle\Lambda_\mathrm{SZ}\right\rangle\biggl|_{t\sim T_2^*} &=& \frac{\pi ^2}{18}   \left[6+\sqrt{3}\, \mathrm{arcosh}(2)\right]a n t,
\end{eqnarray}again showing a linear exponential free induction decay,
where the free induction decay time is now $T_2^* = {18}/\left[{\pi ^2 a n \left(6+\sqrt{3} \cosh ^{-1}(2)\right)}\right]  = 1.63\,\mu$s.
For times shorter than $t_d$, we again see a quadratic scaling in the decoherence function, given by
\begin{eqnarray}
   \left\langle\Lambda_\mathrm{SZ}\right\rangle\biggl|_{0<t\ll T_2^*} &\sim& \frac{\pi  a^2 n t^2}{3 l^3}\nonumber\\
   &=& \left(\frac{t}{607\,ns}\right)^2.
\end{eqnarray}
Finally, for the case of a typical single realisation of the SZ spin bath distribution, we find,
\begin{eqnarray}
  \left\langle\Lambda_\mathrm{SZ}\right\rangle\biggl|_{t\sim T_2^*}^\mathrm{(single)}   &=& \frac{4}{9} \pi ^2 a^2 n^2 t^2\nonumber\\
   &=& \left(\frac{t}{3.53\,\mu\mathrm{s}}\right)^2
\end{eqnarray}Both the long and short-time analytic scalings are plotted together with the numerical results in Fig.,\,\ref{SBAFIDDecoherenceFunction}(b), showing excellent agreement.

It is interesting to note that in the single-realisation case, taking the ratio of the FID times for the high and low field cases gives
\begin{eqnarray}
  \frac{T_{2,\mathrm{ZS}}^*}{T_{2,\mathrm{SZ}}^*} &=&  \sqrt\frac{5}{2}\approx1.58,
\end{eqnarray}in agreement with the results of ref.\,\onlinecite{Maz12}, whereas for the ensemble case we have
\begin{eqnarray}
  \frac{T_{2,\mathrm{ZS}}^*}{T_{2,\mathrm{SZ}}^*} &=&  \frac{3}{8} \left[2 \sqrt{3}+\cosh ^{-1}(2)\right]\approx1.80,
\end{eqnarray}showing the ensemble FID times experience a greater enhancement from an increased magnetic field strength than those of a single realisation of the bath impurity distribution.

\subsection{Spin-echo decoherence}\label{SpinEchoDecoherence}
We now move on to consideration of the spin-echo decoherence due to all spin clusters in the environment. Using Eq.\,\ref{LogApprox}, the leading order behaviour of the
decoherence functions for the ZSE, SZE and SEZ regimes are given by
\begin{eqnarray}
  \left\langle\Lambda_\mathrm{ZSE} \right\rangle &\sim& \left\langle\sin ^2\left(\frac{B t}{4}\right)\sin ^2\left(\frac{\Delta_z}{4} t\right)\right\rangle\nonumber\\
   \left\langle\Lambda_\mathrm{SZE} \right\rangle &\sim&\left\langle\sin ^2\left(\frac{B t}{4}\right) \sin ^2\left(\frac{\Delta }{4}\,t \right)\right\rangle\nonumber\\
 \left\langle \Lambda_\mathrm{SEZ} \right\rangle &\sim& \left\langle\frac{8}{15} \sin ^2\left(\frac{3 B t}{4}\right)
 \left[\sin ^2\left(\frac{A t}{2} \right)+\sin ^2\left(\frac{A t}{4} \right)\right]\right\rangle\nonumber\\
 \label{SEEnsAveBare}
\end{eqnarray}respectively.

\begin{figure*}
  \includegraphics[width=\textwidth]{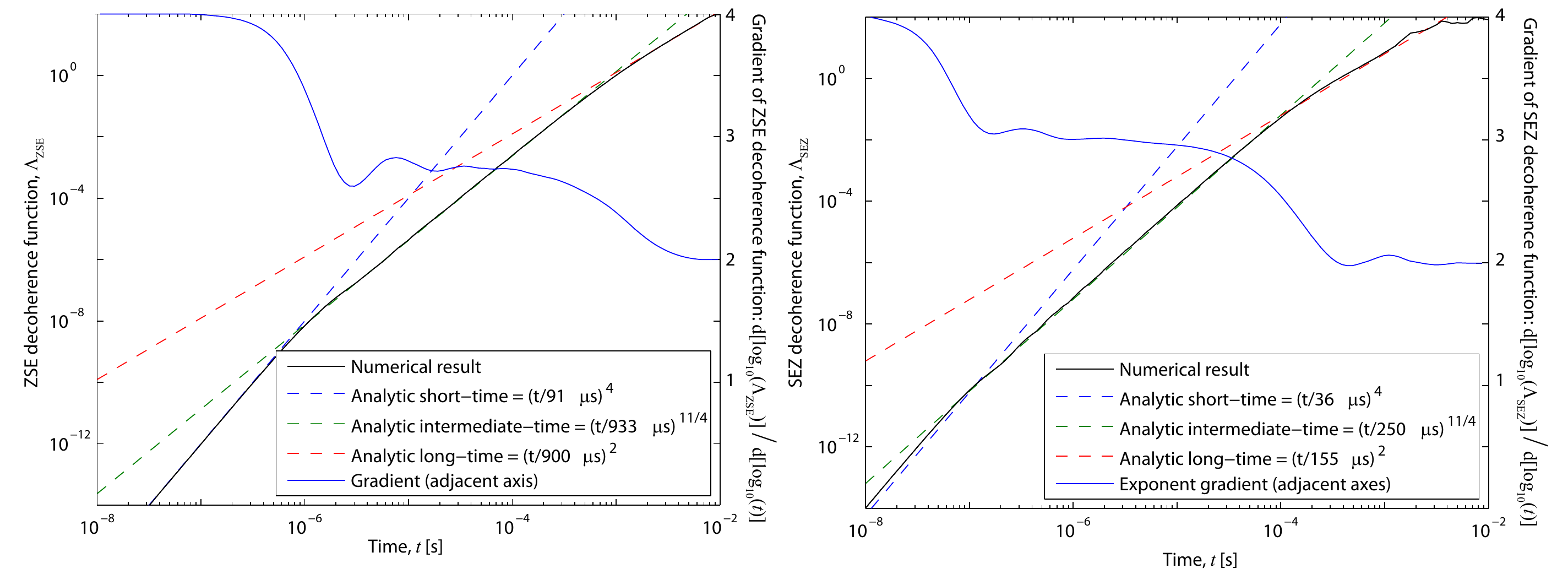}\\
  \caption{Plot showing the agreement between numerical and analytic results for the decoherence of an NV centre spin in a 1.1\% $^{13}$C nuclear spin bath. (a) Results for the ZSE regime. The black curve shows the numerical result for $\left\langle\Lambda_\mathrm{ZSE} \right\rangle$ computed using a $7^\mathrm{th}$ order cluster expansion method averaged over $10^6$ realisations of the surrounding impurity distribution. Analytic results for the short, intermediate and long-time scalings are given by the blue, green and red dashed curves respectively. The scaling exponent, found by numerically computing $k = \mathrm{d}\left[\ln\left(\left\langle\Lambda_\mathrm{ZSE} \right\rangle\right)\right]/\mathrm{d}\left[\ln\left(t\right)\right]$ is plotted as the blue solid curve (adjacent axis), and shows the transition from 4 to 2.75 to 2, as consistent with the analytic results. (b) As in (a) but for the SEZ regime. }\label{SBALambdaZSELattice}
\end{figure*}

\subsubsection{Spin echo decay at high magnetic fields (ZSE)}
At magnetic fields in excess of a few hundred Gauss, every $^{13}$C nuclear spin exists in
the ZSE regime. Thus, to compute the ensemble averaged decoherence function
for the high field case, we integrate $\Lambda_{\mathrm{ZSE}}$ over the spatial
distributions of the environmental spins. A numerical calculation using the cluster expansion method to 7$^\mathrm{th}$ order over $10^6$ realisations of the impurity distribution is shown in Fig.\,\ref{SBALambdaZSELattice}\,(a), from which we see a number of complex features. In particular, the scaling of $\Lambda_{\mathrm{ZSE}}$ with $t$ changes significantly between $t=0.1$ and 1\,ms from $\Lambda_{\mathrm{ZSE}}\sim \mathcal{O}\left(t^{2.75}\right)$ to $\Lambda_{\mathrm{ZSE}}\sim \mathcal{O}\left(t^{2}\right)$. At very short times, where $t\ll T_2^*$, we find $\Lambda_{\mathrm{ZSE}}\sim \mathcal{O}\left(t^{4}\right)$. The analytic origins of scaling are discussed in what follows.

We initially consider the long-time limit, where the decoherence function exhibits a quadratic scaling. Since dipolar interactions on these timescales correspond to impurity separations of 0.3\,nm, enclosing some 28 lattice sites, the environmental distribution essentially resembles a continuum in which spin impurities may adopt any position within the lattice. Such timescales are still much shorter than the environmental correlation time however, meaning that we may expand $\Delta_{z}$ for small $r$, giving
\begin{eqnarray}
  \Delta_z &\sim& \frac{3a r}{R^4} \biggl[\sin (\theta ) \sin (\Theta ) \left(1-5 \cos ^2(\Theta )\right) \cos (\phi -\Phi )\nonumber\\
  &&\,\,\,\,\,\,\,\,\,\,\,\,\,\,+\cos (\theta ) \cos (\Theta ) \left(3-5 \cos ^2(\Theta )\right)\biggr]\nonumber\\
  &\equiv& \frac{a_\alpha r}{R^4}.\label{DefOfaalpha}
\end{eqnarray} Integration of $\Lambda_{\mathrm{ZSE}}$ over
$0\leq r,\,R\leq\infty$ yields (see appendix\,\ref{AppZSE} for details)
\begin{eqnarray}
 \left\langle\Lambda_\mathrm{ZSE} \right\rangle\biggl|_{t\sim T_2}  
   &\sim& \pi\,\left(a n t\right)^{3/4} (b n t)^{5/4},\label{DecFnZSELong}
\end{eqnarray}
giving a spin-echo coherence time of
\begin{eqnarray}
  T_2^\mathrm{ZSE} &=& \left[\pi\,\left(a n \right)^{3/4} (b n )^{5/4}\right]^{-1/2} = 900\,\mu\mathrm{s},\label{LambdaZSECTS}
\end{eqnarray}
in excellent agreement with the numerical results of \cite{Ren12}.

 Prior to this quadratic scaling, $\Lambda_{\mathrm{ZSE}}$ exhibits a scaling of $\sim \mathcal{O}\left(t^{11/4}\right)$, which, as we detail below, is the result of spin impurities only being able to adopt discreet positions within the lattice. Such effects become important at short timescales, where the correspondingly small separation distances begin to approach the atomic spacing in the crystal.
We can reproduce the effect of this spacing by choosing a lower cutoff for $r$ of the diamond bond length, $l = 1.54\,{\mathrm{\AA}}$. We note that the integral of $\Lambda_{\mathrm{ZSE}}$ over $r$ is intractable for arbitrary integration terminals, however we may integrate from 0 to $\infty$ as above, and subtract the contribution 0 to $l$ by expanding the integrand for small $r$ as follows (see appendix\,\ref{AppSE} for details),
\begin{eqnarray}
   \left\langle\Lambda_\mathrm{ZSE} \right\rangle\biggl|_{T_2^*<t\ll T_2}
        &\sim& \frac{a^{3/4} b^2 n^2 t^{11/4}}{l^{9/4}}\nonumber\\
        & =& \left(\frac{t}{933\,\mu\mathrm{s}}\right)^{11/4}.\label{DecFnZSEInt}
\end{eqnarray}

This expression shows perfect agreement with the numerical calculation in terms of both scaling and magnitude, as depicted in Fig.\,\ref{SBALambdaZSELattice}. At longer timescales, the magnitude of $\left\langle\Lambda_\mathrm{ZSE} \right\rangle$ becomes much larger than the discreet correction term, and we simply recover the expression given in Eq.\,\ref{LambdaZSECTS}. This is to be expected: at long timescales, dipole-dipole interactions from spin impurities occupying adjacent sites essentially average out due to their high-frequency behaviour, whereas the more long range interactions become important. As the separation distance increases, the number of sites available for occupation essentially approaches that of a continuum.

Finally, to deduce the short-time quartic scaling, we again integrate over $R$ and $r$ from l to $\infty$, and and compute the formal short-time expansion, valid for $t:\,at/l^3\ll1\,\,\mathrm{and}\,\,bt/l^3\ll1$.

As the smallest possible separation distance is $l$, we are only justified in making this expansion for $t<50\,$ns in the ensemble case. The resulting expression is, to leading order (see appendix\,\ref{AppSE} for details),
\begin{eqnarray}
  \left\langle\Lambda_\mathrm{ZSE} \right\rangle\biggl|_{t\ll T_2^*} &=& \frac{1}{80}\left(\frac{\pi nab t^2}{l^3}\right)^2 \left[1-\frac{  \sqrt[6]{3} \pi  l  \left(4\pi n\right)^{1/3}}{360 \Gamma
   \left(\frac{4}{3}\right)}\right]\nonumber\\
   &=& \left(\frac{t}{91\,\mu\mathrm{s}}\right)^4.\label{DecFnZSEShort}
\end{eqnarray}

We note that some variation will exist between individual realisations of the impurity distribution, as most NV centres will not have spin impurities on adjacent lattice sites, meaning that the quartic scaling may persist for longer than in the ensemble case. To show this, we instead perform the $R$ integral from a lower cutoff, $R_0 = \left[3/(4\pi n)\right]^{(1/3)}$, defining the radius of a spherical volume in which we would expect to find less than one impurity on average for the ensemble case, meaning we would not expect impurities at distances closer than this in most individual cases. On the other hand, even the case of an individual distribution involves the NV coupling to many clusters, hence there is a large enough sampling of possible cluster configurations to justify an average over these configurations.
Integrating $\left\langle\Lambda_\mathrm{ZSE} \right\rangle$ over $R$ from $R_0$ to $\infty$, and over $r$ from $l$ to $\infty$, we find
\begin{eqnarray}
  \left\langle\Lambda_\mathrm{ZSE} \right\rangle\biggl|^\mathrm{(single)}_{t\ll T_2^*}  &=& \frac{1}{90}\pi^4a^2b^2n^4t^4 \left(\frac{ \sqrt[3]{6}   }{ l (\pi n)^{1/3}}-\frac{4 \sqrt{3} \pi }{9 \Gamma \left(\frac{4}{3}\right)} \right)\nonumber\\
   &=& \left(\frac{t}{393\,\mu\mathrm{s}}\right)^4,
\end{eqnarray}again showing a quartic scaling of $\left\langle\Lambda_\mathrm{ZSE} \right\rangle$ with $t$, but one that persists for some 10-100\,$\mu$s, as opposed to the 50\,ns for the ensemble case.

\subsubsection{Spin echo decay at moderate magnetic fields (SZE)}
At magnetic fields between 0.01\,G and 100\,G, every $^{13}$C nuclear spin
exists in the SZE regime. The procedure to compute the ensemble averaged
decoherence function is the same as for that above, however, we simply make
the substitution $\Delta_z\mapsto\Delta$, leading to what is essentially a redefinition of $a_\alpha$:
\begin{eqnarray}
  a_\alpha &\mapsto&  \frac{1}{\sqrt{3 \cos ^2(\Theta )+1}}\biggl(4 \cos (\theta ) \cos ^3(\Theta ) \nonumber\\
  &&+ \sin (\theta ) \left[2 \sin (\Theta )+\sin (3 \Theta )\right] \cos (\phi -\Phi )\biggr).\,\,\,
\end{eqnarray}
All subsequent results scale accordingly, the most important of which is $T_2^\mathrm{SZE} = 780\,\mu\mathrm{s}$. Other notable properties that emerge in this regime are the electron spin-echo envelope modulation (ESEEM) peaks, which manifest as periodic decays and revivals at half the Larmor frequency of the NV. As these effects do not represent any true decoherence, we defer their discussion until section\,\ref{ESEEMEns}.

\begin{figure*}
  \includegraphics[width=\textwidth]{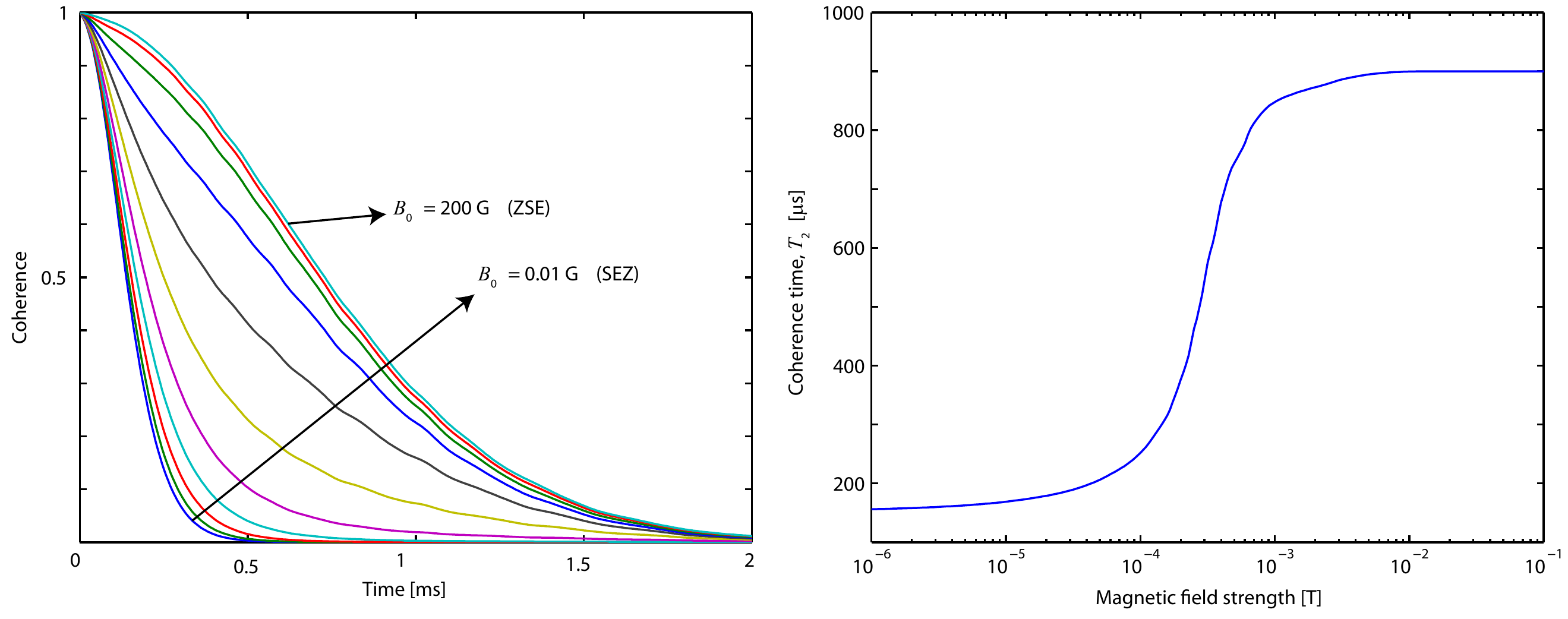}
  \\
  \caption{Magnetic field dependence of the NV spin coherence. (a) Coherence envelopes of an NV centre coupled to
  a naturally occurring 1.1\,\% $^{13}$C nuclear spin bath corresponding to external magnetic field strengths from 10\,$\mu$T
  to 10\,mT.
  (b) Plot showing the dependence of the coherence time on the strength of an external
  magnetic field.   }\label{SBAT2vsB0}
\end{figure*}

\subsubsection{Spin echo decay at low magnetic fields (SEZ)}
At magnetic fields below 0.01\,G, every $^{13}$C nuclear spin exists in the
SEZ regime. The transition of the fluctuation amplitude from $\Delta$ to $A$ effectively decouples the S-E from the E-E evolution in $\Lambda_{\mathrm{SEZ}}$, drastically changing the nature of the resulting decoherence. This allows for a convenient separation of the contribution of the hyperfine and dipolar coupling to the overall decoherence. A calculation of the decoherence function using a numerical cluster expansion to 7$^\mathrm{th}$ order is shown in Fig.\ref{SBALambdaZSELattice}\,(b) for $10^6$ realisations of the impurity distribution, from which a number of features are evident. As with the ZSE and SZE regimes, we see quartic and quadratic scalings at short and long times respectively, but in contrast to the other regimes, we se a cubic scaling at intermediate times. Furthermore, the coherence times exhibited by the SEZ regime are effectively an order of magnitude shorter than the other two regimes. The analytic origins of these features are explained in what follows.

As with the ZSE and SZE cases, we begin with the consideration of the long time dynamics of the SEZ regime. As before we need not worry about the discretised lattice at these timescales, and we therefore integrate $\Lambda_\mathrm{SEZ}$ over both $R$ and $r$ from 0 to $\infty$ (see Appendix\,\ref{AppSE} for details), giving
\begin{eqnarray}
  \bigl\langle\Lambda_\mathrm{SEZ}\bigr\rangle\biggl|_{t\sim T_2} &\sim& \frac{2}{15}\,ab\left(\pi ^2   n t \right)^2 \nonumber\\
  &=& \left(\frac{t}{155\,\mu\mathrm{s}}\right)^2\label{DecFnSEZLong}
\end{eqnarray}
Again, we see an overall Gaussian behaviour at long timescales, but a very different dependence on the hyperfine and dipolar dynamics of the environment ($\Gamma_\mathrm{ZSE,SZE}\sim a^{3/4}b^{5/4}$ vs. $\Gamma_\mathrm{SEZ}\sim ab$). This is a consequence of the changes in behavior of the environmental autocorrelation function as the environment transitions from secular to non secular dynamics. The corresponding coherence time of the SEZ regime is $T_2^\mathrm{SEZ} = 155\,\mu\mathrm{s}$.

 To derive the intermediate cubic scaling of $\bigl\langle\Lambda_\mathrm{SEZ}\bigr\rangle$, we note that, as was the case with $\bigl\langle\Lambda_\mathrm{ZSE}\bigr\rangle$ and $\bigl\langle\Lambda_\mathrm{SZE}\bigr\rangle$, the integral of $\sin ^2\left(\frac{3 B t}{4}\right)$ over the cluster size distribution (Eq.\,\ref{SBANKdist}) from $l$ to $\infty$ has no closed form. As such, we simply expand $\mathrm{P}(r)$ for small r, as detailed in Appendix\,\ref{AppSE}, giving
 \begin{eqnarray}
  \bigl\langle\Lambda_\mathrm{SEZ}\bigr\rangle_{T_2^*<t\ll T_2}  &\sim& \frac{\pi ^3 a b^2 n^2 t^3 \left[2 (4 \gamma -5) \pi  l^3 n+3\right]}{15 l^3}\nonumber\\
  &=& \left(\frac{t}{250\,\mu\mathrm{s}}\right)^3.\label{DecFnSEZInt}
\end{eqnarray}

To determine the short time scaling, we again integrate $\Lambda_\mathrm{SEZ}$ over $R$ and $r$ from l to $\infty$, and use the same short-time expansion as employed in the ZSE case. As the smallest possible separation distance is $l$, we are only justified in making this expansion for $t<50\,$ns in the ensemble case. The resulting expression is, to leading order (see appendix\,\ref{AppSE} for details), yielding an initially quartic dependence on time, given by
\begin{eqnarray}
  \left\langle \Lambda_\mathrm{SEZ} \right\rangle\biggl|_{t<T_2^*}  &\sim& \frac{5  }{48 }\left(\frac{ \pi a b n t^2}{ l^3}\right)^2 \left[2 (4 \gamma -5) \pi  l^3 n+3\right] \nonumber\\
   &=& \left(\frac{t}{36\,\mu\mathrm{s}}\right)^4,\label{DecFnSEZShort}
\end{eqnarray} where $\gamma\approx0.577$ is the Euler-Mascheroni constant.

\subsubsection{Full magnetic field dependence}
We now consider the full magnetic field dependence of the coherence time of
an NV centre exposed to a 1.1\%\,$^{13}$C nuclear spin bath. The full spin
echo envelope is the product of contributions from the 6 parameter regimes,
with the dominant contribution coming from the ZSE, SZE and SEZ regimes,
$ S_\mathrm{C13} \approx S_\mathrm{ZSE}S_\mathrm{SZE}S_\mathrm{SEZ}$,
which implies the full decoherence function is given by the sum
of decoherence functions due to each region,
\begin{eqnarray}
  \bigl\langle\Lambda_{\mathrm{C}13}\bigr\rangle &=& \bigl\langle\Lambda_\mathrm{ZSE}\bigr\rangle_{r>r_{\mathrm{Z}}}^{R>R_{\mathrm{Z}}} + \bigl\langle\Lambda_\mathrm{SZE}\bigr\rangle_{r>r_{\mathrm{Z}}}^{R<R_{\mathrm{Z}}} + \bigl\langle\Lambda_\mathrm{SEZ}\bigr\rangle_{r<r_{\mathrm{Z}}}^{R<R_{\mathrm{Z}}},\nonumber
\end{eqnarray}where the $R_z$ and $r_z$ quantities denote the Zeeman dependent integration domains in $r-R$ space.

Fig.\,\ref{SBAT2vsB0}(a) shows gradual transition of the decoherence
envelopes from the SEZ regime through to the ZSE regime with increasing
magnetic field. The full dependence of the corresponding coherence times, $T_2$, is
shown in Fig.\,\ref{SBAT2vsB0}(b), where we see that coherence times of an NV spin
coupled to the naturally occurring 1.1\% $^{13}$C nuclear spin bath can be
almost 1\,ms for magnetic fields in excess of 100\,G. Our results show excellent
agreement with the extensive numerical investigation conducted in ref.\,\onlinecite{Ren12}. The persistent Gaussian shape predicted by our theory is a
radical departure from currently accepted theories in the literature claiming
either a $\Lambda\sim\left(t/T_2\right)^3$ and
$\Lambda\sim\left(t/T_2\right)^4$ dependence irrespective of the physical
origin of the spin bath. We have shown here that the former is not valid for
the case of an NV centre immersed in a $^{13}$C nuclear spin bath, except
where spin densities are well below those currently realised experimentally. The latter is only valid in the short time limit, and may be explained as follows.

\subsubsection{Analysis of the electron spin echo envelope modulation (ESEEM)}\label{ESEEMEns}
\begin{figure*}
  \includegraphics[width=\textwidth]{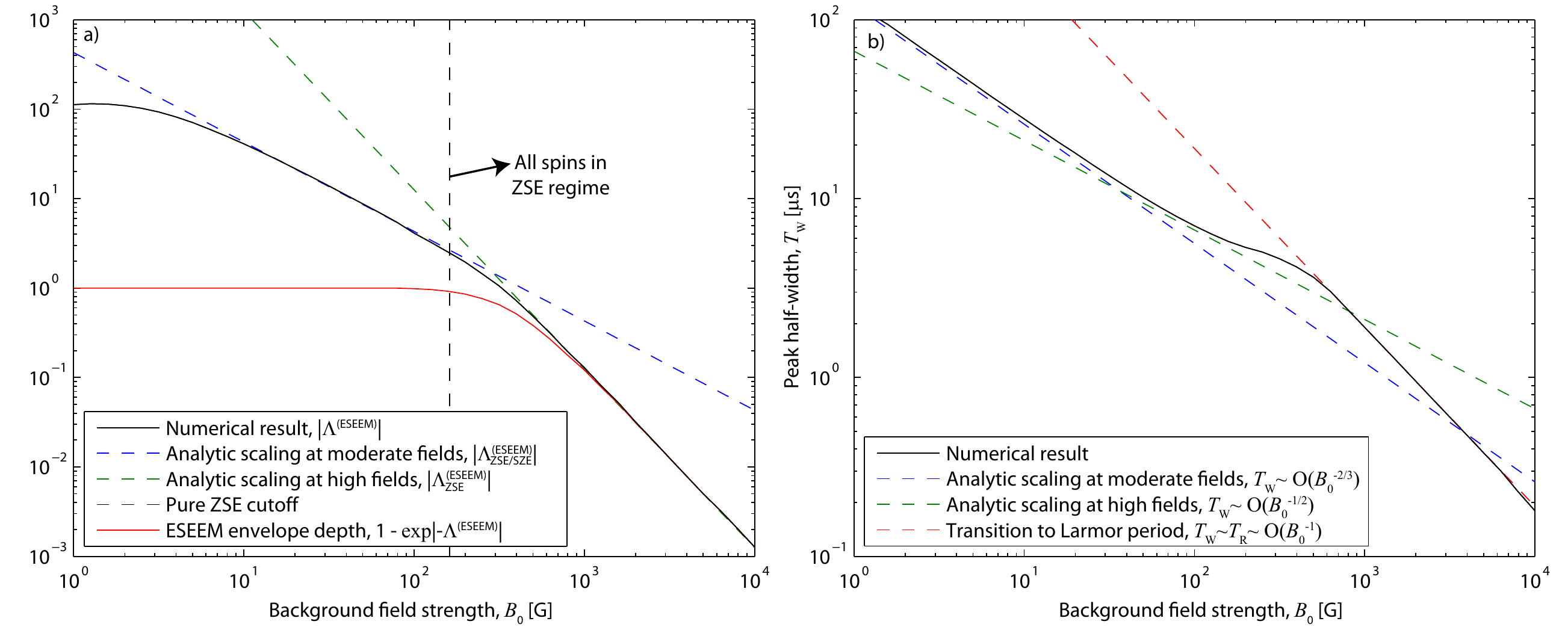}\\
  \caption{Magnetic field dependence of ESEEM features. (a) Plot detailing the magnitude of the modulation of the decoherence function. The numerical result is plotted in black, whereas the moderate and high field analytic limits are plotted in blue and green respectively. The saturation of the numerical result at low fields arises because the revival rate has decreased below the decoherence rate, which also saturates at low field. The red curve shows the magnitude of the corresponding spin-echo envelope contrast due to ESEEM, $1-L = 1-\exp\left(-\Lambda^\mathrm{(ESEEM)}\right)$, which saturates at unity when the decoherence function is greater than unity. (b) Plot detailing the dependence of the width of the ESEEM peaks on the magnetic field strength. The numerical result is plotted in black, and the low, moderate and high field analytic results are plotted in blue, green and red respectively. In the case of the high field result, the decay depth is less than unity, hence the width of the peaks is defined by the revival period instead of the decay time.}\label{SBAESEEM}
\end{figure*}

One of the key features observed in experiments conducted on NV spins in ultra-pure diamond is the emergence of decays and revivals in the spin-echo envelope at half the Larmor frequency of the $^{13}$C spins, an effect referred to as electron spin echo envelope modulation, or ESEEM. To this point we have only concerned ourselves with the decoherence arising from flip-flop processes in the bath and have ignored the ESEEM contribution to the evolution. In this section, we analyse this effect and show how resulting properties such as the revival frequency, decay depth and revival peak width depend on the background magnetic field strength. For background field strengths below approximately $B_0 = 1\,G$, the revival frequency, $\omega_\mathrm{R} = \frac{1}{2}\gamma_c B_0 = 1.75$\,kHz, is lower than the decoherence rate. As such, the notion of a decay depth no longer makes sense, as there will be no subsequent revival before decoherence has occurred. In the following, we derive the analytic origins of these scalings.

 Where the oscillations in the FID envelope occured at the Larmor frequency, $\omega_0$, the revival frequency during a spin-echo sequence is one half of the Larmor frequency, $\omega_\mathrm{R} = \frac{1}{2}\gamma_\mathrm{E}B_0 = 17.5\,\mathrm{MHz\,T^{-1}}\,B_0$. Some broadening of this effect will occur due to the distribution of axial dipolar couplings in the bath, leading to a perceived increase in the decoherence rate, however this effect will be addressed in the following section.

Next we detail the dependence of the depth of the decay valleys on the magnetic field strength. A numerical calculation of the maximum amplitude of the ESEEM component of the decoherence function, $\left|\Lambda^{\mathrm{(ESSEM)}}(t)\right|_\mathrm{max}$ is plotted in Fig.\,\ref{SBAESEEM}\,(a). From this, we see that the decay depths scale with the inverse-square of the magnetic field strength at high fields, but only with the inverse at moderate field strengths.

Recall from Eq.\,\ref{SEZPS} that in the ZSE limit, the ESEEM correction to $\Lambda_\mathrm{ZSE}$ due to a single nuclear spin is given by
\begin{eqnarray}
  \Lambda^{(\mathrm{ESEEM})}_\mathrm{ZSE} &=&    4\frac{ A_{x}^2+A_{y}^2 }{\omega ^2}\sin ^2\left[  \left(A_{z}+\omega \right)\frac{t}{4}\right] \sin ^2\left(\frac{t \omega }{4}\right).\nonumber\\\label{ZSEESEEM}
\end{eqnarray}
As this correction has been calculated in the $\omega\gg A$ limit, we must distinguish between cases where all spins are in the ZSE regime; and cases where distant spins from the NV are in the ZSE regime, but closer spins are in the SZE regime due to their dominant hyperfine interaction. For the former case, we have that the Zeeman coupling is greater than the most strongly coupled nuclear spin (ie where $B_0 > 162\,$G), giving
\begin{eqnarray}
  \left| \Lambda_\mathrm{ZSE}^\mathrm{(ESEEM)}\right|_\mathrm{max} &=& -4\left\langle\frac{ A_{x}^2+A_{y}^2 }{\omega ^2} \right\rangle\nonumber\\
    &=& -\frac{2}{15} \left(8\pi\frac{ a n}{\omega}\right)^2\nonumber\\
    &=& -\left(\frac{354\,\mathrm{G}}{B_0}\right)^2,
\end{eqnarray}thus reproducing the $\sim \mathcal{O}\left(B_0^{-2}\right)$ scaling of the numerical result, as plotted in Fig.\,\ref{SBAESEEM}\,(a).

For the case where the field strength is low enough to have spins in both the ZSE and SZE regimes, we must determine the contribution from both. The ZSE contribution to the decay may be determined by integrating Eq.\,\ref{ZSEESEEM} over only the spins in this regime. To determine the SZE contribution, we expand the ESEEM terms for $\omega\ll A$ as given in Eq.\,\ref{SEPZS},
\begin{eqnarray}
\Lambda^{(\mathrm{ESEEM})}_\mathrm{SZE}&=&4\frac{
   A_{x,1}^2+A_{y,1}^2}{A_1^2}\left[1
   -\frac{2 \omega  A_{z,1}}{A_1^2}\right]\nonumber\\
   &&\times\sin ^2\left(\frac{t \lambda _1}{4}\right) \sin ^2\left(\frac{t \omega }{4}\right),\label{SZEESEEM}
\end{eqnarray}
and integrate over only the spins in the SZE regime. The sum of these two contributions gives
\begin{eqnarray}
   \left|\Lambda_\mathrm{ZSE/SZE}^\mathrm{(ESEEM)}\right|_\mathrm{max} 
&=& \frac{32 \pi  a n}{5 \omega }+ \frac{448 \pi  \left(5 \sqrt{3} \pi -27\right) a n}{135 \omega }\nonumber\\
 &=& \frac{860\,\mathrm{G}}{B_0},
\end{eqnarray}again in agreement with the numerical result (see Fig.\,\ref{SBAESEEM}\,(a)).

Finally, we analyse the dependence of the decay widths, $T_\mathrm{w}$ on the magnetic field. The numerical results in plotted in Fig.\,\ref{SBAESEEM}\,(b) show these width to scale as $T_\mathrm{w}\sim B_0^{-0.67}$ at moderate fields, as consistent with the scaling of $B_0^{-0.63}$ in the numerical results of Ref.\,\onlinecite{Ren12}. At high fields, our numerical results show a slight change in this scaling for a brief period, with $T_\mathrm{w}\sim B_0^{-1/2}$. If the magnetic field is increased further, the decay depths will be less than unity (see Fig.\,\ref{SBAESEEM}\,(a)), meaning the widths will be effectively characterised by half the revival period, $T_\mathrm{W}\sim\frac{1}{2}T_\mathrm{R}$, showing an inverse linear dependence on the magnetic field strength, again consistent with the numerical results. We do not consider the low field regimes, as revivals are not visible prior to the onset of decoherence. The analytic origins of these results are discussed in the following.

In the high field (ZSE) limit, we expand Eq.\,\ref{ZSEESEEM} about any of the revival peaks, giving
\begin{eqnarray}
  \left\langle\Lambda^{(\mathrm{ESEEM})}_\mathrm{ZSE}\right\rangle &=& \left\langle\frac{1}{64} t^4 \left({A_{x,1}}^2+{A_{y,1}}^2\right) ({A_{z,1}}+\omega )^2\right\rangle\nonumber\\
  &\sim& \frac{1}{30} \pi ^2 a^2 n^2 t^4 \omega ^2,
\end{eqnarray}giving a decay width of $T_\mathrm{W} =  {61\,\mu\mathrm{s\,G^{1/2}}}/{\sqrt{B_0}}$

To find the decay widths at low fields, we integrate Eq.\,\ref{SZEESEEM}. This puts us in a regime where $\omega\ll an$, meaning we must integrate the resulting expression over $\mathbf{R}$, and then expand for $1/an\ll t\ll1/\omega$, giving
\begin{eqnarray}
  \left\langle\Lambda^{(\mathrm{ESEEM})}_\mathrm{\mathrm{SZE}}\right\rangle &=& \left\langle\frac{\omega ^2 t^2 \left({A_{x,1}}^2+{A_{y,1}}^2\right) \sin ^2\left(\frac{1}{4}A_1 t \right)}{4
   A_1^2}\right\rangle\nonumber\\
  &\sim& \frac{1}{192} \pi ^2 a n t^3 \omega ^2 \left(18-5 \sqrt{3} \cosh ^{-1}(2)\right)\nonumber,
\end{eqnarray}showing the revival peaks to have a cubic shape. The resulting peak width is then
\begin{eqnarray}
  T^{(\mathrm{SZE})}_\mathrm{W} 
  &=& \frac{121\,\mu\mathrm{s\,G^{2/3}} }{B_0^{2/3}}.
\end{eqnarray}Whilst this result is consistent with both the numerical work of our own, and that of ref.\,\onlinecite{Ren12}, it differs from the analysis given in ref.\,\onlinecite{Chi06}, which claims a quartic shape for the peaks, leading to a $T^{(\mathrm{SZE})}_\mathrm{W} \sim \mathcal{O}\left(B_0^{-1/2}\right)$ dependence at moderate fields. This analysis was performed using a short-time expansion with respect to both Zeeman and hyperfine couplings, however such an expansion is not valid in the SZE regime where short-time with respect to $T_\mathrm{R}\sim1/\omega$ is still long compared with $T_2^*\sim1/an$.

\subsubsection{Additional dephasing due to inhomogeneous broadening of the nuclear Larmor frequency}
In addition to the NV spin decoherence resulting from its entanglement with $^{13}$C spins, and their subsequent mutual interaction, there will be a dephasing effect due to the broadening of the $^{13}$C Larmor frequencies due to both their Zeeman interaction, and the distribution of their axial couplings to all other spins in the bath. Whilst this is technically not representative of any true decoherence process, it does nevertheless give the illusion of additional decoherence due to an increase in the spin-echo envelope decay rate.

To treat the `background' coupling of a given nuclear spin to other nuclei outside its associated cluster, we note that the total axial frequency shift will be a sum over that due to a large number of spins. As such, we expect the Larmor frequency, $\omega_i$, of a given spin to be normally distributed with mean $\omega_{0}$, due to the background magnetic field, and variance $\sigma$,
due to the axial couplings to all other spins in the bath,
\begin{eqnarray}
  \mathrm{P}\left(\omega\right) &=& \frac{1}{\sigma\sqrt{2\pi}} \exp\left(-\frac{\left(\omega-\omega_0\right)^2}{2\sigma^2}\right).\label{StaticDist}
\end{eqnarray}
Of course, a given realisation of this distribution only applies for a snapshot in time, however, it will only change on timescales of the order of $T_{\mathrm{S}}$ (see section\,\ref{EnsAutoCorrFns}), meaning that we can regard realisations of this distribution as being static on timescales of $T_2$ and shorter. Higher order corrections to this approximation can be made by employed the associated autocorrelation function of the bath spins, however there will be no resulting correction to leading order.

To show the additional dephasing effect of this broadening, we return to the ESEEM contribution to the ZSE regime, as detailed in Eq.\,\ref{ZSEESEEM}. Integrating this expression over Eq.\,\ref{StaticDist} with $1/\omega^2\sim1/\omega_0^2$, and noting that we may ignore terms like $\sin ^2\left(\frac{ \omega_0 t}{4}\right)$ since they do not contribute to the dephasing, we find the broadened ESEEM contribution to the ZSE regime to be
\begin{eqnarray}
\int_{-\infty}^\infty  \mathrm{P}(\omega)\Lambda^{(\mathrm{ESEEM})}_\mathrm{ZSE}\,\mathrm{d}\omega &=& 4\frac{ A_{x}^2+A_{y}^2 }{\omega ^2}\left(\frac{\sigma t}{4}\right)^2 \sin ^2\left(\frac{A_{z} t}{4}\right).\nonumber\\
\end{eqnarray}Summing over the hyperfine contribution from all spins as before, we find
\begin{eqnarray}
 \left\langle \Lambda^{(\mathrm{ESEEM})}_\mathrm{ZSE}\right\rangle &=& \frac{4 \pi ^2 a^2 n^2 t^2 \sigma ^2}{15 \omega _0^2}.
\end{eqnarray}To evaluate $\sigma^2$, we integrate over all axial nuclear-nuclear couplings, giving
\begin{eqnarray}
  \sigma^2 &\sim&  \frac{64}{45} \pi ^2 b^2 n^2,\nonumber\\
  &=& \left(89\,\mathrm{Hz}\right)^2
\end{eqnarray}This gives an additional Gaussian contribution to the decoherence envelope, with a corresponding decay rate of
\begin{eqnarray}
 \Gamma_\mathrm{ZSE}^\mathrm{ESEEM}  &=& \frac{3.1\,\mathrm{kHz\,G}}{B_0},
\end{eqnarray}showing a negligible contribution to the overall dephasing in the ZSE regime.

We apply the same approach to the ESSEM correction of the SZE decoherence function, as given in Eq.\,\ref{SZEESEEM}.
\begin{eqnarray}
  \int_{-\infty}^\infty  \mathrm{P}(\omega)\Lambda^{(\mathrm{ESEEM})}_\mathrm{SZE}\,\mathrm{d}\omega  &=& 4\frac{A_{x}^2+A_{y}^2 }{ A^2} \left(\frac{ \sigma t }{4} \right)^2\sin ^2\left(\frac{A t}{4}\right)\nonumber\\
\end{eqnarray}Again, summing over the hyperfine contribution from all spins as before, we find the resulting contribution to the full SZE decoherence function to be
\begin{eqnarray}
 \left\langle \Lambda^{(\mathrm{ESEEM})}_\mathrm{SZE}\right\rangle &=& \frac{\pi ^2}{3}  a n t \left(\frac{\sigma t}{16}\right)^2 \left[18-5 \sqrt{3} \mathrm{arcosh}(2)\right],\nonumber\\
\end{eqnarray}which gives an additional cubic contribution to the decoherence envelope, with a corresponding decay rate of
\begin{eqnarray}
 \Gamma_\mathrm{SZE}^\mathrm{ESEEM}  &=& 713\,\mathrm{Hz}.
\end{eqnarray}This additional cubic component of the decoherence function, despite not representing any true decoherence, gives a perceived reduction in the coherence time, reducing it slightly from $T_2^\mathrm{SZE} = 780\,\mu$s to $T_2^\mathrm{SZE} = 724\,\mu$s. As couplings between distant nuclei are not accounted for, this effect is generally not observed in numerical simulations.

\section{On the question of whether the quantum spin bath may be modeled as a classical magnetic field}\label{NoClassical}
 In treating the influence of the surrounding spin bath on a central spin, one commonly adopted approach\cite{Sou03,Sou03b,Tay08,Maz08b,Han08,Dob09,Hal10a,Lan10,Lan11,Wan12} is to replace the collective hyperfine field felt by the NV spin with a semi-classical magnetic field whose internal dynamics are dictated by the autocorrelation functions discussed above. This field, the operator of which
is denoted $\mathcal{B}(t)$,  will produce a time dependent Zeeman shift
given by $\mathcal{H}_z = \vec{\mathcal{B}} \cdot \vec{\mathcal{S}}\equiv
\mathcal{S}_zB(t)$ and a corresponding free-time evolution operator of
\begin{eqnarray}
  \mathcal{U}_f(t',t'') &=& e^{-i\phi(t',t'')}\bigl|1\bigr\rangle \bigl\langle 1 \bigr| + \bigl|0\bigr\rangle \bigl\langle 0 \bigr|
\end{eqnarray}where $\phi(t',t'') = \int_{t'}^{t''}B(t)\,dt$. Such an approach is potentially problematic, as it ignores the effect of the hyperfine couplings on the evolution of the nuclei, which as we have shown, are a critical component of this evolution.

The time evolution operator for a spin echo
experiment is
\begin{eqnarray}
\mathcal{U} &=&  \mathcal{U}_f(t/2,t)\mathcal{F} \mathcal{U}_f(0,t/2),
\end{eqnarray}
and for an arbitrary pulse sequence with pulses applied at $t_k =
\left\{t_1,t_2,\ldots,t_n\right\}$, we have
\begin{eqnarray}
  \mathcal{U}(t) &=&  \mathcal{U}(t_{n},t)\mathcal{F}\ldots\mathcal{F}\mathcal{U}(t_1,t_2)\mathcal{F}\mathcal{U}(0,t_1)\nonumber\\
        &\equiv& e^{-i\phi_1}\bigl|1\bigr\rangle \bigl\langle 1 \bigr|+e^{-i\phi_0}\bigl|0\bigr\rangle \bigl\langle 0 \bigr|,
\end{eqnarray}
where
\begin{eqnarray}
 \phi_1 &=& \phi(0,t_1)+\phi(t_2,t_3)+\ldots,\nonumber\\
 \phi_0 &=& \phi(t_1,t_2)+\phi(t_3,t_4)+\ldots
\end{eqnarray}are the accumulated phases of the $\bigl|1\bigr\rangle$ and $\bigl|0\bigr\rangle$ states respectively.

Using this semi-classical approach for an initial probe spin state of $\left|\psi_0\right\rangle =
\frac{1}{\sqrt2}\bigl(|0\rangle + |1\rangle\bigr)$, we find the in-plane
projection of the magnetisation to be
\begin{eqnarray}
  L &=& \mathrm{Tr}\left\{\left(\mathcal{S}_x + i\mathcal{S}_y\right)\mathcal{U}(t)\rho_0\mathcal{U}^\dag(t)\right\}\nonumber\\
  &=&\frac{1}{2}\exp\bigl[i\left(\phi^*_1(t)-\phi_0(t)\right)\bigr]\nonumber
      \end{eqnarray}This quantity is an average over the quantum degrees of
      freedom in the system, but we have not yet addressed the statistics of
      the field $B$. Firstly, we note that the amplitude of $B(t)$ at any
      given $t$ is a sum over a large number of sources and is therefore
      normally distributed. Furthermore, at room temperature, thermal
      energies are much larger than the coupling of environmental spins to
      static background fields, $k_BT \gg \omega\sqrt{S(S+1)}$, implying that
      $\langle B\rangle=0$ and hence $\left\langle\varphi\right\rangle=0$. To compute the ensemble average,
      $\bigl\langle S \bigr\rangle$, we make the substitution to the normally
      distributed variable $\varphi = \phi^*_1(t)-\phi_0(t)$, which, by
      definition, has standard deviation $\sqrt{\left\langle\varphi^2\right\rangle -\left\langle\varphi\right\rangle^2}$,
      giving $\bigl\langle S \bigr\rangle = \frac{1}{2}\exp\left(-\frac{1}{2}\left\langle\varphi^2\right\rangle\right)$
where
\begin{eqnarray}
  \varphi 
    &=&\sum_{k=0}\left[\int_{t_{2k}}^{t_{2k+1}}-\int_{t_{2k+1}}^{t_{2k+2}}\right]B(t')\,dt' 
\end{eqnarray}
and
\begin{eqnarray}
  \left\langle\varphi^2\right\rangle &=&
 \sum_{k=0}\sum_{j=0}\left[\int_{t_{2k}}^{t_{2k+1}}\,dt'-\int_{t_{2k+1}}^{t_{2k+2}}\,dt'\right]\nonumber\\
 &&\times\left[\int_{t_{2j}}^{t_{2j+1}}\,dt''-\int_{t_{2j+1}}^{t_{2j+2}}\,dt''\right] \bigl\langle B(t')B(t'')\bigr\rangle.\nonumber\\
 \label{PhaseVar}
\end{eqnarray}We therefore define the semi-classical analogue of the decoherence function, $\Lambda$, via
\begin{eqnarray}
  \Lambda &\equiv& \frac{1}{2}\left\langle\varphi^2\right\rangle.
\end{eqnarray}The problem of determining $\Lambda$ then reduces to finding an expression
for the autocorrelation function of the effective magnetic field, as was detailed in sections\,\ref{SingleClusterCorrFns}\,\,and\,\,\ref{EnsAutoCorrFns}.

The phase shift of the central spin will always depend on the pulse sequence
employed, but a certain degree of abstraction is achieved if we consider the
second integral of the environmental autocorrelation function, $G$, defined by
\begin{eqnarray}
  \frac{d^2}{dt^2}G(t) &=& \bigl\langle B(t')B(t'')\bigr\rangle
\end{eqnarray}It then becomes a simple exercise to show, using Eq.\,\ref{PhaseVar}, the
pulse sequence-specific decoherence functions are given by appropriate linear
combinations of dilated $G$ functions,
\begin{eqnarray}
    \Lambda_{\mathrm{fid}}&=&\frac{1}{2}\bigl\langle\phi^2_{\mathrm{fid}}(t)\bigr\rangle = \bigl\langle G(t)\bigr\rangle,\nonumber\\
        \Lambda_{\mathrm{se}}&=&\frac{1}{2}\bigl\langle\phi^2_{\mathrm{se}}(t)\bigr\rangle = 4\bigl\langle G({t}/{2})\bigr\rangle - \bigl\langle G(t)\bigr\rangle,
\end{eqnarray}and so on, showing that $G$ essentially plays the role of a classical `generalised decoherence function'.

Using the secular autocorrelation function\,(Eq.\,\ref{SecAcn}), we find the corresponding semiclassical spin-echo decoherence function for the ZSE regime to be
\begin{eqnarray}
  \Lambda &\sim& \frac{128}{45}\frac{ \sqrt[3]{2} \left(4-\sqrt[3]{2}\right) \pi ^{14/3} (ant)^2   (b n t)^{5/3}}{ 3^{2/3} \Gamma
   \left(\frac{14}{3}\right)}\nonumber\\
   &&-\frac{64 \pi ^5 (ant)^2 (bnt)^2 }{135 \sqrt{3} \Gamma \left(\frac{4}{3}\right)}\nonumber\\
   &=&\left(\frac{t}{120\,\mu s}\right)^{11/3}-\left(\frac{t}{180\,\mu s}\right)^{4},
\end{eqnarray}giving a coherence time of $127\,\mu$s. Notice that the effective magnetic field emanating from the lateral components of the nuclear spins have been suppressed to order $A^2_{x,y}/\omega^2$ by virtue of double integration with respect to $t$ of terms involving $\cos(\omega t)$, leaving only $z-z$ components of the effective field in the $\mathrm{Z}>\mathrm{E}$ limit. This is consistent with the suppression of lateral components seen in the transition from SZE to ZSE regimes in the quantum mechanical analysis of this work.

Similarly, using the non-secular autocorrelation function\,(Eq.\,\ref{NonSecAcn}), the semiclassical spin-echo decoherence function corresponding to the SEZ regime is
\begin{eqnarray}
  \Lambda &=& \frac{64}{243} \pi ^4 (ant)^2 bnt,
\end{eqnarray}which has an associated coherence time of $45\,\mu$s.
\begin{figure}
  \includegraphics[width=\columnwidth]{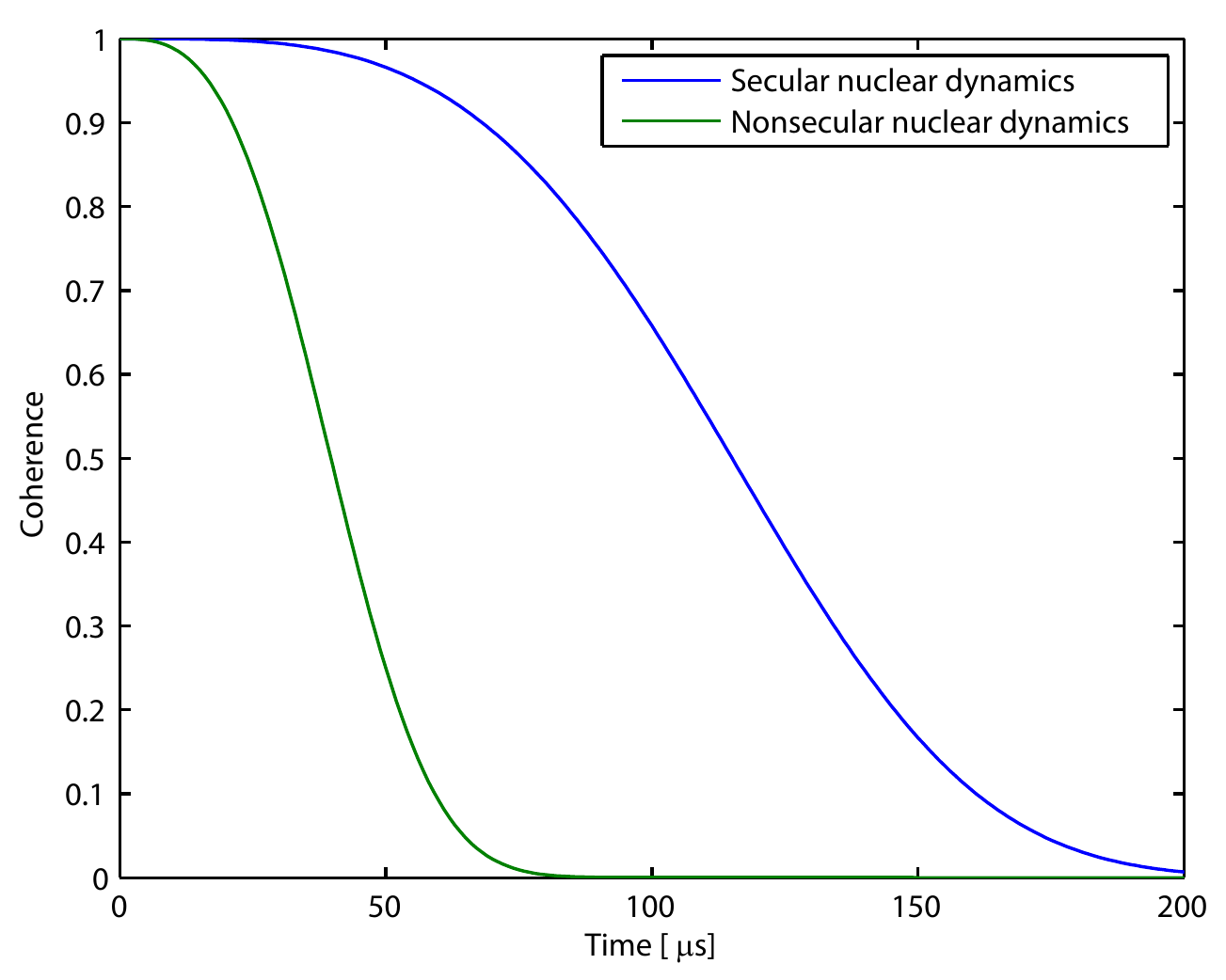}\\
  \caption{Plots showing the decoherence envelopes calculated using a semi-classical approach based on the determination of the autocorrelation function of the effective magnetic field from the hyperfine coupling of the environmental nuclei to the central spin. Qualitatively, this approach reproduces the effect of increasing the magnetic field, in that the decoherence rates are much faster for a non-secular environmental regime than those of a secular regime.
 However, the resulting coherence times are nearly an order of magnitude shorter than those computed with the quantum mechanical approach developed in this work, resulting from a mistreatment of the hyperfine couplings, ultimately showing that a semi-classical treatment of this problem is not adequate.}\label{SBASpinEchoFromACFn}
\label{SBASpinEchoFromACFn}
\end{figure}

The resulting spin-echo envelopes from this semi-classical analysis are plotted in Fig.\,\ref{SBASpinEchoFromACFn}. From these results, we see that the associated coherence times are almost an order of magnitude shorter than those deduced using the quantum mechanical approach developed in this work. We can attribute this discrepancy to the semi-classical approach not taking into account the back-action of the central spin on the environment, leading to a number of consequences.

Firstly, this allows for the environment to evolve freely under its own influence at all times, irrespective of the spin state (and hence the projected hyperfine field) of the NV, essentially doubling the effective fluctuation rate of the semi-classical spin bath field. In previous work\cite{Hal09}, we have shown that this increases the spin-echo decoherence rate for systems that exist in a slowly fluctuating regime, as is the case for the nuclear spin bath considered here.

Secondly, the semi-classical approach over estimates the dependence of the scaling of the temporal scalings of the resulting decoherence functions, leading to scalings of $\Lambda\sim t^{11/3}$ and $\Lambda\sim t^3$ associated with the secular and non-secular nuclear dynamics. This is again in contrast to the quantum mechanical results, which show quadratic scalings for the three parameter regimes applicable to this problem.

The third consequence is more critical. Whereas in the quantum mechanical analysis, the hyperfine coupling entered into the decoherence function as $\sin^2\left(A_z t/4\right)$, in the semiclassical case the hyperfine coupling manifests as $\left(A_zt\right)^2$, showing the latter to correspond to the short-time limit of the former. This means the two approaches only agree on timescales that are shorter than the FID time, implying that the semiclassical approach is not valid in analysing the spin-echo decay of an NV centre coupled to a nuclear spin.

\section{Conclusion}
In this work we have developed a purely quantum mechanical methodology by which to treat the decoherence of an NV centre spin coupled to a nuclear spin environment. This approach, based on the spatial statistics of environmental impurities, affords a natural decomposition of the bath into 6 distinct parameter regimes as defined by the relative strengths of the hyperfine, Zeeman and mutual dipolar coupling of the environmental spins. Such a rigorous treatment of this problem negates the need for any unjustified ad-hoc assumptions regarding the environmental NV-nuclear or nuclear-nuclear dynamics, allowing us to definitively resolve the analytic scalings of the associated decoherence functions in each of the relevant regimes. This has, consequently, allowed us to analytically derive the dependence of quantities such as coherence times and characteristic ESEEM features on the strength of a background magnetic. In doing so we have demonstrated excellent agreement with existing numerical and experimental work, whilst simultaneously correcting a number of analytic results in the literature.

\begin{acknowledgments}
The authors would like to thank D.A. Simpson, C.D. Hill, J. Wrachtrup, H. Fedder, F. Jelezko, L.P. McGuinness, B. Naydenov and V. Jaques for helpful
discussions.
This work was supported by the Australian Research Council under the Centre of Excellence scheme (project number CE110001027).
\end{acknowledgments}


\newpage
\appendix
\begin{widetext}
\section{Exact form of single cluster decoherence envelopes}\label{APPL}
For completeness, we include the full spin-echo decoherence envelope of the NV spin due to a 2-spin cluster undergoing a secular flip-flop process. This expression is exact, and the ZSE and SZE analytic limits have been employed in the main text. This envelope will contain contributions from flip-flop (lateral and longitudinal), precession (lateral only) and simultaneous flip-flop and precession processes. For clarity, we outline these contributions separately, whence
\begin{eqnarray}
  L_\mathrm{sec} &=&  1 + L_\mathrm{FF}+ L_\mathrm{P}+L_\mathrm{FF-P}.
\end{eqnarray}
The precession component is responsible for the decays and revivals at moderate magnetic fields, and is given by
\begin{eqnarray}
  L_\mathrm{P} &=&  -2\frac{ A_{x,1}^2+A_{y,1}^2}{\lambda _1^2} \sin ^2\left(\frac{ \lambda _1t}{4}\right) \sin ^2\left(\frac{ \omega t}{4}\right)
  -2\frac{ A_{x,2}^2+A_{y,2}^2}{\lambda _2^2}\sin ^2\left(\frac{ \lambda_2t}{4}\right) \sin ^2\left(\frac{ \omega t}{4}\right) \nonumber\\
  &&  +4\frac{  \left(A_{x,1}^2+A_{y,1}^2\right) \left(A_{x,2}^2+A_{y,2}^2\right)}{\lambda _1^2 \lambda _2^2}\sin ^2\left(\frac{ \lambda _1 t}{4}\right) \sin ^2\left(\frac{ \lambda_2 t}{4}\right) \sin ^4\left(\frac{ \omega t}{4}\right),\label{APPESEEMLP}
\end{eqnarray}
the flip-flop processes are responsible for the decoherence of the NV spin,
\begin{eqnarray}
   L_\mathrm{FF} &=& \frac{ A_{x,1} A_{x,2}+A_{y,1} A_{y,2}+\Omega _1 \Omega_2}{2 \lambda _1 \lambda _2}\sin ^2\left(\frac{B t}{4}\right) \sin \left(\frac{t \lambda _1}{2}\right) \sin \left(\frac{t \lambda _2}{2}\right)-\frac{1}{2} \sin ^2\left(\frac{B t}{4}\right) \left[1-\cos \left(\frac{t \lambda _1}{2}\right) \cos \left(\frac{t \lambda _2}{2}\right)\right]\nonumber\\
  &&  -2\frac{ \left(A_{x,2} A_{y,1}-A_{x,1} A_{y,2}\right){}^2+\left(\Omega _1 A_{x,2}-\Omega _2 A_{x,1}\right){}^2+\left(\Omega _1 A_{y,2}-\Omega _2 A_{y,1}\right){}^2}{\lambda _1^2
   \lambda _2^2}\sin ^2\left(\frac{B t}{4}\right) \sin ^2\left(\frac{t \lambda _1}{4}\right) \sin ^2\left(\frac{t \lambda _2}{4}\right),\nonumber\\
   &&\label{APPESEEMLF}
\end{eqnarray}
and the hybrid processes are described by
\begin{eqnarray}
 L_\mathrm{FF-P} &=&2\frac{  \left(\Omega _1 A_{x,2}-\Omega _2 A_{x,1}\right){}^2+\left(\Omega _1
   A_{y,2}-\Omega _2 A_{y,1}\right){}^2}{\lambda _1^2 \lambda _2^2}\sin ^2\left(\frac{B t}{4}\right) \sin ^2\left(\frac{t \lambda _1}{4}\right) \sin
   ^2\left(\frac{t \lambda _2}{4}\right) \sin ^2\left(\frac{t \omega }{4}\right)\nonumber\\
   &&-\frac{ A_{x,1} A_{x,2}+A_{y,1} A_{y,2}}{\lambda _1 \lambda
   _2}\sin ^2\left(\frac{B t}{4}\right) \sin
   \left(\frac{t \lambda _1}{2}\right) \sin \left(\frac{t \lambda _2}{2}\right) \sin ^2\left(\frac{t \omega }{4}\right)\nonumber\\
   &&+2\frac{
   A_{x,1}^2+A_{y,1}^2}{\lambda _1^2} \sin ^2\left(\frac{B t}{4}\right) \sin ^2\left(\frac{t \lambda _1}{4}\right) \cos ^2\left(\frac{t \lambda _2}{4}\right) \sin ^2\left(\frac{t \omega }{4}\right)\nonumber\\
   &&+2\frac{  A_{x,2}^2+A_{y,2}^2}{\lambda _2^2}\sin ^2\left(\frac{B t}{4}\right) \sin ^2\left(\frac{t \lambda _2}{4}\right) \cos ^2\left(\frac{t \lambda _1}{4}\right) \sin
   ^2\left(\frac{t \omega }{4}\right).
  \label{APPESEEMLPF}
\end{eqnarray}

\section{Exact forms of collective autocorrelation functions}\label{AppMandN}
The secular autocorrelation function for a two spin cluster is proportional to the difference in hyperfine couplings of the two nuclei. The leading order behaviour, corresponding to the high frequency limit associated with smaller cluster sizes comes from expanding these quantities for small $r$. In the ZSE regime, the magnitude of the fluctuating component only depends on the difference in the $z-z$ components of the respective hyperfine couplings,
\begin{eqnarray}
\Delta_z &=& \left|A_{z,1}-A_{z,2}\right|\nonumber\\
   &\sim& \frac{3 a r }{4 R^4}\biggl(\sin (\theta ) \bigl[\sin (\Theta )+5 \sin (3 \Theta )\bigr] \cos (\phi -\Phi )+\cos (\theta ) \bigl[3 \cos (\Theta )+5 \cos (3 \Theta )\bigr]\biggr),
\end{eqnarray}
whereas in the SZE limit, this magnitude depends on all couplings to the axial component of the NV spin,
\begin{eqnarray}
\Delta &=& \sqrt{A_{x,1}^2+A_{y,1}^2+A_{z,1}^2}-\sqrt{A_{x,2}^2+A_{y,2}^2+A_{z,2}^2}\nonumber\\
   &\sim& \frac{6 a r}{R^4}\frac{ \sin (\theta ) \bigl[2 \sin (\Theta )+\sin (3 \Theta )\bigr] \cos (\phi -\Phi )+4 \cos (\theta ) \cos ^3(\Theta )}{ \sqrt{6 \cos (2 \Theta )+10}}.
\end{eqnarray}

Employing these expansions and averaging over the spatial degrees of freedom using Eq.\,\ref{SBANKdist}, we find the collective secular autocorrelation function to be that given in Eq.\,\ref{SecAcn}, with the secular magnetisation function given by
\begin{eqnarray}
  M(t) &=& \frac{1}{3} \left(3
   \Gamma \left(\frac{2}{3}\right)+\sqrt[6]{6} \pi ^{11/6} (b n t)^{5/6} \left[\left(\sqrt{3}-3\right)
   \text{ber}_{\frac{5}{3}}\left(4 \sqrt{\frac{\pi }{3}} \sqrt{b n t}\right)-\left(3+\sqrt{3}\right) \text{bei}_{\frac{5}{3}}\left(4 \sqrt{\frac{\pi }{3}} \sqrt{b n t}\right)\right.\right.\nonumber\\
   &&\,\,\,\,\,\,\,\,\left.\left.+2 \sqrt{3}
   \text{bei}_{-\frac{5}{3}}\left(4 \sqrt{\frac{\pi }{3}} \sqrt{b n t}\right)-2 \sqrt{3} \text{ber}_{-\frac{5}{3}}\left(4 \sqrt{\frac{\pi }{3}} \sqrt{b n t}\right)\right]\right),\label{DefOfM}
\end{eqnarray}where $\mathrm{ber}(x)$ and $\mathrm{bei}(x)$ are the Kelvin functions, defined by the real and imaginary parts of $J_\nu\left(xe^{3\pi i/4}\right)$ respectively, and $J_\nu(x)$ is the $\nu^\mathrm{th}$ order Bessel function of the first kind.

Similarly, the collective non-secular autocorrelation function is given by Eq.\,\ref{NonSecAcn}, with the non-secular magnetisation function given by
\begin{eqnarray}
  N(t) &=& \frac{2}{9} \left[\pi  b n t\, \mathfrak{G}_{0,4}^{3,0}\left(\frac{1}{9} b^2 n^2 \pi ^2 t^2|
\begin{array}{c}
 -\frac{1}{2},0,\frac{1}{2},0
\end{array}
\right)-3\right],
\end{eqnarray}where $\mathfrak{G}$ is the Meijer G-function.

\section{Exact analytic forms of collective decoherence functions}
\subsection{Free induction decay}\label{AppFIDZS}
To obtain the FID decoherence functions for the ZS and SZ regimes, we start with the single-spin decoherence function as given by Eq.\,\ref{FIDDecoherenceFunctions}. Integration over $R$ gives
\begin{eqnarray}
  \int_l^\infty 4\pi n R^2 \Lambda_\mathrm{ZS}\,\mathrm{d}R &=& \frac{2}{3} \pi  l^3 n \left[\, _1\mathfrak{F}_2\left(-\frac{1}{2};\frac{1}{2},\frac{1}{2};-\frac{a^2 t^2 (3 \cos (2 \Theta )+1)^2}{64 l^6}\right)-1\right],\\
    \int_l^\infty 4\pi n R^2 \Lambda_\mathrm{SZ}\,\mathrm{d}R &=& \frac{2}{3} \pi  l^3 n \left[\, _1\mathfrak{F}_2\left(-\frac{1}{2};\frac{1}{2},\frac{1}{2};-\frac{a^2 t^2 (3 \cos (2 \Theta )+5)}{32 l^6}\right)-1\right],
\end{eqnarray}where $\mathfrak{F}$ is the generalised hypergeometric function.
Expanding these expressions to leading order for short and long times, and integrating over the angular degrees of freedom gives the collective decoherence functions discussed in section\,\ref{SectionCollFID} of the main text.

\subsection{Spin-echo}\label{AppSE}
\subsubsection{Analytic limits of the decoherence function in the ZSE regime}\label{AppZSE}
To obtain the ZSE decoherence function in the long time limit, we integrate $\left\langle\Lambda_\mathrm{ZSE}\right\rangle$ (Eq.\,\ref{SEEnsAveBare}) over the spatial degrees of freedom, $R$ and $r$, using Eq.\,\ref{SBANKdist},
\begin{eqnarray}
 \left\langle\Lambda_{\mathrm{ZSE}}\right\rangle  &=&  \frac{(-1)^{15/16} \left(\sqrt[4]{-1}-1\right) \pi ^{19/8} a_\alpha^{3/4}
   \sqrt[4]{b} n t \Gamma \left(-\frac{3}{4}\right)}{16 \sqrt[8]{2} 3^{3/8}} (b n t)^{3/8}
   \left[\left((-1)^{3/8}+i\right) \text{ber}_{-\frac{5}{4}}\left(2 \sqrt{\frac{2 \pi }{3}} \sqrt{b nt}\right)\right.\nonumber\\
  && +\left(\sqrt[8]{-1}+(-1)^{3/4}\right) \text{ber}_{\frac{5}{4}}\left(2 \sqrt{\frac{2 \pi }{3}} \sqrt{b n t}\right)
  +\left(1+(-1)^{7/8}\right) \text{bei}_{-\frac{5}{4}}\left(2 \sqrt{\frac{2 \pi}{3}} \sqrt{b n t}\right)\nonumber\\
   &&\left.+\left(\sqrt[4]{-1}+(-1)^{5/8}\right) \text{bei}_{\frac{5}{4}}\left(2 \sqrt{\frac{2 \pi }{3}} \sqrt{b n t}\right)\right]+\left(\frac{2 \pi }{3}\right)^{3/4} \sin \left(\frac{\pi }{8}\right) \Gamma \left(\frac{5}{4}\right)^2 (a_\alpha n t)^{3/4},
\end{eqnarray}where $a_\alpha$ is as defined in Eq.\,\ref{DefOfaalpha} of the main text. Expanding this expression for $t\gg1/an$ and $t\ll1/bn$, and integrating over the angular degrees of freedom gives the decoherence function for $t\sim T_2$ (Eq.\,\ref{DecFnZSELong}).

To obtain the behaviour of the decoherence function at intermediate times, we must make a correction for the diamond bond length to the nuclear-nuclear component of the evolution, whilst integrating over the hyperfine dynamics as above. The associated integral is generally intractable for arbitrary limits of $r$, however, as $r\ll n^{-1/3}$ we may approximate the probability distribution, Eq.\,\ref{SBANKdist}, by its leading order behavior, $\mathrm{P}(r)\sim4\pi n r^2$, giving \begin{eqnarray}
   \left\langle\Lambda_\mathrm{ZSE} \right\rangle &\sim& \left( \int_0^\infty\,\mathrm{P}(r) - \int_0^{l}\,4\pi r^2\right)\frac{2\pi  n}{3 } \sin \left(\tfrac{\pi }{8}\right) \Gamma \left(\tfrac{1}{4}\right) \left(\frac{a_\alpha r t}{2}\right)^{3/4}\sin ^2\left(\frac{3 B t}{4}\right)\,\mathrm{d}r\nonumber\\
   &=& \frac{ \pi ^2}{20} \sqrt{2 \left(\sqrt{2}-1\right)} l^{15/4} n^2  \Gamma \left(\frac{1}{4}\right) (a t)^{3/4} \left[\, _1\mathfrak{F}_2\left(-\frac{5}{8};\frac{3}{8},\frac{1}{2};-\frac{b^2 t^2}{16
   l^6}\right)-1\right].
\end{eqnarray}Expanding this expression for $t\gg1/an$ and $t\ll1/bn$, gives the expression for the ZSE decoherence function at intermediate times (Eq\,\ref{DecFnZSEInt}).

To obtain the behaviour of the decoherence function at short times, we make a similar adjustment for the bond length in the hyperfine interaction, and then expand for $t\ll1/an$ and $t\ll1/bn$, as given by Eq.\,\ref{DecFnZSEShort} of the main text.

\subsubsection{Analytic limits of the decoherence function in the SEZ regime}\label{AppSEZ}
To obtain the analytic limits of the SEZ decoherence function, we follow the same progression as in the ZSE limit above. The SEZ limit is somewhat simpler owing to the fact that the hyperfine and dipole-dipole processes are decoupled from one another in the single-cluster SEZ decoherence function (Eq.\,\ref{SEEnsAveBare}). Integration over the hyperfine component from $l<R<\infty$ yields
\begin{eqnarray}
   \int_l^\infty\left[\sin ^2\left(\frac{A t}{2} \right)+\sin ^2\left(\frac{A t}{4} \right)\right]\,R^2\mathrm{d}R &=&  \frac{2}{3}   n \left[l^3 \left(\, _1\mathfrak{F}_2\left(-\frac{1}{2};\frac{1}{2},\frac{1}{2};-\frac{a^2 t^2}{16 l^6}\right)+\cos \left(\frac{a t}{l^3}\right)-2\right)+a t \text{Si}\left(\frac{a
   t}{l^3}\right)\right],\nonumber\\\label{LambdaSEZoSE}
\end{eqnarray}

Integration over $r$ from $l$ to $\infty$ in the dipolar interaction gives
\begin{eqnarray}
  \int_l^\infty  \mathrm{P}(r) \sin ^2\left(\frac{3 B t}{4}\right)\mathrm{d}r &=&\frac{1}{4} \left[2-\pi  b n t\, \mathfrak{G}_{0,4}^{3,0}\,\left(\left.\frac{b^2 n^2 \pi ^2 t^2}{4}\right|
\begin{array}{c}
 -\frac{1}{2},0,\frac{1}{2},0
\end{array}
\right)\right]\nonumber\\
&&-\frac{1}{6} \pi  n \left[-6 b t \text{Si}\left(\frac{3 b t}{2 l^3}\right)-4 l^3 \cos \left(\frac{3 b t}{2 l^3}\right)+3 \pi  b t+4 l^3\right],\label{LambdaSEZoEE}
\end{eqnarray}
where $\mathrm{Si}$ is the Sine integral function, define by $\mathrm{Si}(x) = \int_0^xt^{-1}\sin(t)\,\mathrm{d}t$. Taking the relevant limits of the dipolar and hyperfine components, integrating over the angular degrees of freedom, and substituting into the definition of $\left\langle\Lambda_\mathrm{SEZ}\right\rangle$ (Eq.\,\ref{SEEnsAveBare}), we find the long, intermediate and short-time limits of the SEZ decoherence to be as given in Eqs.\,\ref{DecFnSEZLong}, \ref{DecFnSEZInt} and \ref{DecFnSEZShort} respectively.

\end{widetext}

\end{document}